\documentclass[prd,preprint,showpacs,preprintnumbers,amsmath,amssymb]{revtex4}
\usepackage{dcolumn}
\usepackage{bm}
\usepackage{epsfig}
\usepackage{amsmath}
\usepackage{subfigure}
\usepackage{amsfonts}
\usepackage{amssymb}
\usepackage{graphicx}
\usepackage{pstricks}
\usepackage{array}
\usepackage{hyperref}
\usepackage{multirow}
\usepackage{caption}
\usepackage{float}
\usepackage{color}
\usepackage{xcolor}

\DeclareSymbolFont{AMSa}{U}{msa}{m}{n}
\DeclareSymbolFont{AMSb}{U}{msb}{m}{n}
\let\Box\relax
\DeclareMathSymbol{\Box}{\mathord}{AMSa}{"03}

\newcommand{\be}{\begin{equation}}
\newcommand{\ee}{\end{equation}}
\newcommand{\bea}{\begin{eqnarray}}
\newcommand{\eea}{\end{eqnarray}}

\newcommand{\eq}[1]{\begin{equation}\begin{split} #1 \end{split}\end{equation}}

\newcommand{\Sum}{\displaystyle\sum\limits}
\newcommand{\ds}{\displaystyle}

\def \abs#1{\left\vert#1\right\vert}

\usepackage{pslatex}
\usepackage[T1]{fontenc}
\usepackage{graphicx}
\usepackage{epsfig}

\usepackage{calc}
\usepackage{ifthen}
\usepackage{subfigure}

\newcommand{\lh}{\lambda_h}
\newcommand{\ls}{\lambda_s}
\newcommand{\lhs}{\lambda_{hs}}

\begin{document}

\title{Higgs inflation and cosmological electroweak phase transition with N scalars in the post-Higgs era}
\author{Wei Cheng}
\affiliation{Department of Physics, Chongqing University, Chongqing 401331, China}
\author{~Ligong Bian}
\email{lgbycl@cqu.edu.cn}
\affiliation{Department of Physics, Chongqing University, Chongqing 401331, China\\
 Department of Physics, Chung-Ang University, Seoul 06974, Korea}

\begin{abstract}

We study inflation and cosmological electroweak phase transitions utilizing the Standard model augmented by $N$ scalars respecting a global $O(N)$ symmetry. 
We observe that the representation of the global symmetry is restricted by the inflationary observables and the condition of a strongly first order electroweak phase transition. 
Theoretical constraints including the stability, perturbativity and unitarity are used to bound the model parameter space.  The Electroweak precision observables and Higgs precisions limit the representation of the symmetry. We evaluate the possibility to simultaneously address the inflation and the dark matter after considering the experimental constraints from the future leptonic colliders. When the $O (N)$ symmetry respected by the N-scalar is spontaneously broken to the $O (N-1)$ symmetry, both the one-step and two-step SFOEWPT can occur within the inflation viable parameter regions, which will be tested by the future CEPC, ILC and FCC-ee. 
The relation between the number of Goldstones and the SFOEWPT condition depends on phase transition patterns. The situation of Goldstone faking neutrinos and contributing to the dark radiation are investigated.
 \end{abstract}

\maketitle
\preprint{}

\section{Introduction}

To our knowledge, the Standard model of particle physics (SM) is incapable to explain the three long-standing problems of particle physics and cosmology, i.e., the horizon, flatness and monopole problems of the Universe, the baryon asymmetry of the Universe (BAU), and the existence of the dark matter though the nature of which is unknown to us.
The cosmic inflation~\cite{Guth:1980zm,Starobinsky:1980te,Linde:1981mu} solves the first one successfully. The primordial density fluctuations generated during
the inflation can explain the formation of large scale structure of the universe observed by CMB~\cite{Ade:2015lrj}.
The inflation scenario is attractive when the inflaton field can play an important role in particle physics. A fascinating scenario is
the Higgs inflation~\cite{Bezrukov:2007ep,Barvinsky:2008ia,Xianyu:2014eba,Ellis:2016spb,Ellis:2014dxa}\footnote{For other non-minimally
coupled models see Ref.~\cite{Marzola:2016xgb,Artymowski:2016dlz,Racioppi:2018zoy}.}, where the inflaton is the SM Higgs being observed by LHC ~\cite{Aad:2012tfa,Chatrchyan:2012xdj}. There is a lot of debate on whether the Higgs inflation suffers from the unitarity problem at high scale around $\sim\mathcal{O}(10^{13})$ GeV~\cite{Burgess:2009ea,Barbon:2009ya,Barvinsky:2009ii,Bezrukov:2010jz,Burgess:2010zq,Hertzberg:2010dc,Calmet:2013hia,Burgess:2014lza,Bezrukov:2014ipa,Fumagalli:2016lls,Enckell:2016xse}\footnote{For a resolution of the controversy see Ref.~\cite{Ren:2014sya}. Ref.~\cite{Xianyu:2014eba} gave the first natural minimal SM Higgs inflation without invoking
nonminimal coupling and any new particles below UV-scale under asymptotical safe approach.}, which is beyond the scope of this paper.
Among various mechanisms to explain the BAU, the electroweak baryogenesis mechanism(EWBG) raises peoples interest due to the two essential ingredients of which are able to be tested at experiments. A strongly first order electroweak phase transition (SFOEWPT) as an essential ingredient usually requires the extension of the Higgs sector of the SM~\cite{Morrissey:2012db,DOnofrio:2014rug}, and the modified scalar potential could be detected at hadronic and leptonic colliders~\cite{Arkani-Hamed:2015vfh}. The additional CP violation, as another essential ingredient for the EWBG, can be probed indirectly with the electric dipole moment experiments. The CP violation study is beyond the scope of this paper though it may affect the phase transition.

To realize a SFOEWPT, one simplest and extensively studied approach is extending the SM with an additional real singlet scalar~\cite{Espinosa:2011ax,Profumo:2014opa,Ghorbani:2017jls,Ghorbani:2017lyk,Ghorbani:2018yfr,Alves:2018jsw,Alves:2018oct} or complex singlet scalar~\cite{Jiang:2015cwa,Chiang:2017nmu} through the Higgs portal. For the Higgs inflation with assistance of singlet scalars utilizing Higgs-portal interactions, we refer to Ref.~\cite{Lerner:2011ge,Ballesteros:2016euj,He:2014ora}\footnote{For the Higgs inflation with no-scale SUSY GUT, we refer to Ref.~\cite{Ellis:2016spb,Ellis:2014dxa}. }.  
For the Higgs inflation in the Higgs-portal scenarios, the typical quartic scalar couplings are required to be around $\sim\mathcal{O
}(10^{-1})$~\cite{Lerner:2011ge,Ballesteros:2016euj,Cheng:2018ajh}. 
To obtain a one-step type SFOEWPT, a relatively large Higgs portal quartic coupling is required~\cite{Curtin:2014jma}, which might lead to an unexpected theoretical problem, i.e.,
breaking the perturbativity, and unitarity at high scale. Therefore, largeness of the quartic couplings can not accommodate the successful inflation.  For the previous attempts to connect cosmic inflation and Electroweak phase transition (EWPT) in this case we refer to Ref. \cite{Tenkanen:2016idg,Enqvist:2014zqa,Cheng:2018ajh}. 

The straightforward approach to ameliorate the situation can be extending the SM by singlet scalars that respect $O(N)$ symmetry, then the one-step SFOEWPT can be realized with a lower magnitude of the Higgs-portal interaction $|H|^2 S_iS_i$($i=1,...,N$)~\cite{Espinosa:2007qk,Espinosa:2008kw,Kakizaki:2015wua,Hashino:2016rvx}. 
Previous studies of N-scalars with an exact $O(N)$ symmetry suggest that a one-step SFOEWPT can be realized with a relatively large N, which results in detectable gravitational wave signals with typical frequency of $\sim \mathcal{O}(10^{-3}-10^{-1})$Hz~\cite{Espinosa:2008kw,Kakizaki:2015wua,Hashino:2016rvx} and a substantial triple Higgs couplings deviation to be probed by the future colliders~\cite{Kakizaki:2015wua,Hashino:2016rvx}.
Therefore, we introduce an additional N hidden scalars that respect the $\mathcal{O}(N)$ symmetry, and investigate the possibility to accommodate inflation together with a SFOEWPT. Here, the additional $N$ hidden scalars might also alleviate the hierarchy problem through positive contributions to  radiative corrections of the Higgs boson mass and therefore satisfy the Veltman
conditions~\cite{Bian:2013wna,Bian:2014cja,Endo:2015ifa}.  As an additional benefit, the scalars can saturate DM candidate\footnote{See Ref.~\cite{Ge:2016xcq} for a study on dark matter and Higgs inflation.}. 
The previous studies of Ref.~\cite{Espinosa:2008kw} indicate that a one-step SFOEWPT cannot be addressed together with a correct DM relic density unless the quartic couplings of $|H|^2 S_iS_i$ and masses of $S_i$ are non-universal\footnote{The WIMP DM situation in the classical scale invariant N-scalars model with $O(N)$ symmetry~\cite{Endo:2015ifa} are ruled out for $N>4$ even for rather large quartic coupling~\cite{Endo:2015nba}. }. 
 In this work, we study both one-step and two-step EWPT, and evaluate the DM together with the inflation explanation. Our study shows that the Electroweak precision observables (EWPOs) constraints invalidate the inflation explanation when $N>4$ for the $\mathcal{O}(N)$ scalars, that shout down the window to accommodate a SFOEWPT. We further explore the scenario wherein the $\mathcal{O}(N)$ symmetry is spontaneously broken to the $\mathcal{O}(N-1)$ symmetry. There are $N-1$ Goldstones that can fake the effective neutrinos.
 The possibility of Goldstones contributing to dark radiations provided they gain masses from non-renormalizable gravity effects will be estimated.
Our results demonstrate that, after considering the theoretical constraints and the current Higgs precisions, the inflation explanation and a SFOEWPT can be reached in certain parameter spaces for both one- and two- step phase transitions.

The paper is organized as follows:
In Section.~\ref{sec:mod}, we introduce the model including the case of the scalars respecting the $O(N)$ symmetry and the scenario where the $\mathcal{O}(N)$ is spontaneously broke to the $\mathcal{O}(N-1)$, the relevant theoretical constraints and the Higgs precision constraints are explored.
Cosmological implications to be studied including inflation, electroweak phase transition, dark matter and dark radiations are given in Sec.~\ref{sec:cosimp}. The numerical results for both the $\mathcal{O}(N)$ and $\mathcal{O}(N-1)$ scenarios are presented in Sec.~\ref{sec:infptdm}. We conclude with Sec.~\ref{sec:conc}.

\section{The Models}
\label{sec:mod}
In this work, we study two scenarios of $N$ singlet scalars extended SM.
In the case of the $N$ singlet scalars ($S$) with $O(N)$ symmetry, the $O(N)$ symmetry might break at finite temperature and restore at the zero temperature.
Another scenario is that the $O(N)$ is spontaneously broken to $O(N-1)$ at zero temperature, we use ``$s$" rather than ``$S$" to differentiate it from the $O(N)$ scenario.

 For the $O(N)$ scenario, the zero temperature tree-level potential
is given by
\begin{eqnarray}
V_0(H, S) &=& -\mu _h^2 H^\dag H+\lambda_h |H^\dag H|^2+\frac{\mu_s^2 }{2}S_iS_i+\frac{\lambda_s }{4}(S_iS_i)^2+\frac{1}{2} \lambda_{hs}|H|^2 S_iS_i\;,\label{eq:lag}
\end{eqnarray}
with $H^T=( G^+,
(v+ h + iG^0)/\sqrt{2})$. After the spontaneously symmetry breaking of the Electroweak symmetry, the mass term of $S_i$ is given as $m_{S_i}^2=\mu_s^2+\lambda_{hs}v^2/2$.
For the $O(N\to N-1)$ scenario,
the minimization conditions of the potential can be obtained when EW symmetry is broke and the $O(N)$ being broke along the direction of $s$, with other directions being $s_i$ ($i=1,...,N-1$)\footnote{The breaking of $O(N)$ can happen in any direction, means we can have any of $s_i$ with $i=1,2,...,N$ obtain VEV, here we assume $O(N)$ breaks in $s_N$ direction.},
\begin{eqnarray}
\frac{dV_0(h,s,A)}{dh}\big{|}_{h=v}=0,~\frac{dV_0(h,s,A)}{ds}\big{|}_{s=v_s}=0\;,
\end{eqnarray}
which give rise to $\mu_h^2=\lambda_h v^2+\lambda _{hs} v_s^2/2,\mu_s^2= -(\lambda_{hs} v^2/2+\lambda_s v_s^2)$. The mass matrix is given by
\begin{eqnarray}
\mathcal{M}^2=
\bigg(\begin{array}{*{20}{c}}
{2{v^2}\lambda_h}&{vv_s\lambda_{hs}}\\
{vv_s\lambda_{hs}}&2{v_s^2\lambda_s}
\end{array}\bigg)\; .\label{eq:mM}
\end{eqnarray}
In order to diagonalize the mass matrix, we introduce the rotation matrix $R= \left( (\cos\theta ,\sin\theta ) ,( -\sin\theta ,\cos\theta ) \right)$ with $\tan2\theta  =  - {\lambda_{hs}vv_s}/{(\lambda_h{v^2} - \lambda_s v_s^2)}$ to relate the mass basis and field basis,
\begin{eqnarray}
\bigg(\begin{array}{*{20}{c}}
h\\
s
\end{array}\bigg)=
\bigg(\begin{array}{*{20}{c}}
\cos\theta&\sin\theta\\
-\sin\theta&\cos\theta
\end{array}\bigg)\bigg(\begin{array}{*{20}{c}}
h_1\\
h_2
\end{array}\bigg)\;.\label{eq:rescale}
\end{eqnarray}
The mass squared eigenvalues are
\begin{eqnarray}\label{eq:msg}
m_{{h_1,h_2}}^2 =\lh v^2+\ls v_s^2\mp \frac{ \ls v_s^2 - \lh v^2}{\cos 2 \theta}\;.
\end{eqnarray}
Identify the $h_1$ being the 126~GeV SM-like Higgs boson, and requiring the $h_2$ is dominated by $s$ set $\cos\theta>1/\sqrt{2}$. Here, we note that the situation of $m_{h_2}>m_{h_1}$ and $\abs{\theta}<\pi/4$ correspond to $\lambda_s v_s^2>\lambda_h v^2$.
The quartic couplings can be expressed as functions of the Higgs masses, $v$, $v_{s}$ and the mixing angle $\theta$,
\begin{eqnarray}
\lambda_h&=&\frac{ m_{h_2}^2\sin^2\theta+m_{h_1}^2\cos^2\theta}{2 v^2},\\
\lambda_s&=&\frac{  m_{h_2}^2\cos^2\theta+m_{h_1}^2\sin^2\theta}{2 v_s^2},\\
\lambda_{hs}&=&\frac{\left(m_{h_2}^2-m_{h_1}^2\right)\sin2\theta}{2 v v_s}.\label{eq:coupling}
\end{eqnarray}
We note that in our parameterization, the positiveness of the squared Higgs mass eigenvalues (given in Eq.~\ref{eq:msg}) is justified when the determinant of the Hessian matrix (Eq.\ref{eq:mM}) is positive, which leads to $4\lambda_h\lambda_s-\lambda_{hs}^2>0$, and results in a condition of $m_{h_2}^2m_{h_1}^2/(v^2v_s^2)>0$ after considering the Eq.~\ref{eq:coupling}.

The number of scalars($N$) or Goldstones ($N-1$), and scalar quartic couplings $\lambda_{s,hs,h}$ in the interaction basis for the $O(N)$ and $O(N\to N-1)$ scenarios will be constrained by the perturbativity and unitarity, stability, and Higgs precisions, as well as EWPOs.
The parameter spaces will be further restricted by the inflationary observables and the condition of SFOEWPT, which will be studied in the following sections. As will be studied in the following sections,
in both the $O(N)$ and the $O(N\to N-1)$ scenarios, the number of $N$ will be bounded by the condition of the slow-roll inflation and a SFOEWPT. For the $O(N\to N-1)$, it means the number
of Goldstones is bounded. As will be shown later, for the light extra Higgs mass, the dark radiation set bounds on the number of Goldstones in the $O(N\to N-1)$ scenario.

\subsection{Theoretical constraints}
\label{sec:thcons}

Firstly, due to the additive property of the scalar quartic couplings contribution to the beta functions (see Appendix.~\ref{sec:RGEs}), one need to aware the possible perturbativity problem at high scale when one performs the inflation analysis.
We impose the following conditions to preserve the perturbativity,
\bea
 |\lambda_h|<1,~ |\lambda_s|<\sqrt{4\pi},~ |\lambda_{hs}|<\sqrt{4\pi}\;.
\eea
The perturbative unitarity condition is obtained by requiring the absolute value of the s-wave $2\to2$ scattering amplitudes among longitudinal gauge bosons and scalars being smaller than $1/2$.
Which set bounds on scalar quartic coupling of the tree-level potential for $O(N)$ and $O(N\to N-1)$ scenarios as follows~\cite{Hashino:2016rvx},
\bea
\frac{1}{32\pi}\left(3\lambda+(N+2)\lambda_s+\sqrt{(3\lambda-(N+2)\lambda_s)^2+4 N\lambda_{hs}^2}\right)<\frac{1}{2}\;.
\eea
To prevent the unbounded from bellow of the scalar potential, the vacuum
stability conditions should be satisfied,
 \bea
 \lambda_h > 0,~ \lambda_s> 0, ~ \lambda_{hs} >0 {~\rm or} ~ \lambda_{hs}> - 2\sqrt{\lambda_h\lambda_s}\;.
 \eea
 Here, for completeness, the last condition include both the scenarios of $\lambda_{hs}>0$ and $\lambda_{hs}<0$. The third equation of the Eq.~\ref{eq:coupling} indicates that the scenario of $\lambda_{hs}>0$ corresponds to $m_{h_2}>m_{h_1}$ in the parameter region of $0<\theta<\pi/4$. 

\begin{figure}[!htp]
\begin{center}
\includegraphics[width=0.4 \textwidth]{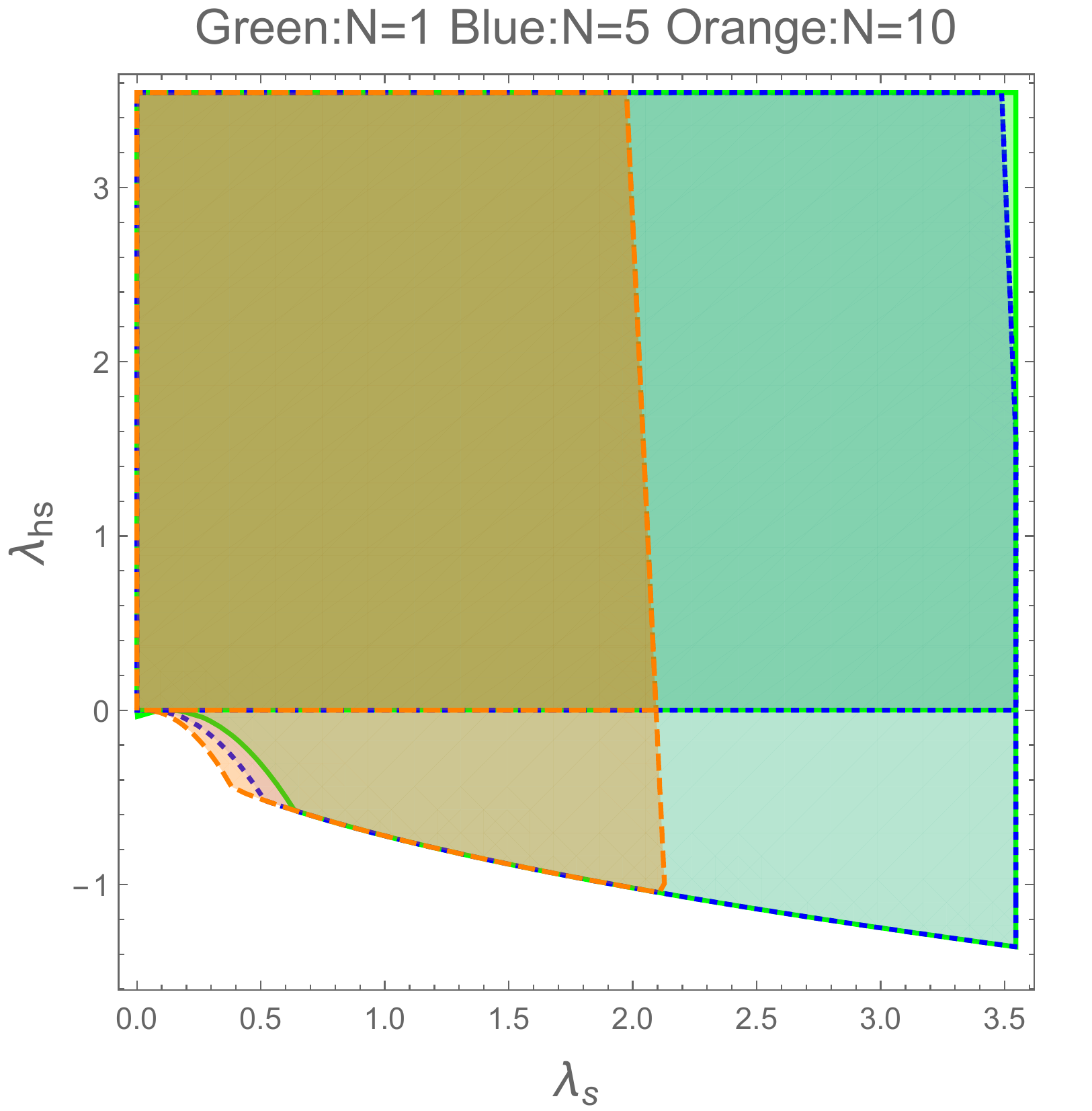}
\end{center}
\caption{Parameters regions allowed by Perturbativity+Unitarity+Stability in the $O(N)$ scenario at EW scale. Both the $\lambda_{hs}>0$ and $\lambda_{hs}> - 2\sqrt{\lambda_h\lambda_s}$ are shown.}\label{fig:ON_pus}
\end{figure}

\begin{figure}[!htp]
\begin{center}
\includegraphics[width=0.4 \textwidth]{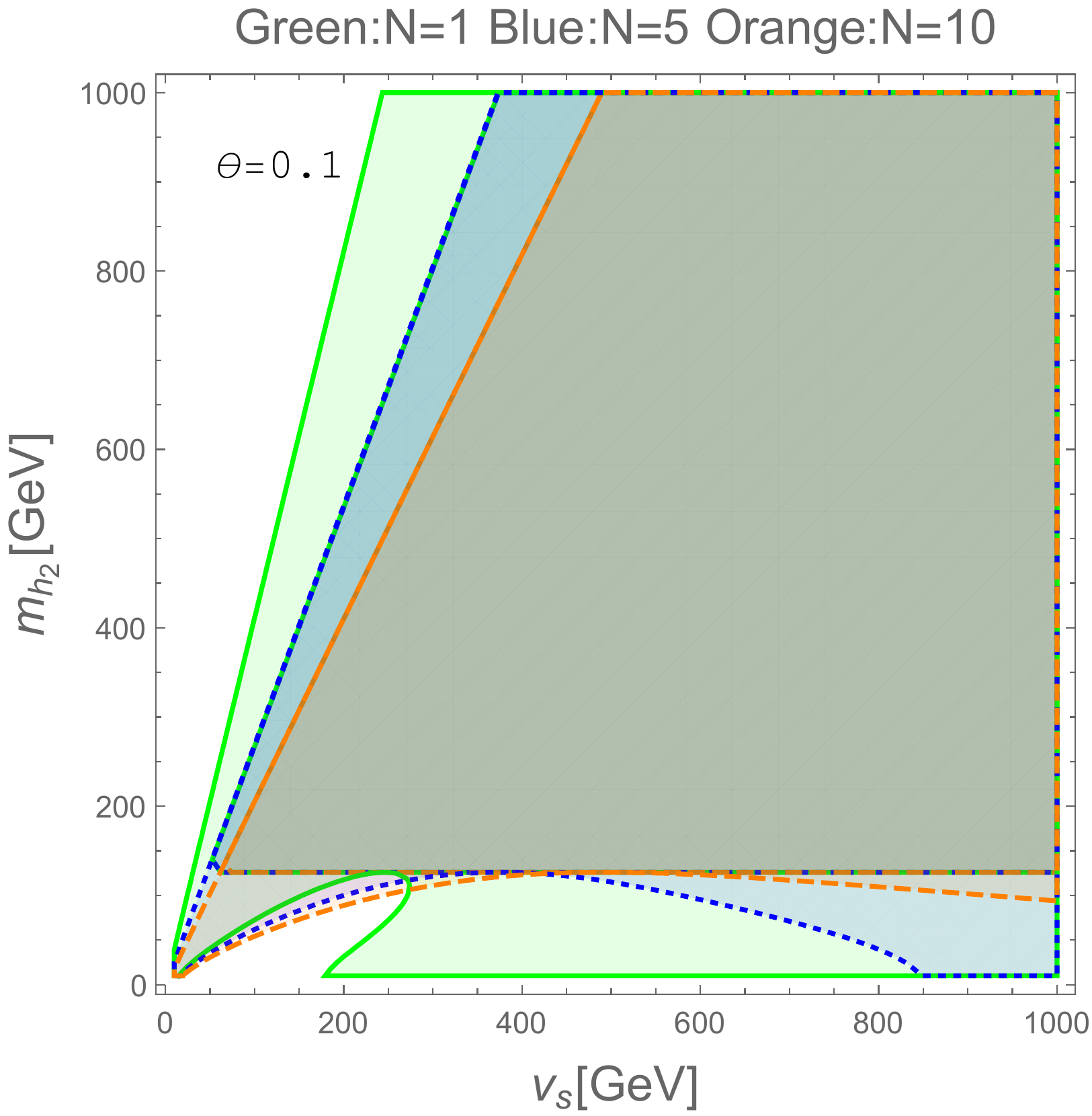}
\includegraphics[width=0.4 \textwidth]{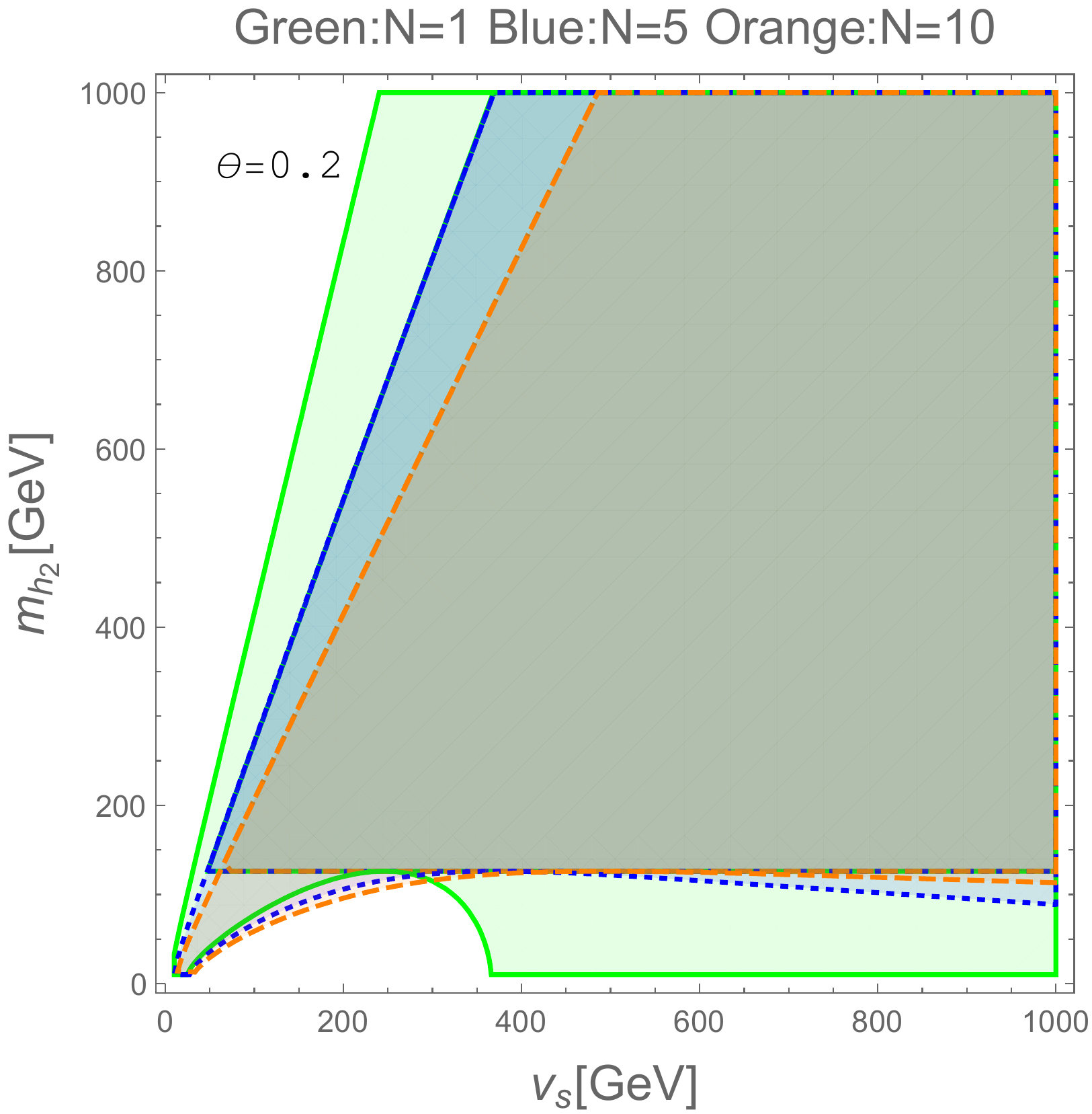}
\end{center}
\caption{Parameters regions allowed by Perturbativity+Unitarity+Stability in the $O(N\to N-1)$ scenario at EW scale.}\label{fig:ONTON1_pus}
\end{figure}

A simple analysis of these theoretical limits on the scalar quartic couplings at the Electroweak scale is given in Fig.~\ref{fig:ON_pus} and Fig.~\ref{fig:ONTON1_pus} for $O(N)$ and $O(N\to N-1)$ scenarios.
The perturbativity roughly sets the upper limit of $\lambda_{hs,s}$, the shape of the boundary is set by the unitarity bounds. The lower bound of the quartic couplings $\lambda_{hs,s}$ is given by
the stability conditions where more parameter spaces are allowed by $\lambda_{hs}> - 2\sqrt{\lambda_h\lambda_s}$ in comparison with $\lambda_{hs}>0$. In the Fig.~\ref{fig:ONTON1_pus}, it's converted to the bounds on the $m_{h_2}$ and $v_s$ correspondingly.

For the study of inflation and EWPT, we implement three conditions of the perturbativity, unitarity, and the stability of the inflationary potential from the Electroweak (EW) scale to Planck scale,
which are evaluated with the renormalization group equations list in Appendix.\ref{sec:RGEs}.

\subsection{Higgs precisions}

Integrating out the heavy scalar fields results in the dimension-six operators,
\bea
\mathcal{L}\supset \frac{c_H}{\Lambda^2} \mathcal{O}_H+\frac{c_6}{\Lambda^2} \mathcal{O}_6\;,
\eea
with $O_H\equiv\frac{1}{2}(\partial |H^{\dag} H |)^2$ and $\mathcal{O}_6\equiv |H^{\dag} H |^3$.
Here, the operator $\mathcal{O}_H$ can lead to the universal shift of Higgs couplings by the Higgs field redefinition or Higgs wavefunction renormalization\footnote{See Ref.~\cite{He:2015spf,Ge:2016zro} for the collider studies on the new physics that yielding the two operators.}. The operator $\mathcal{O}_6$ can alleviate the triple Higgs coupling and is crucial for the realization of a SFOEWPT~\cite{Grojean:2004xa}.
For the Wilson coefficients generated at tree-level, we have~\cite{Henning:2014gca,Henning:2014wua},
\begin{eqnarray}\label{eq:cnh}
&&c^N_H=N\frac{\lambda_{hs}^2}{2\lambda_s}\; ,\qquad \quad c^N_6=0 \; ,\\
&&c^{N\to N-1}_H=\frac{\lambda_{hs}^2}{2\lambda_s}\;,\quad~c^{N\to N-1}_6=0\; ,
\end{eqnarray}
with the $\Lambda\approx \mu_s$ for the two scenario. For $O(N)$ case one have $ \mu_s\approx m_{S_i}$, and in the $O(N\to N-1)$ case $\mu_s\approx m_{h_2}$ for the small mixing angle limit.
 The loop-level induced dim-6 operator Wilson coefficients are
\bea
c^N_H=\frac{N\lambda_{hs}}{48\pi^2 },~c^N_6=-\frac{N \lambda_{hs}^3}{48\pi^2}\;,
\eea
for  the $O(N)$ case.  These reduces to
\bea
c^{N\to N-1}_H=\frac{\lambda_{hs}}{48\pi^2 },~c^{N\to N-1}_6=-\frac{\lambda_{hs}^3}{48\pi^2}\;.\label{eq:wco}
\eea
for $O(N\to N-1)$ scenario.
We note that the same as the tree-level induced dim-6 operator, the factor of ``$N$" in $O(N\to N-1)$ doesn't appear in Eq.~\ref{eq:wco} because there is only one heavy scalar and the other $s_{N-1}$ scalars are massless Goldstones.
For the quartic coupling $\lambda_{hs}\sim \mathcal{O}(10^{-1})$ being required by the inflation, we can safely ignore the loop-level induced operator
effects on the Higgs precision.
In the small mixing angle limit of the $O(N\to N-1)$ scenario, we have: $1-\cos\theta\approx c^{N\to N-1}_H v^2/(2 m_{h_2}^2)\approx c^{N\to N-1}_H v^2/2 \mu_{s}^2$.

Now, we explore the EWPOs in the two scenarios.
The operator $\mathcal{O}_H$ induces the operator combinations $O_W + O_B$ and $O_T$ operators through RGE~\cite{Henning:2014gca,Craig:2013xia}, which results in the S and T parameters
\bea
&&\Delta S=\frac{1}{12}c_H \frac{v^2}{m_{S(h_2)}^2}\log(\frac{m_{S(h_2)}^2}{m_W^2})\;,\\
&&\Delta T = -\frac{3}{16\pi c_W^2} c_H\frac{v^2}{m_{S(h_2)}^2}\log(\frac{m_{S(h_2)}^2}{m_W^2})\;.
\eea
We set bounds on $m_{S(h_2)}$ and the ``$N$" using the electroweak fit in Ref. \cite{Baak:2014ora} ,
\bea\label{eq:ST}
S = 0.06 \pm0.09 ,~ T = 0.10 \pm 0.07 \;,
\eea
with the correlation coefficient between the S and T parameters being $+0.91$.
In the case of $O(N\to N-1)$, the parameter spaces are more strictly constrained by T parameter rather than S parameter, which set stringent bounds on the mixing angle and the masses of the heavy Higgs. When the heavy Higgs is not highly decoupled~\footnote{Indeed, if one want the heavy Higgs take part in the EWPT process, it cannot be highly decoupled.}, i.e., $m_{h_2}\sim m_{h}$, one can obtain the oblique parameter T following Ref.~\cite{Barger:2007im},
\begin{eqnarray}
\nonumber
 \label{eq:Txsm1}
T & = & -\left(\frac{3}{16\pi s_W^2}\right)\Biggl\{
 \cos^2\theta\, \Bigl [\frac{1}{c_W^2}\left(\frac{m_{h_1}^2}{m_{h_1}^2-M_Z^2}\right)\, \ln\, \frac{m_{h_1}^2}{M_Z^2}-\left(\frac{m_{h_1}^2}{m_{h_1}^2-M_W^2}\right)\, \ln\, \frac{m_{h_1}^2}{M_W^2}\Bigr]+ \sin^2\theta\, \Bigl[\frac{1}{c_W^2}\left(\frac{m_{h_2}^2}{m_{h_2}^2-M_Z^2}\right)\nonumber \\
&&\times \ln\, \frac{m_{h_2}^2}{M_Z^2} - \left(\frac{m_{h_2}^2}{m_{h_2}^2-M_W^2}\right)\, \ln\, \frac{m_{h_2}^2}{M_W^2}\Bigr ]\Biggr\} \; .
\end{eqnarray}

\begin{figure}[!htp]
\begin{center}
\includegraphics[width=0.4 \textwidth]{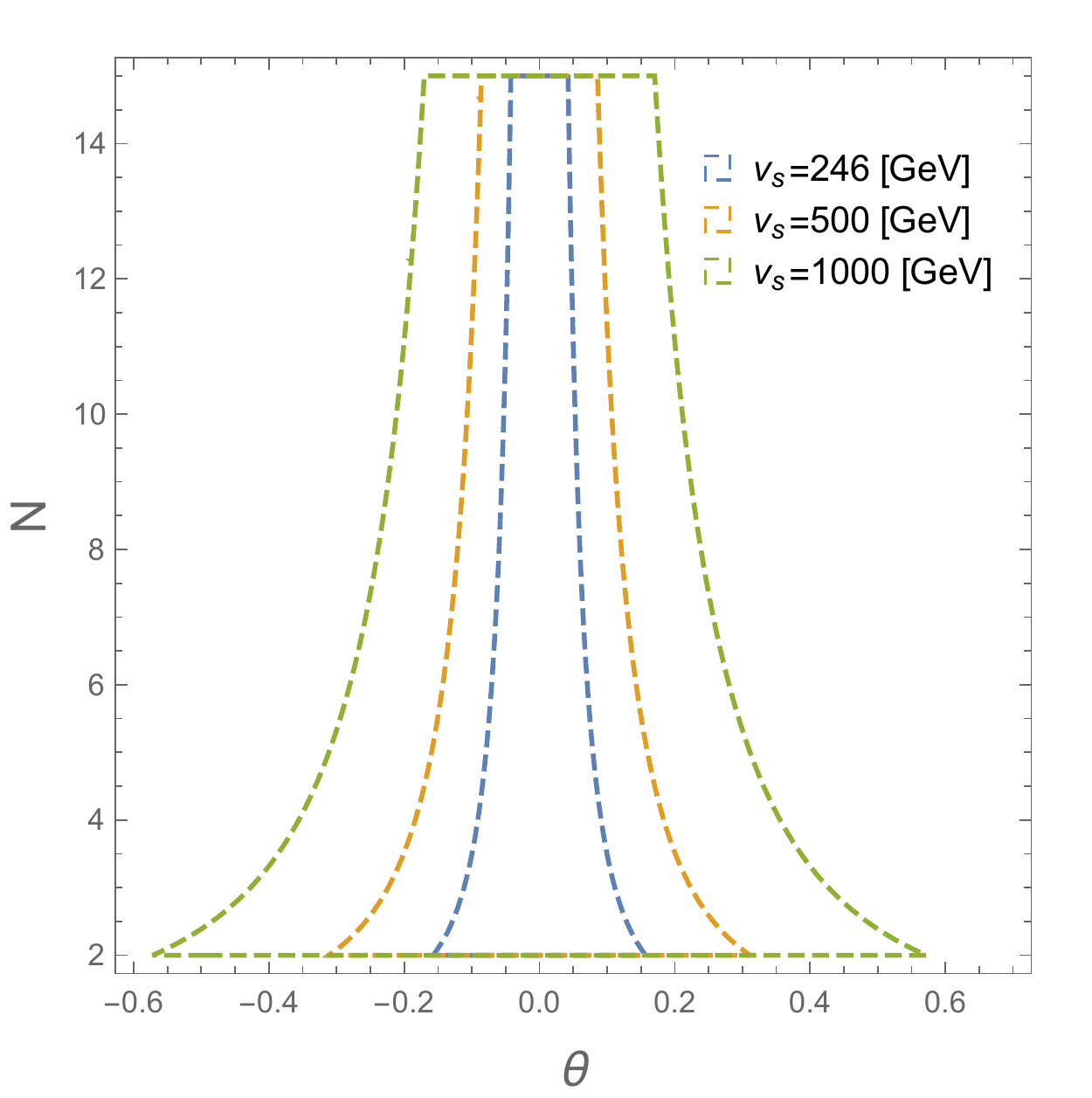}
\end{center}
\caption{Invisible decay bounds on $N$ and mixing angle $\theta$ in $O(N \to N-1)$ model coming from LHC~\cite{Khachatryan:2016vau}.}\label{fig:ONTON1_hinv}
\end{figure}

For the $O(N\to N-1)$ scenario,
the mixing angle and the heavy Higgs masses are subjective to the bounds coming from the LHC Higgs data, which limit the mixing angle $\theta$ to be $\abs{\cos\theta}\geq 0.84 $~\cite{Profumo:2014opa}. After including the current LHC and High-luminosity LHC Higgs production rates together with the EWPOs, a moderate of $\theta\sim \sqrt{\lambda_{hs}^2v^2/(4\lambda_s m_{h_2}^2)}=0.2$ can be safety~\cite{Carena:2018vpt}.
We firstly perform the Higgs fit without including the change of SM Higgs decay width induced by the
Goldstone, and then constrain the
number of Goldstone bosons $N-1$ with the Higgs invisible decay fit results from ~\cite{Khachatryan:2016vau}: $B_{BSM}<0.34$ at 95\% CL.
For the case of $O(N)$ symmetry being broken to $O(N-1)$ at zero temperature, we have the following Lagrangian to describe the triple scalar interactions,
\begin{eqnarray}
\mathcal{L} \supset \lambda_{h_ih_jh_j}h_ih_jh_j+\lambda_{h_is_{N-1}s_{N-1}}h_is_{N-1}s_{N-1}\; ,
\end{eqnarray}
with $h_{i,j}$ denotes $h_{1,2}$ and $N=2,...,N$, the relevant triple scalar couplings are given bellow,
\begin{eqnarray}
&&\lambda _{{h_2}h_1h_1} = -\frac{m_{h_1}^2}{2{v v_{s}}}\sin (2 \theta )(v_{s}\cos \theta + v\;\sin \theta)(1 + m_{h_2}^2/2m_{h_1}^2)\;,\\
&&\lambda_{h_2 s_{N-1}s_{N-1}}=m_{h_2}^2\cos\theta/(2v_s)\;,\\
&&\lambda_{h_1 s_{N-1}s_{N-1}}=-m_{h_1}^2\sin\theta/(2v_s)\;,\\
&&\lambda _{{h_1}h_2h_2} =\lambda_{h_1 s_{N-1}s_{N-1}}\;.
\end{eqnarray}
From which, when the $m_{h_2}>m_{h_1}$, the decay widths of the SM-like Higgs and the second Higgs are given by
\begin{eqnarray}
\Gamma_{h_2}^{tot}&=&\Gamma_{h_2}({h_2}\to h_1h_1)+\sin^2\theta \Gamma_{h}^{SM}\left|_{m_h \to m_{h_2}}\right.+(N-1)\Gamma_{h_2}(h_2 \to s_{N-1}s_{N-1}) \nonumber\\
&=&\Gamma_{h_2}({h_2}\to h_1h_1)+\sin^2\theta \Gamma_{h}^{SM}\left|_{m_h \to m_{h_2}}\right.+(N-1)\frac{\lambda_{h_2 s_{N-1} s_{N-1}}^2 }{32 \pi m_{h_2}}\;\\
\Gamma_{h_1}^{tot}&=&\cos^2\theta \Gamma_{h}^{SM}+(N-1)\Gamma_{h}(h\to s_{N-1}s_{N-1})\nonumber\\
&=&\cos^2\theta \Gamma_{h}^{SM}+(N-1)\frac{\lambda_{h_1s_{N-1}s_{N-1}}^2 }{32 \pi m_{h_1}}\; ,
\end{eqnarray}
with
\begin{eqnarray}
\Gamma(h_2\to h_1h_1)=\frac{\lambda_{h_2h_1h_1}^2}{32 \pi m_{h_2}}\sqrt{1-4m_{h_1}^2/m_{h_2}^2}\; .
\end{eqnarray}
For the case in which $m_{h_1}>2m_{h_2}$, one need take into account the decay of $h_{1}\to 2h_2$ with the decay width being given by
\begin{eqnarray}
\Gamma(h_1\to h_2h_2)=\frac{\lambda_{h_1h_2h_2}^2}{32 \pi m_{h_1}}\sqrt{1-4m_{h_2}^2/m_{h_1}^2}\; .
\end{eqnarray}
The invisible decay of SM Higgs can be used to set upper bounds to the number of the Goldstones and the mixing angle $\theta$.
At 95\% CL, the LHC (ATLAS+CMS) set $B_{inv}<34\%$~\cite{Khachatryan:2016vau}, see Fig.~\ref{fig:ONTON1_hinv} for the constraints.
With the increase of $v_s$, more parameter space of ($\theta$,$N$) is allowed.

On the other hand, the $\mathcal{O}_H$ leads to the modification of the wavefunction of the Higgs,
 \bea
\mathcal{L}_{eff}\supset(1+\delta Z_h ) \frac{1}{2}(\partial_\mu h)^2\;,
\eea
with $\delta Z_h=2v^2c_H/m_{S(h_2)}^2$ for $O(N)$ ($O(N\to N-1)$) scenarios. Thus one obtains a universal shift of all Higgs couplings.
Which therefore induce the correction to the $e^+ e^- \to hZ$ associated production cross section~\cite{Craig:2013xia}
\bea
\delta \sigma_{Zh}=-2\frac{v^2c_H}{m_{S(h_2)}^2}\; ,
\eea
which has been defined as the fractional change in
the associated production cross section relative to the SM case.
For the $O(N)$ scenario with $m_{S}>m_h$, the Higgs wavefunction renormalization shift the SM-like Higgs couplings to other SM particles by $c^{N}_H v^2/(2 m_{S}^2)\sim N \lambda_{hs}^2\,v^2/(2\lambda_s m_{S}^2)$.
Which results in the constraint on $c_H$ and therefore $N, \lambda_{hs,s}$ from the LHC\cite{Carena:2018vpt} as well as ILC, CEPC, and FCC-ee\cite{Gu:2017ckc}.
The study of Ref.~\cite{Gu:2017ckc} shows that the CEPC, ILC, and FCC-ee can probe the new physics parameter spaces ( through the $e^+ e^- \to hZ$ process ) much better than LHC.

\begin{figure}[!htp]
\begin{center}
\includegraphics[width=0.6 \textwidth]{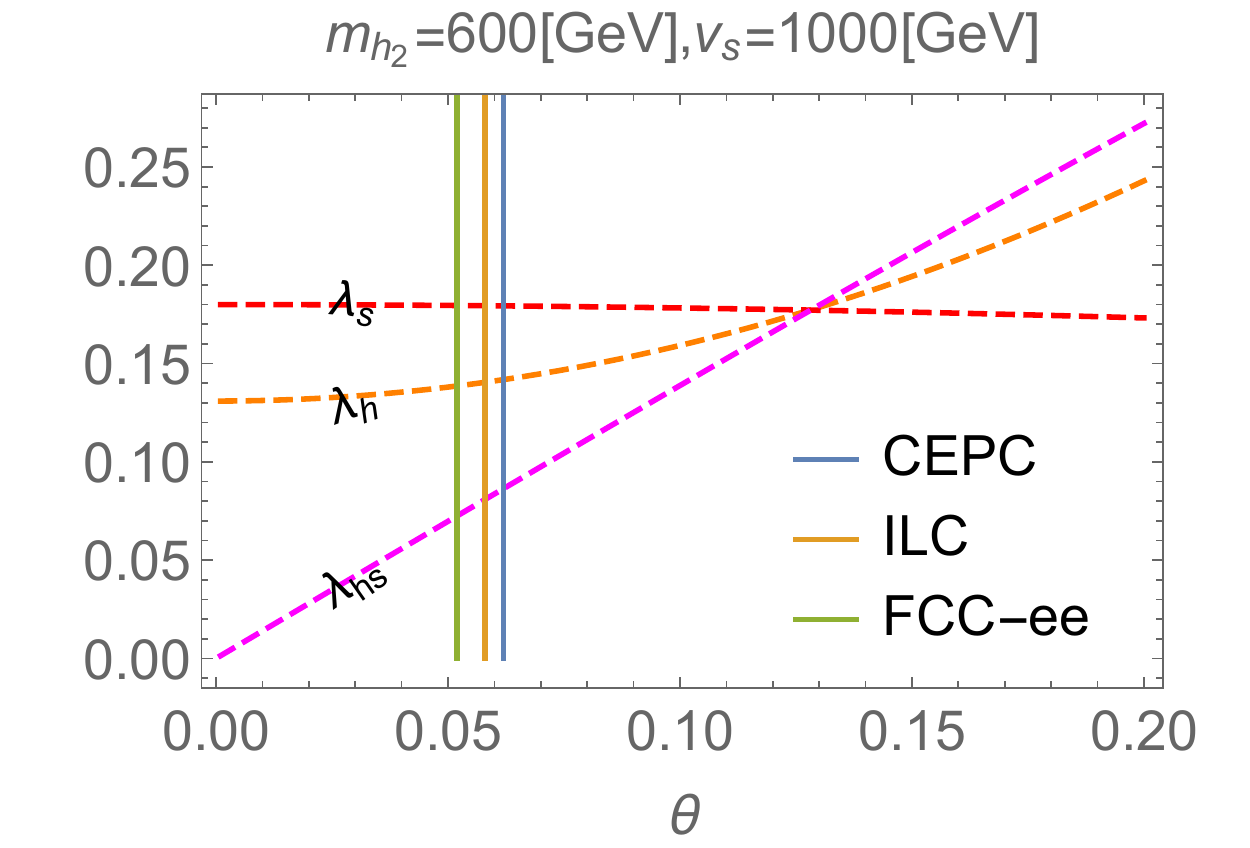}
\caption{The projected sensitivity of the mixing angle for $O(N\to N-1)$ model from lepton colliders .}\label{fig:colliderN1}
\end{center}
\end{figure}

\begin{figure}[!htp]
\begin{center}
\includegraphics[width=0.3 \textwidth]{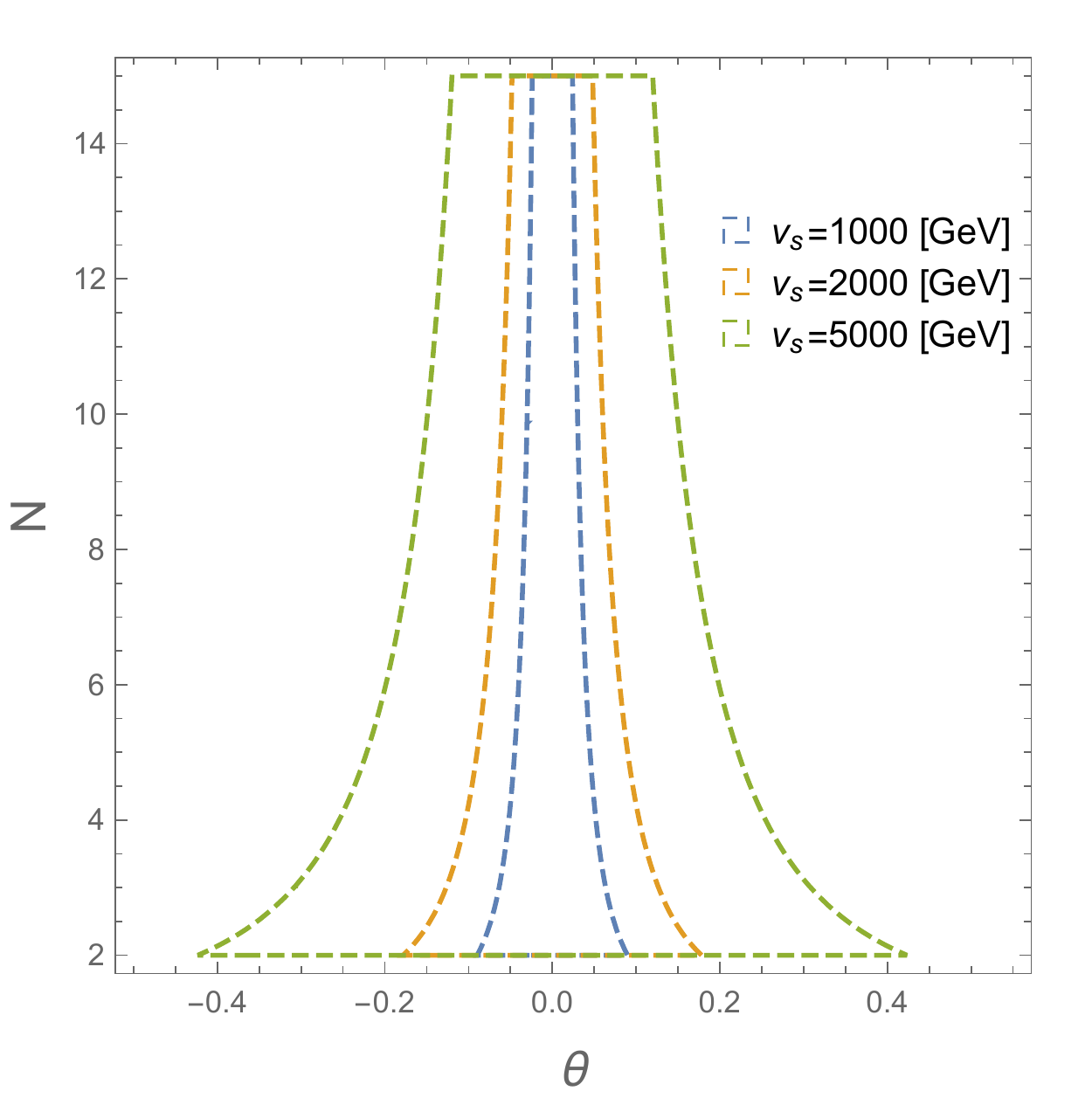}
\includegraphics[width=0.3 \textwidth]{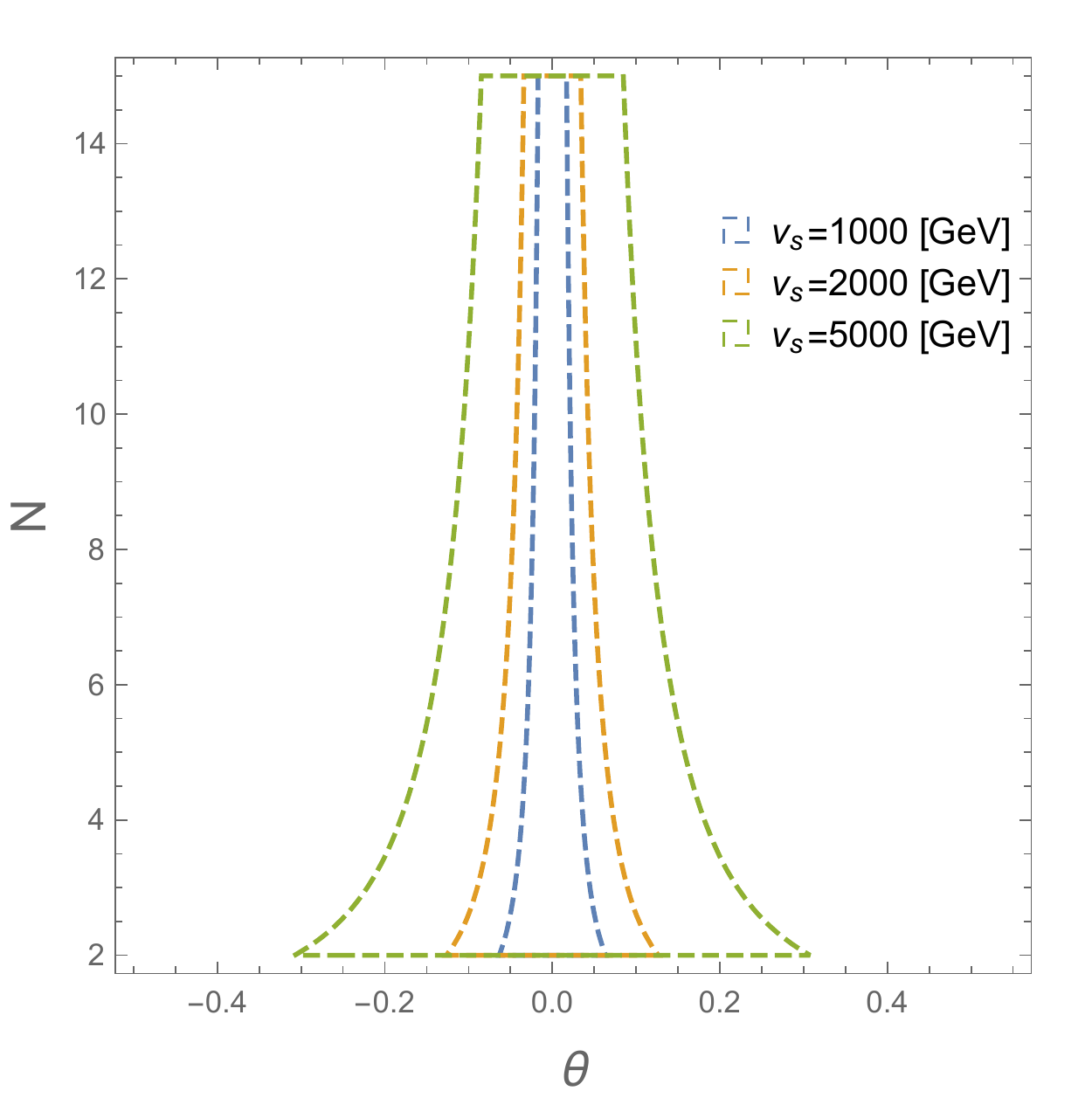}
\includegraphics[width=0.3 \textwidth]{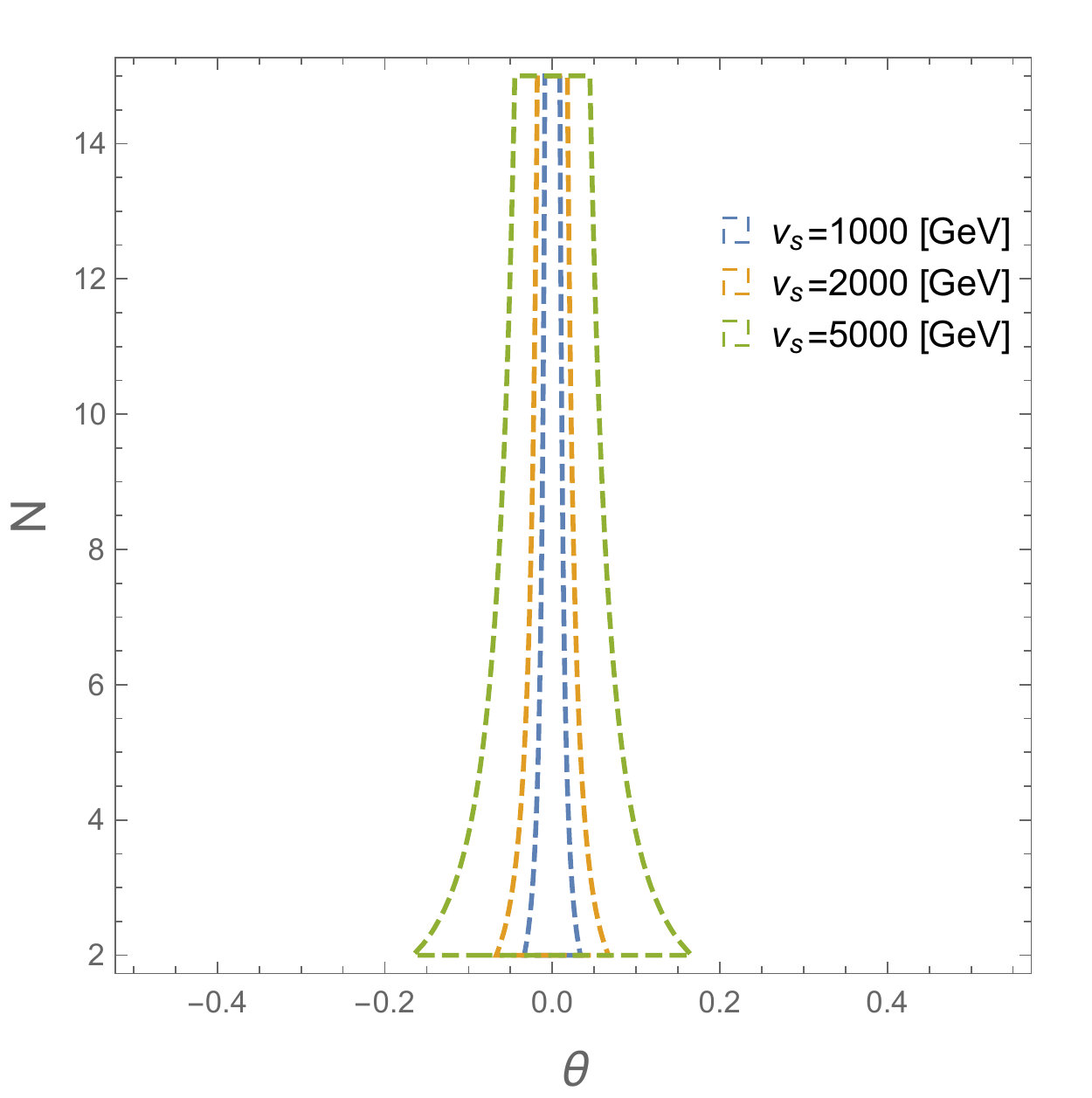}
\end{center}
\caption{Higgs invisible decay bounds on $N$ and $\theta$ in $O(N \to N-1)$ scenario considering the sensitivity of ILC, FCC-ee, and CEPC respectively from left to right panels.}\label{fig:ONTON1hinvep}
\end{figure}

For $N=1$ of the $O(N\to N-1)$ scenario (which is the Higgs-portal 1-singlet scalar case), the CEPC with luminosity of 5 $ab^{-1}$, ILC with all center of mass energies, and FCC-ee with luminosity of
 10 $ab^{-1}$ bound the mixing angle $|\sin\theta|$ to be 0.062, 0.058 and 0.052 at 95\% C.L. ~\cite{Gu:2017ckc}.  We constrain our model parameter spaces with these values, see Fig.~\ref{fig:colliderN1}.
These high sensitivity leptonic colliders set a severe bound on the $\lambda_{hs}$, and therefore
the stability problem can easily preclude the chance to realize the slow-roll Higgs inflation. For the case of $N\geq 2$ of the $O(N\to N-1)$ scenario, using the Higgs Strahlung process, the ILC set $B_{inv}<1\%$~\cite{Fujii:2015jha}, the FCC-ee set $B_{inv}<0.5\%$~\cite{Gomez-Ceballos:2013zzn,dEnterria:2016sca}, and CEPC set 0.14\%~\cite{CEPC-SPPCStudyGroup:2015csa}.  Fig.\ref{fig:ONTON1hinvep} indicates that a large mixing angle $\theta$ is allowed for a large $v_s$, which corresponds to the heavy Higgs decouple cases.
Generally,  to make the Higgs inflation feasible, a relatively large mixing angle $\theta$ is required to enlarge the value of $\lambda_{hs}$ and therefore to ensure the vacuum stability. Firstly, the increasing of $\theta$ can lead to the perturbativity problem of $\lambda_h$, and thus we need a relatively small $m_{h_2}$. Secondly, a large $v_s$ leads to a small $\lambda_{hs}$, therefore to avoid the stability problem we need a large $N$.
The future $e^+e^-$ colliders constraints give a narrow parameter region of the mixing angle $\theta$. In this super-weak couple scenario the SFOEWPT would not occur due to the $\lambda_{hs}$ is too small, thus one needs a much larger $N$ to amplified the effects of the $N-1$ Goldstones in order to obtain a SFOEWPT. On the other hand, the Goldstones faked effective neutrino situation would be changed a lot due to the decouple conditions allowed parameter spaces can be covered by the bounds from $B_{inv}$ at ILC, FCC-ee and CEPC as aforementioned.

\section{Cosmological implications}
\label{sec:cosimp}

We first study the cosmic inflation with large scale fields. With the temperatures of the universe cooling down the low scale physics come to us:
the possibility to obtain a SFOEWPT, and the dark matter physics.

\subsection{The Higgs inflation with N singlet scalars}
\label{sec:ONinf}

The action in the Jordan frame is
\begin{eqnarray}
S_J&=&\int d^4x \sqrt{-g}\Big[-\frac{M_{\rm p}^2}{2} R - \xi_h (H^\dagger H) R-\xi_s S^2 R \nonumber\\
&&+ \partial_\mu H^\dagger \partial^\mu H + (\partial_\mu S)^2- V(H,S)\Big]\;,
\label{action}
\end{eqnarray}
where $\ds{M_{\rm p}}$ is the reduced Planck mass, $R$ is the Ricci scalar, $\xi_{\rm h, \rm {S}}$  define the non-minimal coupling of the $h,S$-field. Here, we drop the subscript to simplify the notation as in Ref.~\cite{Lerner:2009xg}. 
The quantum corrected effective Jordan frame Higgs potential at large field value ($h$) can be written as
\be
V(h) = \frac{1}{4} \lambda_{h} (\mu) h^4 \, ,
\ee
which is evaluated along the higgs axis,
where the scale is $\ds{\mu \sim \mathcal{O}(h) \approx h}$. The potential is the inflationary potential, which will be used to estimate the slow-roll parameters of $\epsilon$ and $\eta$. We impose quantum corrections to the potential following Ref.~\cite{Elizalde:1993ew,Elizalde:2014xva}.
After the conformal transformation,
\be
\label{Omega}
\tilde{g}_{\mu\nu} = \Omega^2 g_{\mu\nu}, \quad \Omega^2\equiv 1+\frac{\xi_{{S}}}{ M_{\rm P}^2 }{S^2} + \frac{\xi_{\rm h} h^2}{M_{\rm P}^2}.
\ee
and a field redefinition
\be
\label{h_chi}
\frac{d\chi_{\rm h}}{dh} = \sqrt{\frac{\Omega^2+6\xi^2_{\rm h}h^2/M_{\rm P}^2}{\Omega^4}}, \quad \frac{d\chi_{\rm {S}}}{d S} = \sqrt{\frac{\Omega^2+6\xi_{{S}}^2{S}^2/M_{\rm P}^2}{\Omega^4}},
\ee
we obtain

\be
\begin{aligned}
S_E =& \int d^4x \sqrt{-\tilde{g}}\bigg(-\frac{1}{2}M_{\rm P}^2R + \frac{1}{2}{\partial}_{\mu}\chi_{\rm h}{\partial}^{\mu}\chi_{\rm h} + \frac{1}{2}{\partial}_{\mu}\chi_{\rm {S}}
{\partial}^{\mu}\chi_{\rm {S}} \\
&+ A(\chi_{\rm {S}}, \chi_{\rm h}){\partial}_{\mu}\chi_{\rm h}{\partial}^{\mu}\chi_{\rm {S}} - U(\chi_{\rm {S}},\chi_{\rm h})  \bigg),
\label{EframeS}
\end{aligned}
\ee
where $U(\chi_{\rm {S}},\chi_{\rm h}) = \Omega^{-4}V(S(\chi_{\rm {S}}),h(\chi_{\rm h}))$ and
\be
A(\chi_{\rm {S}}, \chi_{\rm h}) = \frac{6\xi_{\rm h}\xi_{\rm {S}}}{M_{\rm P}^2\Omega^4}\frac{d{S}}{d\chi_{\rm {S}}}\frac{dh}{d\chi_{\rm h}}h{S}.
\ee
In this work, we consider the Higgs field serves as inflaton for the $O(N)$ model and the $O(N \to N-1)$ model, which is ensured by $\xi_h\gg \xi_S$. We consider $\xi_S=0$ at the Electroweak scale. In this situation, the kinetic terms of the scalar fields are canonical with $A(\chi_{\rm {S}}, \chi_{\rm h}) =0$, and the metric in this case is given by $\ds{\Omega^2 = 1 + (\xi_h h^2 + \xi_{{S}} {S}^2) / M_{\rm pl}^2 \approx 1 + \xi_h h^2 / M_{\rm pl}^2 }$ with  $S\sim 0$~\cite{Clark:2009dc,Lerner:2009xg,Aravind:2015xst}\footnote{We note that Ref.~\cite{Lebedev:2011aq,Lebedev:2012zw} drop the kinematic mixing terms due to suppress of the largeness of the non-minimal coupling combination, and the Higgs inflation conditions there are satisfied in our case.}.

The inflationary action in terms of the canonically normalized field $\chi$ is therefore given as
\eq{
S_{\rm inf} = \int d^4 x \sqrt{\tilde{g}} \left[ \frac{M_{\rm p}^2}{2}R + \frac{1}{2} \left( \partial \chi \right)^2 - U(\chi) \right] \,,
}
with the potential in terms of the canonically normalized field $\chi$ as
\eq{
U(\chi) = \frac{ \lambda_h  \left( h(\chi) \right)^4}{4 \Omega^4}  \,,
}
where the new field $\chi$ are defined by
\eq{
\frac{d \chi}{dh} \approx \left((1 + \xi_{h} h^2/M_{\rm p}^2 + 6 \xi_{h}^2 h^2 / M_{\rm p}^2)/( 1 + \xi_{h} h^2/M_{\rm p}^2 )^2 \right)^{1/2} \,
}
for $h-$ inflations~\cite{Lerner:2009xg}.
Note that $\lambda_{h}$ and $\xi_{h}$ have a scale ($h$) dependence. The potential of $U(\chi)$ at the high scale of $\chi\gg M_P$ should be flat enough to drive the slow-roll inflation.

 The slow-roll parameters used to characterize the inflation dynamics are,
 \bea\label{eq:sr}
\epsilon(\chi) = \frac{M_{\rm p}^2}{2} \left(\frac{dU/d\chi}{U(\chi)} \right)^2 \, , \qquad \eta(\chi) = M_{\rm p}^2 \left( \frac{d^2U/d\chi^2}{U(\chi)} \right)  \, .
\eea
The field value at the end of inflation $\chi_{\rm end}$ is obtained when $\ds{\epsilon= 1}$, and the horizon exit value $\chi_{\rm in}$ can be calculated by assuming
an e-folding number between the two periods,
\be
N_{\rm e-folds} = \int_{\chi_{\rm end}}^{\chi_{\rm in}} d\chi \frac{1}{M_{\rm p} \sqrt{2 \epsilon}}  \, .
\ee
Then, one can relate the inflationary observables of spectrum index $n_s$ and the tensor to scalar ratio of $r$ with the slow-roll parameters at the $\chi_{\rm in}$ ,
\bea\label{eq:delta}
n_s  = 1 + 2 \,\eta - 6\, \epsilon \, ,~
r = 16\, \epsilon \, .
\eea
The Planck results set $n_s = 0.9677 \pm0.0060$ at $1\sigma$ level and $r < 0.11$ at 95\% CL\cite{Ade:2015lrj} for
$N_{\rm e}=60$.
 Meanwhile, the non-minimal gravity couplings $\xi_h$ can be determined using the constraint coming from CMB observations \cite{Ade:2015lrj}, with the amplitude of scalar spectrum fluctuations $\ds{\Delta_\mathcal{R}^2}$ being calculated as
\eq{
\Delta_\mathcal{R}^2 = \frac{1}{24 \pi^2 M_{\rm p}^4}\frac{U(\chi)}{\epsilon} = 2.2\times10^{-9} \, .\label{eq:cmb}
}
With which, one can obtain the slow-roll inflation favored parameter regions of $\lambda_{s}$ and  $\lambda_{hs}$ for the fixed $\lambda_h$ in the $O(N)$ scenario, since the two quartic coupling contribute to the inflation potential indirectly through the RGEs as shown in Appendix.~\ref{sec:RGEs}. For the $O(N\to N-1)$ scenario, the $m_{h_2}, v_s$ and the mixing angle of $\theta$ determine the couplings of $\lambda_{h,hs,s}$
through Eq.~\ref{eq:coupling}.
For the two scenarios, the slow-roll parameters $r$ are all of order $\sim\mathcal{O}(10^{-2})$. Previous inflation studies of
Ref.~\cite{Lerner:2009xg,Aravind:2015xst} shows that the successful implementation of slow-roll Higgs or singlet inflation usually occurs with relatively smaller quartic scalar couplings of order $\mathcal{O}(0.1-1)$.

At last, we comment on the thermal history of the Universe.
One can estimate the reheating temperature when the decay of the inflaton starts competing
with expansion H $\sim\Gamma_{h}$ for Higgs inflation~\cite{Kofman:1994rk},
\begin{equation}
\rho=3H^2 M_p^2=3\Gamma_{h}^2 M_p^2\equiv\frac{\pi^2 g_*}{30}T_R^4\;,
\end{equation}
here g$_* \approx$ 100 is the number of relativistic degrees of freedom in the Universe during the
reheating epoch. The reheating temperature for the Higgs inflation was estimated through 
the parametric resonance of the oscillating Higgs field
to W bosons (via $|H|^2|W|^2$) in Ref.~\cite{Bezrukov:2008ut,GarciaBellido:2008ab}. Ref.~\cite{Bezrukov:2008cq} suggest $T_R\geq(\frac{15\lambda_h}{8\pi^2 g_*})^{1/4}\frac{M_p}{\xi_h}$. Within the inflation viable parameter spaces in $O(N\to N-1)$ scenarios under study, we have $\lambda_h\sim \mathcal {O}(10^{-4}-10^{-1})$ and $\xi_h\sim \mathcal{O}(10^3-10^5)$, therefore the $T_R\geq \mathcal{O}(10^{10}-10^{13})$ GeV. For the $O(N)$ scenario, we have $\lambda_h\sim \mathcal {O}(10^{-2}-10^{-1})$ and $\xi_h\sim \mathcal{O}(10^5)$, thus $T_R\geq \mathcal{O}(10^{13})$ GeV.
Freeze out of cold dark matter requires $x_f\equiv m_{DM}/T_{fs}\approx 20$, and therefore the thermal history can occur as
$T_R>T_C>T_{fs}>T_{BBN}$ to account for the EWPT and reheating as well as the
successful Big Bang Nucleosynthesis (BBN) (with a typical temperature of a few MeV).The freeze-out temperature $T_{fs}$ being smaller than the SFOEWPT temperature $T_C$, set the $m_{DM}<20 T_C\sim 2$ TeV with $T_C$ being around $\sim\mathcal{O}$($10^2$) GeV.

\subsection{Electroweak phase transition}

With the temperature cooling down, the universe can evolve from symmetric phase to the symmetry broken phase. The
behavior can be studied with the finite temperature effective potential with particle physics models~\cite{Dolan:1973qd}.
Through which one can obtain the critical classical field value and temperature being $v_C$ and $T_C$. Roughly speaking, a SFOEWPT can be obtained when $v_C/T_C>1$, then the electroweak sphaleron process is
quenched inside the bubble and therefore one can obtain the net number of baryon over anti-baryon in the framework of EWBG. For the uncertainty of
the value and possible gauge dependent issues we refer to Ref.~\cite{Patel:2011th}.
The finite temperature effective potential includes the tree level scalar potential, the Coleman-Weinberg potential, and the finite temperature corrections~\cite{Arnold:1992rz}. For the finite temperature corrections, we adopt the method of Ref.~\cite{Dolan:1973qd,Bernon:2017jgv} with the Espinosa approach~\cite{Arnold:1992rz}.
Then the critical parameters of EWPT can be calculated when there are two degenerate vacuums with a potential barrier. Due to rich vacuum structures of the potential at finite temperatures, there can be one-step or multi-step phase transitions. A SFOEWPT can occur at the first or the second step in the two-step scenario.
We will investigate one-step and two-step phase transitions in the $O(N)$ and the $O(N\to N-1)$ scenarios.

\begin{figure}[!htp]
\begin{center}
\includegraphics[width=0.4\textwidth]{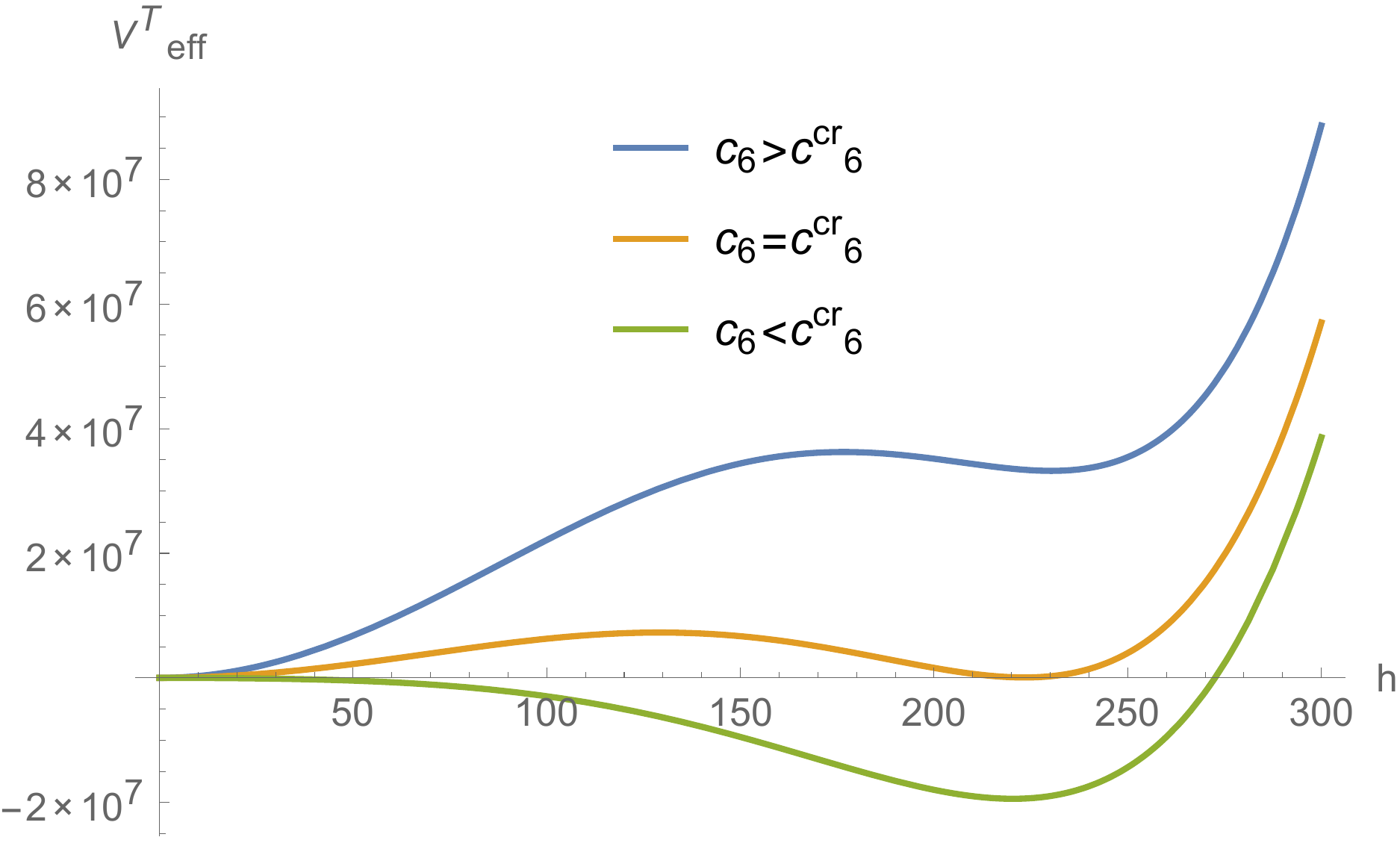}
\includegraphics[width=0.4\textwidth]{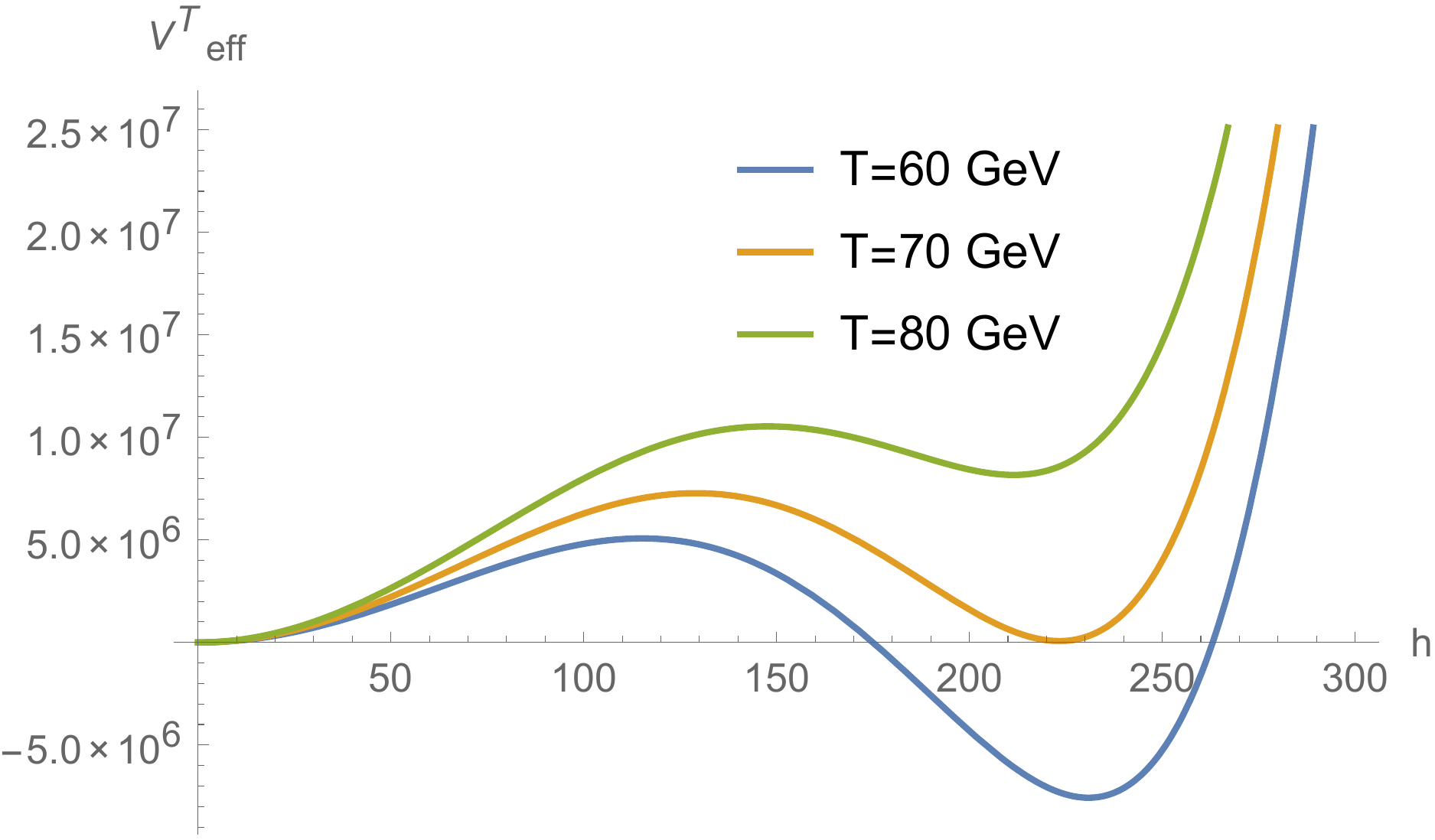}
\caption{The finite temperature effective potential v.s. dim-6 operator Wilson coefficients (temperature) with fixed temperature $T=70$ GeV (fixed $c_6=c_6^{cr}$) for left (right) panel. }
\label{fig:EWPTEFT}
\end{center}
\end{figure}
Before the detailed study, we first warm up by briefly recalling the one-step SFOEWPT condition on Wilson coefficients of the dimensional-six operator~\cite{Huang:2015tdv,Delaunay:2007wb,Grojean:2004xa}\footnote{See Ref.~\cite{Cao:2017oez,Huang:2016odd,Huang:2015izx} for relevant studies with the Gravitational wave and collider searches. },
\bea
\frac{m_h^2}{3v^4}<\frac{c_6}{\Lambda^2}<\frac{m_h^2}{v^4}\;.
\eea
The left panel of Fig.~\ref{fig:EWPTEFT} shows the potential shape at critical temperature with different dimensional six operator Wilson coefficients. Which depicts that a suitable $c_6=c_6^{cr}$ is needed to obtain a proper vacuum barrier to separate two degenerate vacua at critical temperature $T_c$, therefore make the SFOEWPT feasible. On the right panel of Fig.~\ref{fig:EWPTEFT}, we plot the finite temperature effective potential as a function of temperature for fixed $c_6$, one can find that the symmetry will be restored at high temperature and break at temperature lower than the critical temperature $T_c$.
It should be noted that, with the spontaneous symmetry breaking of $O(N\to N-1)$, the two contributions of the $c_6$ from $s h^2$ and $s^3$ terms cancel each other~\cite{Henning:2014wua}, and therefore the tree-level induced dimensional six operator disappears. Which is the same as in the SM+1 singlet case being studied in Ref.~\cite{Carena:2018vpt}. In this case, the dimensional six operator shows up at loop level which is too small to affect the EWPT dynamics. This property can explain why the SM+1
 singlet scalar with $Z_2$ does not prefer one-step SFOEWPT, and here one may need to pursue the two-step types where the DM can be useful for achieving a SFOEWPT~\cite{Curtin:2014jma,Cline:2012hg,Bian:2018bxr,Bian:2018mkl}. In this work, we reconfirm the same property in the $O(N)$ and $O(N\to N-1)$ cases.

\begin{figure}[!htp]
\begin{center}
\includegraphics[width=0.4\textwidth]{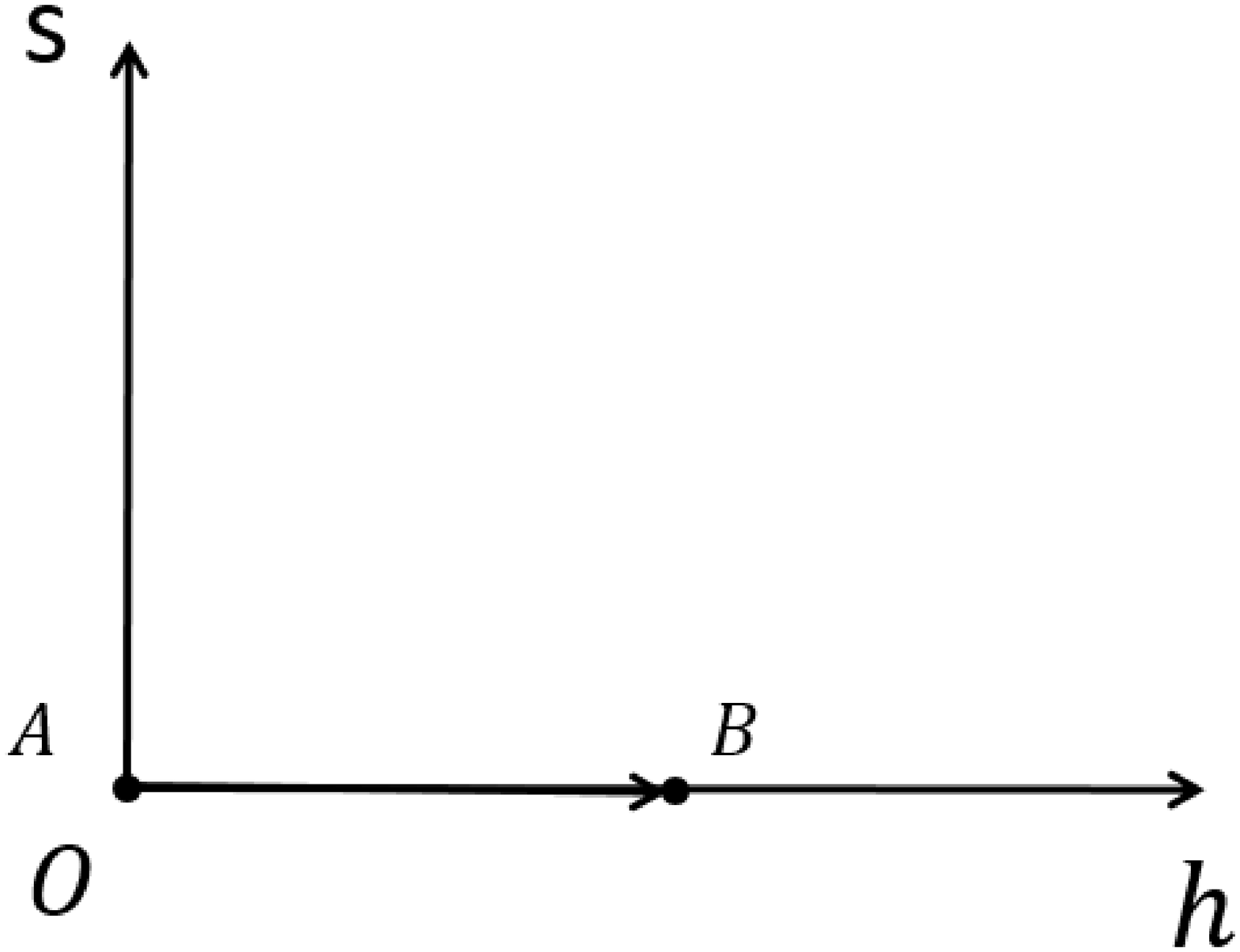}
\includegraphics[width=0.4\textwidth]{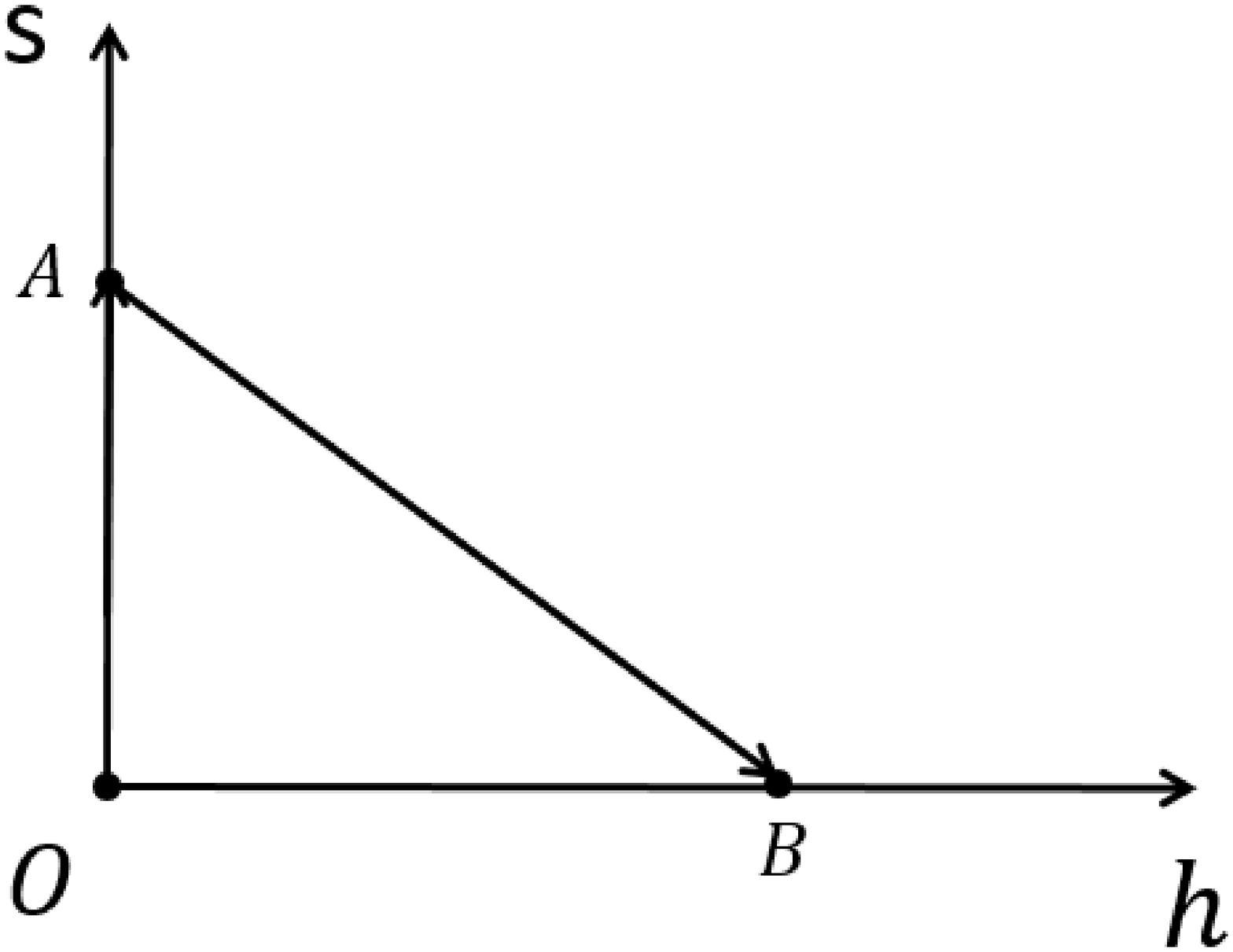}
\caption{One- and two-step EWPT types in O(N) for left and right panels, respectively.}
\label{fig:EWPTtypeON}
\end{center}
\end{figure}

Following the approach of Ref.~\cite{Bernon:2017jgv}, the finite temperature of $V(h,S(s),N,T)=V_0(h,S(s))+V_{CW}(h,S(s),N)+V_T(h,S(s),N,T)+V_{\textit{\ daisy}}(h,S(s),N,T)$ for $O(N)$ $(O(N\to N-1))$ case is adopted to estimate the order parameters of the SFOEWPT, with $V_0$, $V_{CW}$, $V_T$ and $V_{\textit{\ daisy}}$ being zero temperature tree-level potential, one-loop Colemen-Weinberg potential, finite-temperature potential, and Daisy terms. These functions for the $O(N)(O(N\to N-1))$ case are given in Appendix.~\ref{sec:EWPT}.
In the case of $O(N)$ scalars, the corresponding critical temperature and critical field value for one- and two-step EWPT types (see the Fig.\ref{fig:EWPTtypeON}) can be evaluated through the following degeneracy conditions,
\begin{eqnarray}
&&V(0,0,N,T_C)=V(h_C^B,0,N,T_C)\; ,\nonumber\\
&&\frac{dV(h,0,N,T_C)}{dh}|_{h=h_{C}^B}=0\;,
\end{eqnarray}
and
\begin{eqnarray}
&&V(0,s_C^A,N,T_C)=V(h_C^B,0,N,T_C)\; ,\nonumber\\
&&\frac{dV(h,0,N,T_C)}{dh}|_{h=h_{C}^B}=0\; .
\end{eqnarray}
 Here the $s_C$ is the $O(N)$ broken direction, which is analogous to the $O(N\to N-1)$ scenario.
The survey of the one-step EWPT in the $O(N)$ scenario shows that the quartic coupling between the SM Higgs and
the $O(N)$ scalars $S_i$ ( $\lambda_{hs}$ ) should be large enough in order to make the SFOEWPT occurs, which is not favored by the slow-roll Higgs inflation.
Generally, within the parameter spaces of a large $\lambda_{hs}$ where one can have a SFOEWPT and the inflation is invalid, the perturbativity of scalar quartic coupling and unitarity are violated due to the RG running of couplings as explored in Sec.~\ref{sec:infON}.

\begin{figure}[!htp]
\begin{center}
\includegraphics[width=0.4\textwidth]{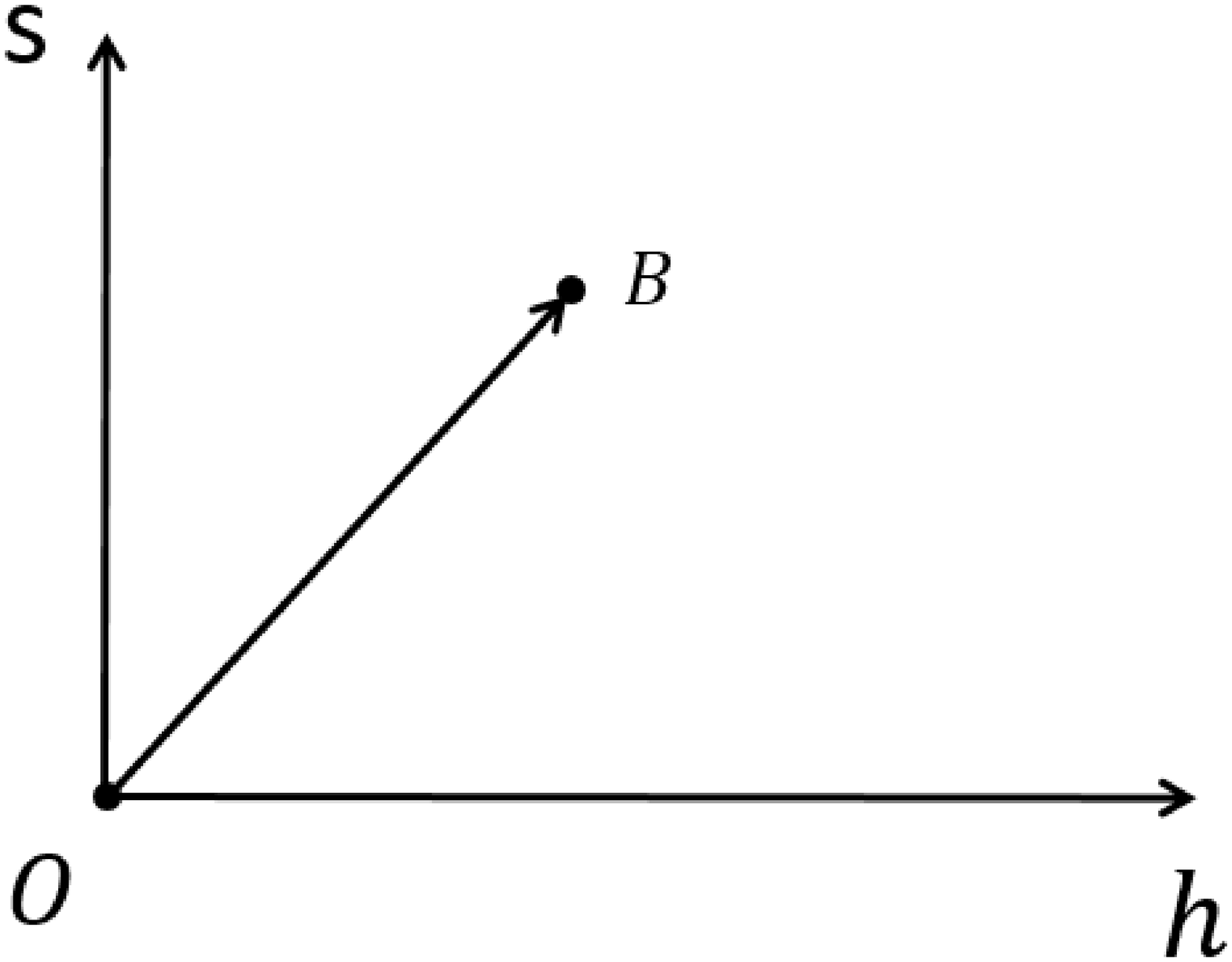}
\includegraphics[width=0.4\textwidth]{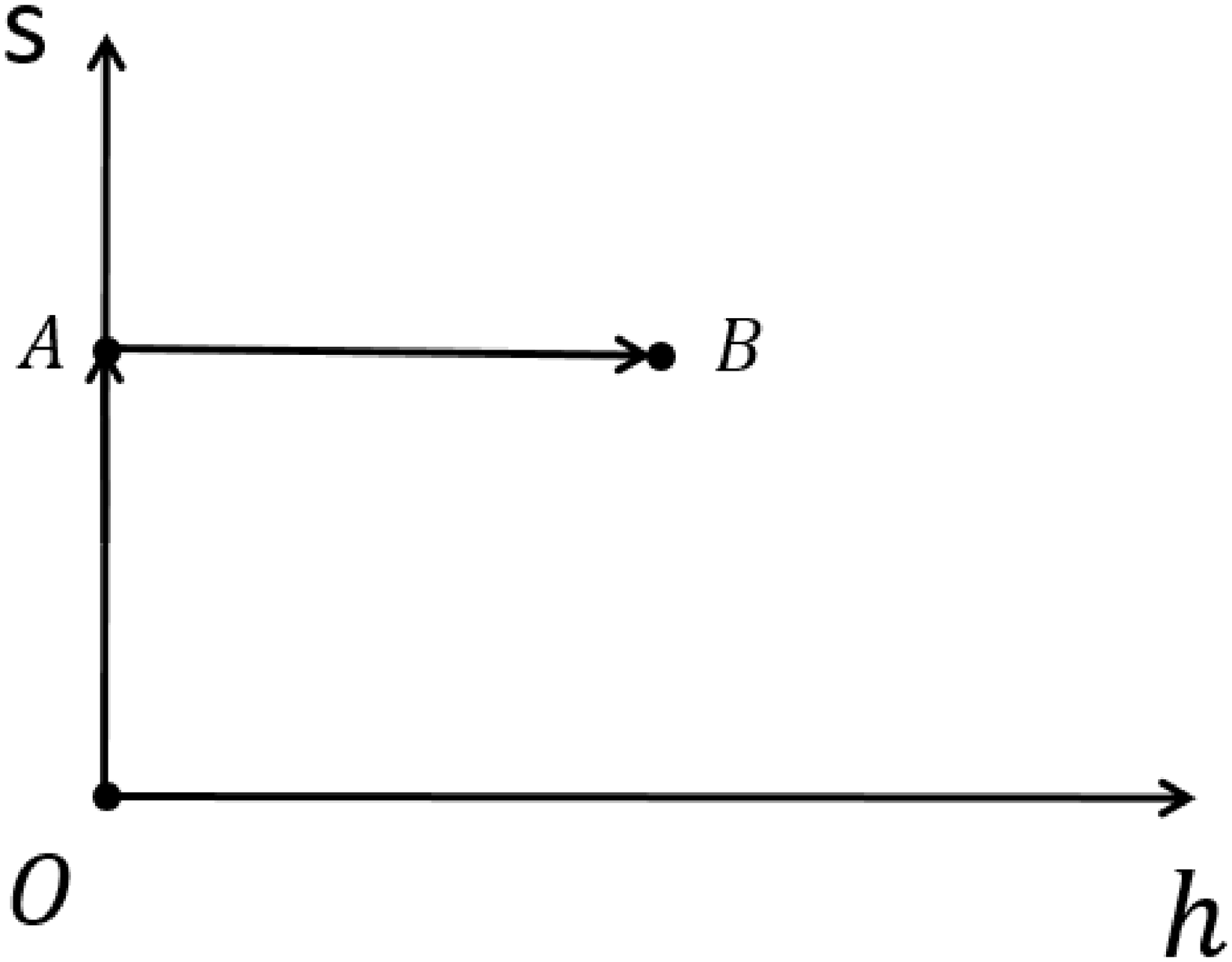}
\caption{One- and two-step phase transition types in $O(N\to N-1)$ scenario for left and right panels.}
\label{fig:EWPTtypeON1}
\end{center}
\end{figure}

We demonstrate the one-step and two-step phase transition patterns in Fig.\ref{fig:EWPTtypeON1}. The one-step EWPT types in $O(N \to N-1)$, occurs along the $\overrightarrow {OB}$ line, and the two-step EWPT occurs through the process of $O\to A\to B$.
With two degenerate vacuums being separated by a potential barrier structures at the critical temperature, the degeneracy conditions can be expressed as Eqs.\ref{ONTON11pt} and Eqs.\ref{ONTON12pt}.

\begin{eqnarray}
&&V(0,0,N,T_C)=V(v_C^B,s_C^B,N,T_C),\nonumber\\
&&\frac{dV(h,s,N,T_C)}{dh}|_{h=h_{C}^B,s=s_{C}^B}=0,~\frac{dV(h,s,N,T_C)}{ds}|_{h=h_{C}^B,s=s_{C}^B}=0.
\label{ONTON11pt}
\end{eqnarray}
\begin{eqnarray}
&&V(0,s_C^A,N,T_C)=V(h_C^B,s_C^B,N,T_C),\nonumber\\
&&\frac{dV(h,s,N,T_C)}{dh}|_{h=h_{C}^B,s=s_{C}^B}=0,~\frac{dV(h,s,N,T_C)}{ds}|_{h=h_{C}^B,s=s_{C}^B}=0.
\label{ONTON12pt}
\end{eqnarray}

\subsection{Dark matter/radiations}
\label{sec:dmdr}

When the $O(N)$ is kept at zero temperature, the N-singlet scalars can all serve as dark matter candidates.
Suppose gravity violates global symmetries, then the Goldstone boson may acquire a mass through nonpertubative gravitational effects \cite{Coleman:1988tj,Akhmedov:1992hi}.
The non-perturbative gravity effects can break the $O(N)$ symmetry at $M_P$ scale through the lowest high dimension operators, i.e., dim-5 operators, induce mass terms
to Goldstone bosons and make $N-1$ majoron like particles,
\begin{eqnarray}
\frac{C_1(H^\dag H)^2s_i}{M_P}+\frac{C_2(H^\dag H)s_i^3}{M_P}+\frac{C_3s_i^5}{M_P}\; .
\end{eqnarray}
For the wilson coefficients $C_i\sim \mathcal{O}(1)$ and the VEV of scalar singlet $v_{s}\sim \mathcal{O}(10^3)$ GeV, one can expect the masses of majoron like particle,
\begin{eqnarray}
m_{s_{1,...,N-1}}=\frac{16 C_1 v^4 + 12 C_2 v^2 v_{s}^2 + 5C_3 v_{s}^4} {2M_P v_{s}}\sim \mathcal{O}(1) {\rm eV}\;.
\end{eqnarray}
In this mass region, we can expect the majoron decaying to diphoton through the non-minimal gravity couple term which breaks the $O(N-1)$ symmetry as in the Ref.~\cite{Cata:2017jar}.
We found the Goldstone bosons here is long-lived, with $\tau \sim \Gamma^{-1}\sim 10^{46} s$, they can survive until the recombination era and may contribute to the Universe radiation density at the time of recombination or BBN.

\section{Numerical results}
\label{sec:infptdm}

\subsection{$ O(N)$ scenario}
\label{sec:infON}

\begin{figure}[!ht]
\begin{center}
\includegraphics[width=0.3 \textwidth]{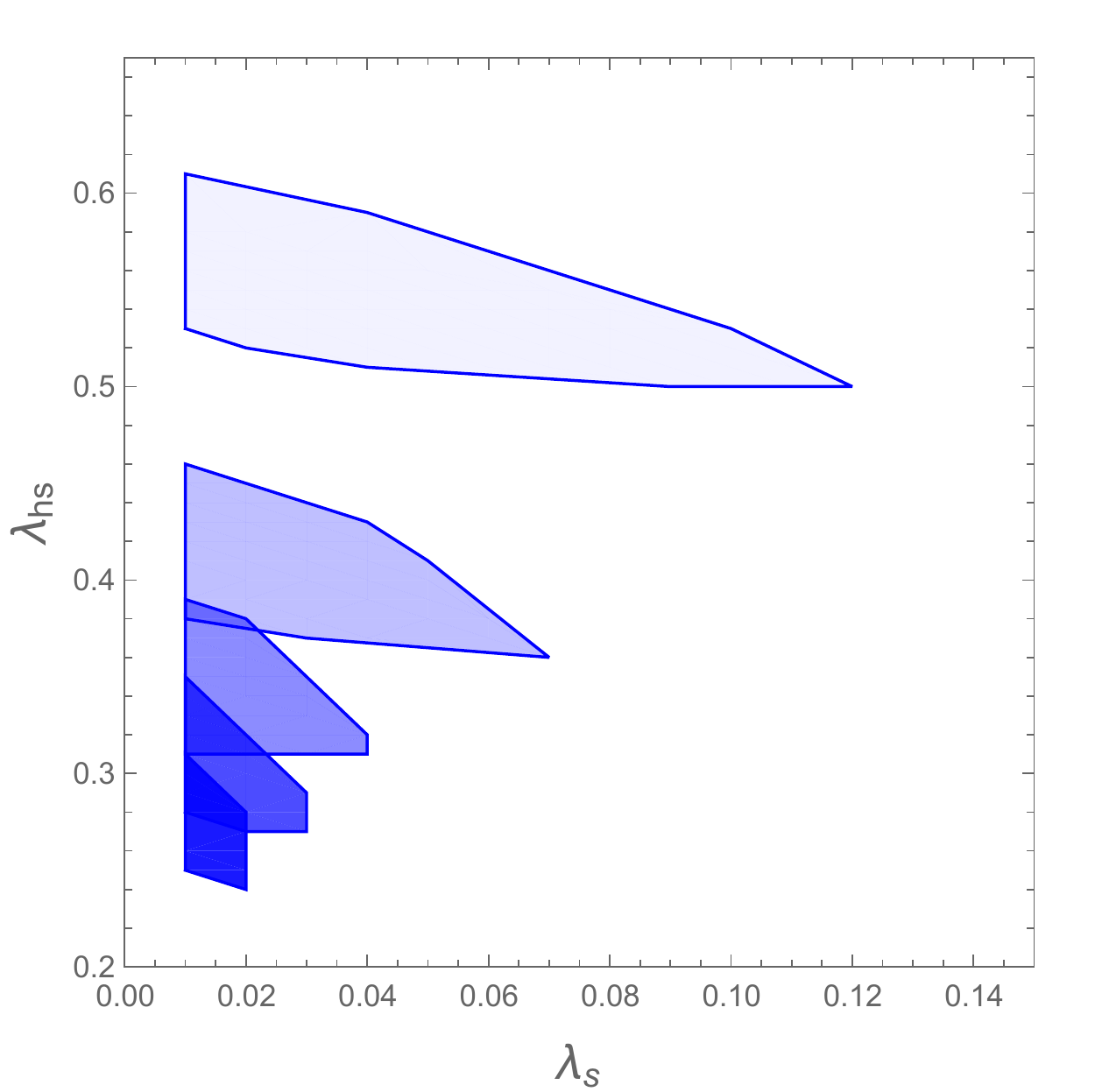}
\includegraphics[width=0.3 \textwidth]{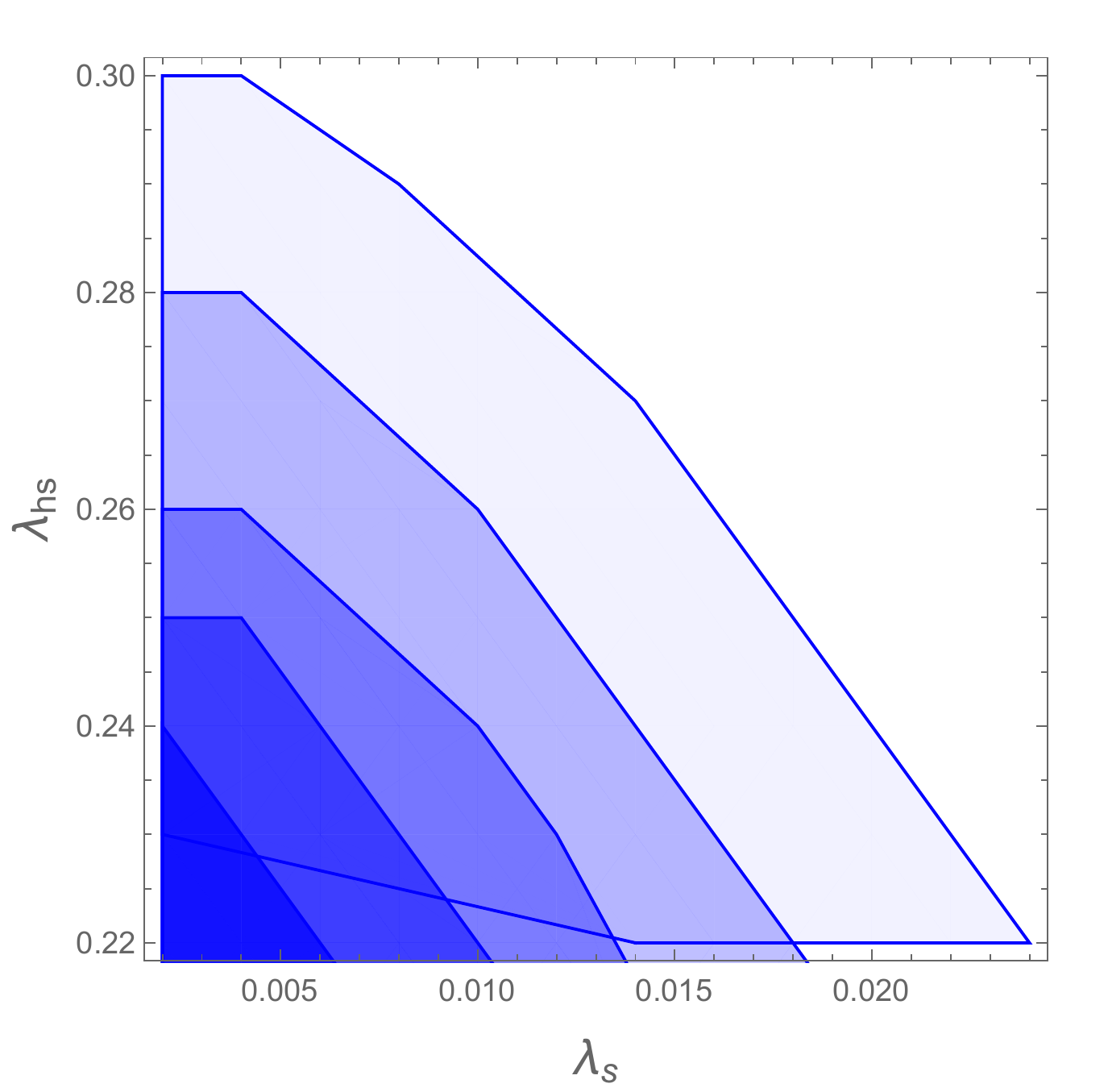}
\includegraphics[width=0.3 \textwidth]{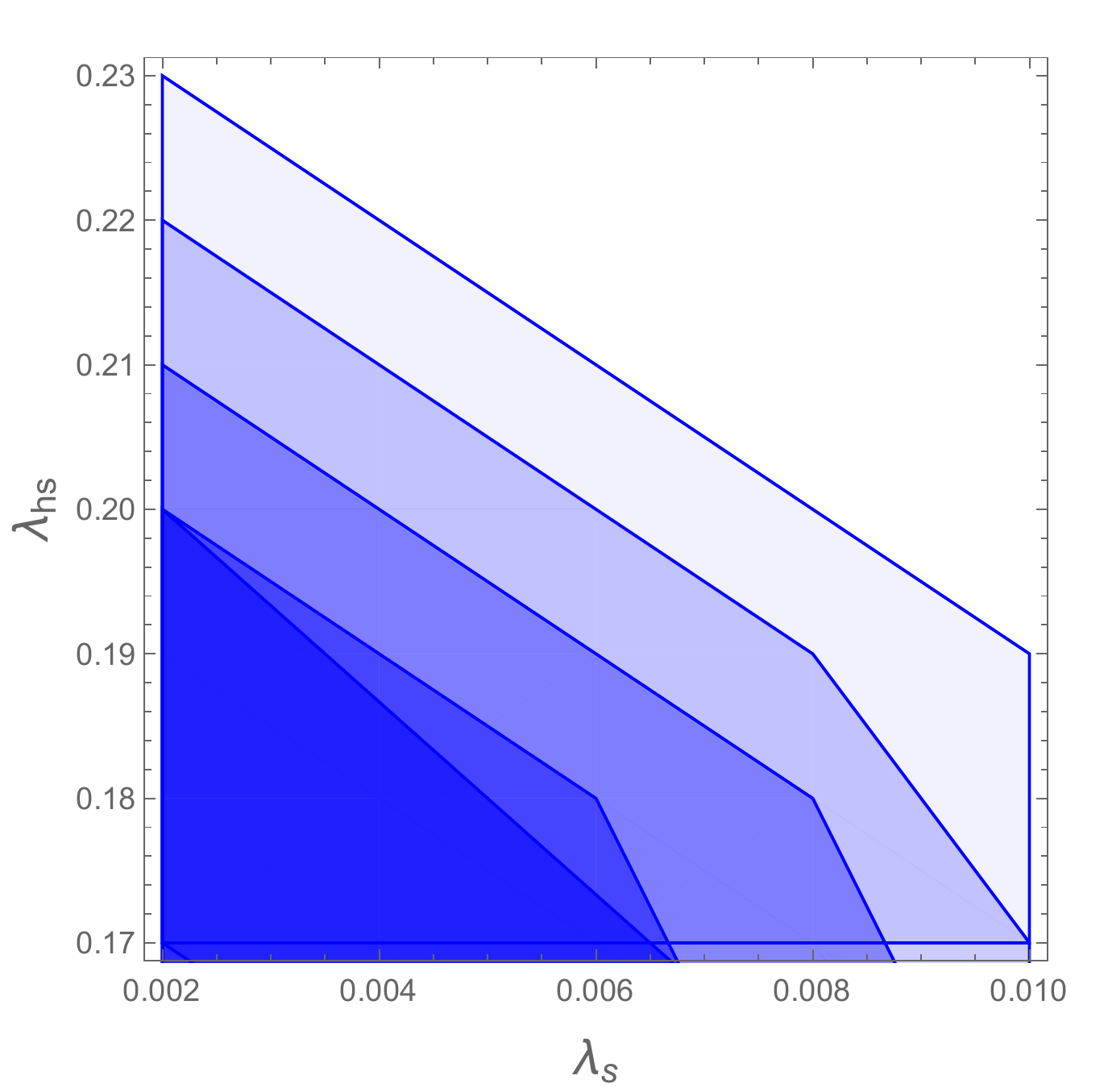}
\caption{Inflation feasible parameters plane of $(\lambda_s,\lambda_{hs})$ for different N within $O(N)$ scalar model, a larger $N$ is shown by a deeper color, and the corresponding $N$ are $1 \to 5$, $6 \to 10$ and $11 \to 15$ for left, middle and right panels, respectively.  Note that the shaded regions are allowed rather than excluded.
 }\label{fig:fig4}
\end{center}
\end{figure}

\begin{figure}[!ht]
\begin{center}
\includegraphics[width=0.4 \textwidth]{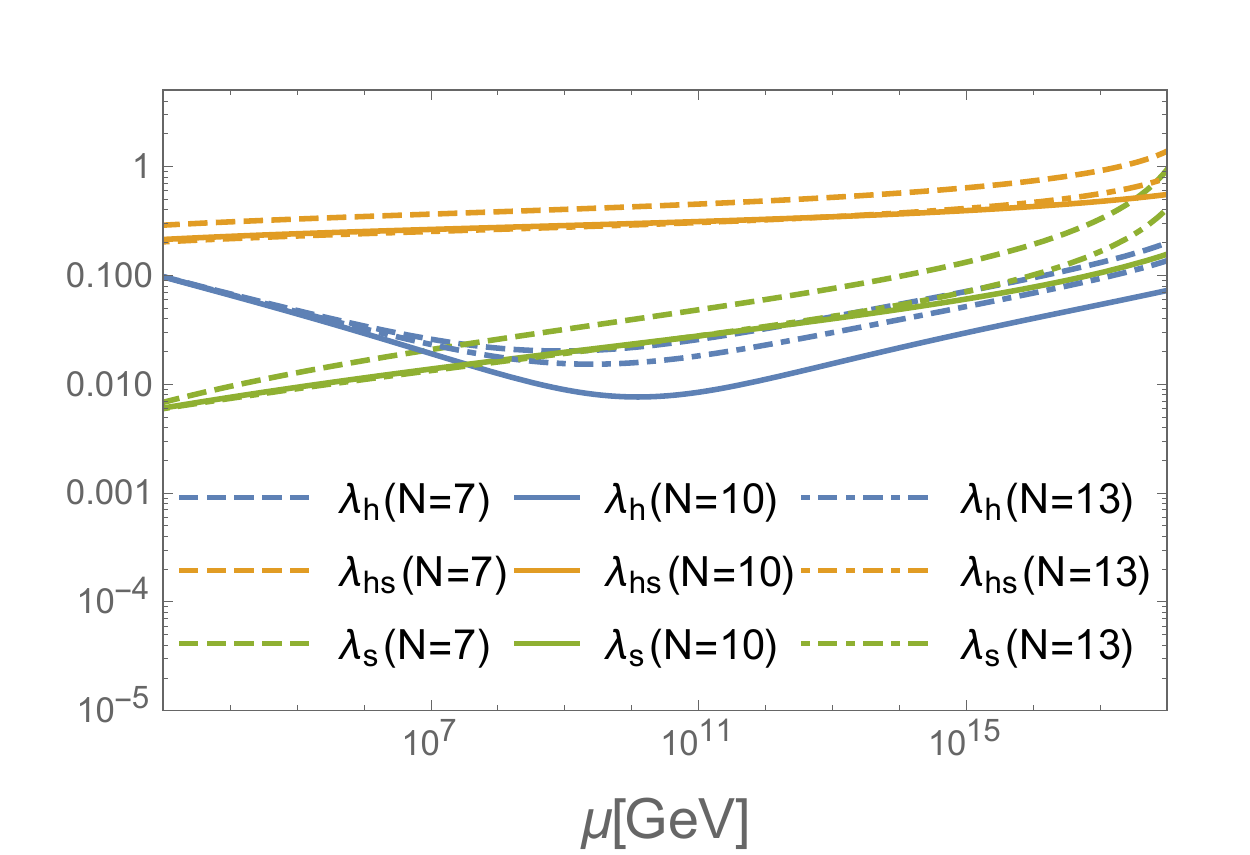}
\includegraphics[width=0.4 \textwidth]{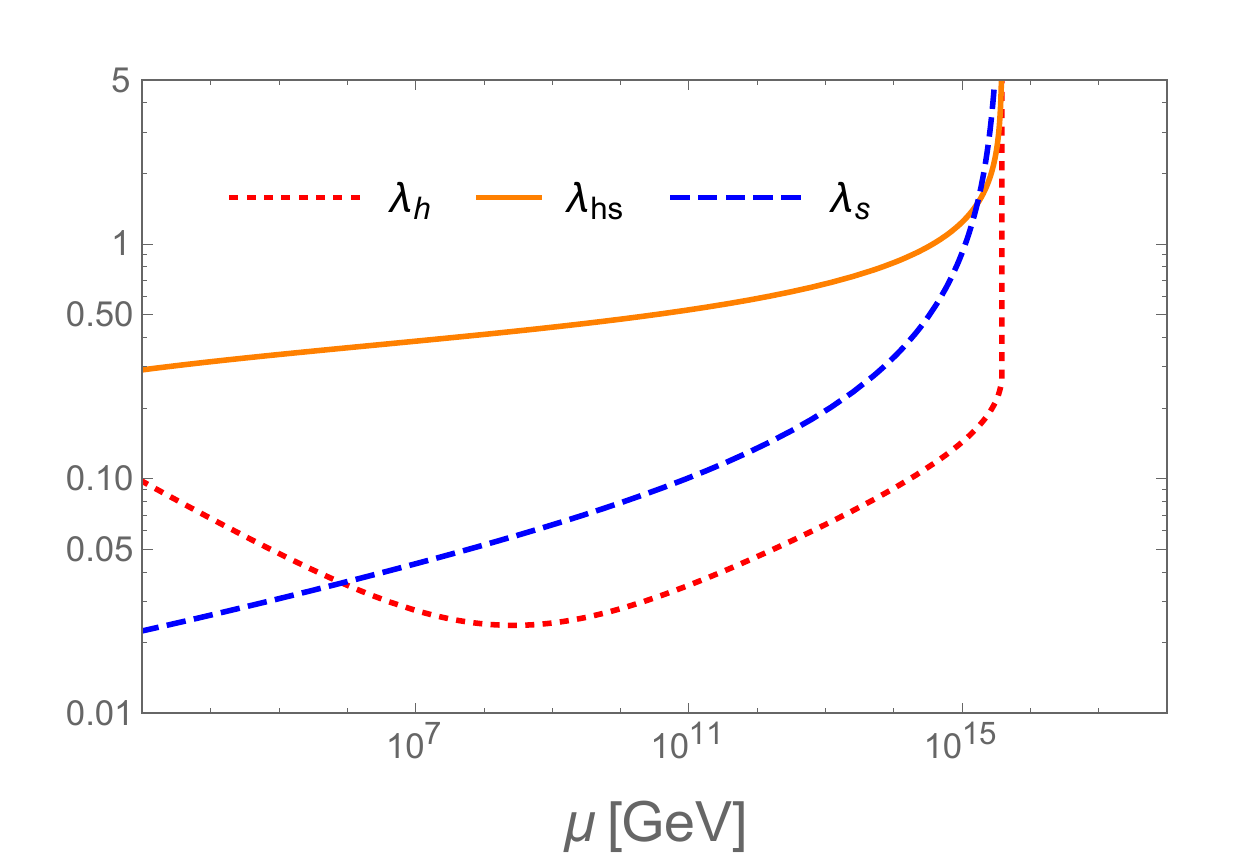}
\end{center}
\caption{RG running of $\lambda_s$, $\lambda_h$ and $\lambda_{hs}$. Left:  the scalar quartic couplings where the inflation is valid. Right:  the scalar quartic couplings lives in the parameter region where the inflation is invalid.  }
\label{fig:fig5}
\end{figure}

We first explore inflation dynamics without taking into account the Higgs precision bounds. In Fig.~\ref{fig:fig4}, we show the Higgs inflation feasible parameter regions in the plane of $(\lambda_s,\lambda_{hs})$ after imposing the theoretical constraints up to Planck scale as aforementioned in Sec.~\ref{sec:thcons}. Where, the upper and lower bounds of $\lambda_{hs}$ are mostly coming from perturbativity and unitarity, and stability conditions.
The inflation feasible range in the plane of $(\lambda_s,\lambda_{hs})$ is largest when $N=1$. The feasible ranges diminish with the increase of $N$ and are overlapped for the two neighbor $N$ expect $N=1$ and $N=2$.
The decrease of the inflation valid area with the increase of $N$ is due to the fact that: a larger N leads to more contributions of $\lambda_{hs}$ to $\lambda_h$ at the inflation scale through RG running (using the RGEs given in Appendix.~\ref{sec:RGEs} ),
and therefore the stability, perturbativity and unitarity set the lower and upper bounds of $\lambda_{hs}$. We plot the RG running of the scalar quartic couplings for the case of $N=7,10,13$ in Fig.~\ref{fig:fig5}.
The perturbativity of quartic couplings and the unitarity can be violated due to RG running of couplings as shown in the right panel of Fig.~\ref{fig:fig5}.

\begin{figure}[!ht]
\begin{center}
\includegraphics[width=0.5 \textwidth]{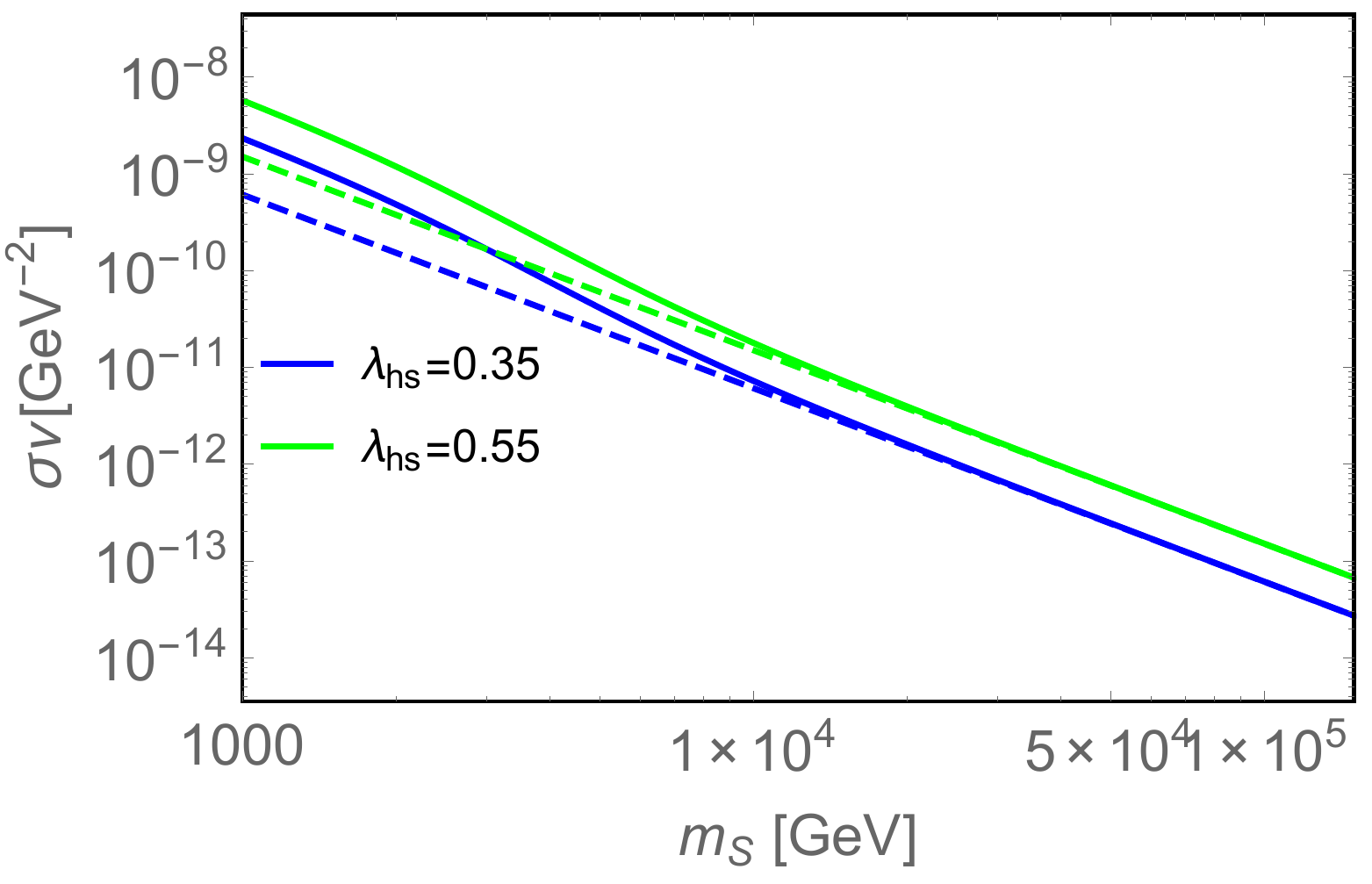}
\end{center}
\caption{DM annihilation cross section with the dashed lines indicating the seagull diagram contribution and the solid lines being the all annihilation channels contributions.}\label{fig:msans}
\end{figure}

With thermal averaged annihilation cross sections being the same as in Ref.~\cite{Lerner:2009xg} for each $S_i$, see
Appendix.~\ref{ap:DM}, and using Lee-Weinberg method~\cite{Kolb:1985nn}, we can estimate $\Omega^{S_i}h^2\sim 1/\sigma v_{rel} \sim m_{S_i}^2/\lambda_{hs}^2$ for a large dark matter mass. Previous studies show that the mass region of Higgs-portal real 1-singlet scalar DM case is excluded up to $\sim $TeV scale by
Xenon1T~\cite{Cline:2013gha,Cheng:2018ajh}.

\begin{figure}[!htp]
\begin{center}
\includegraphics[width=0.3 \textwidth]{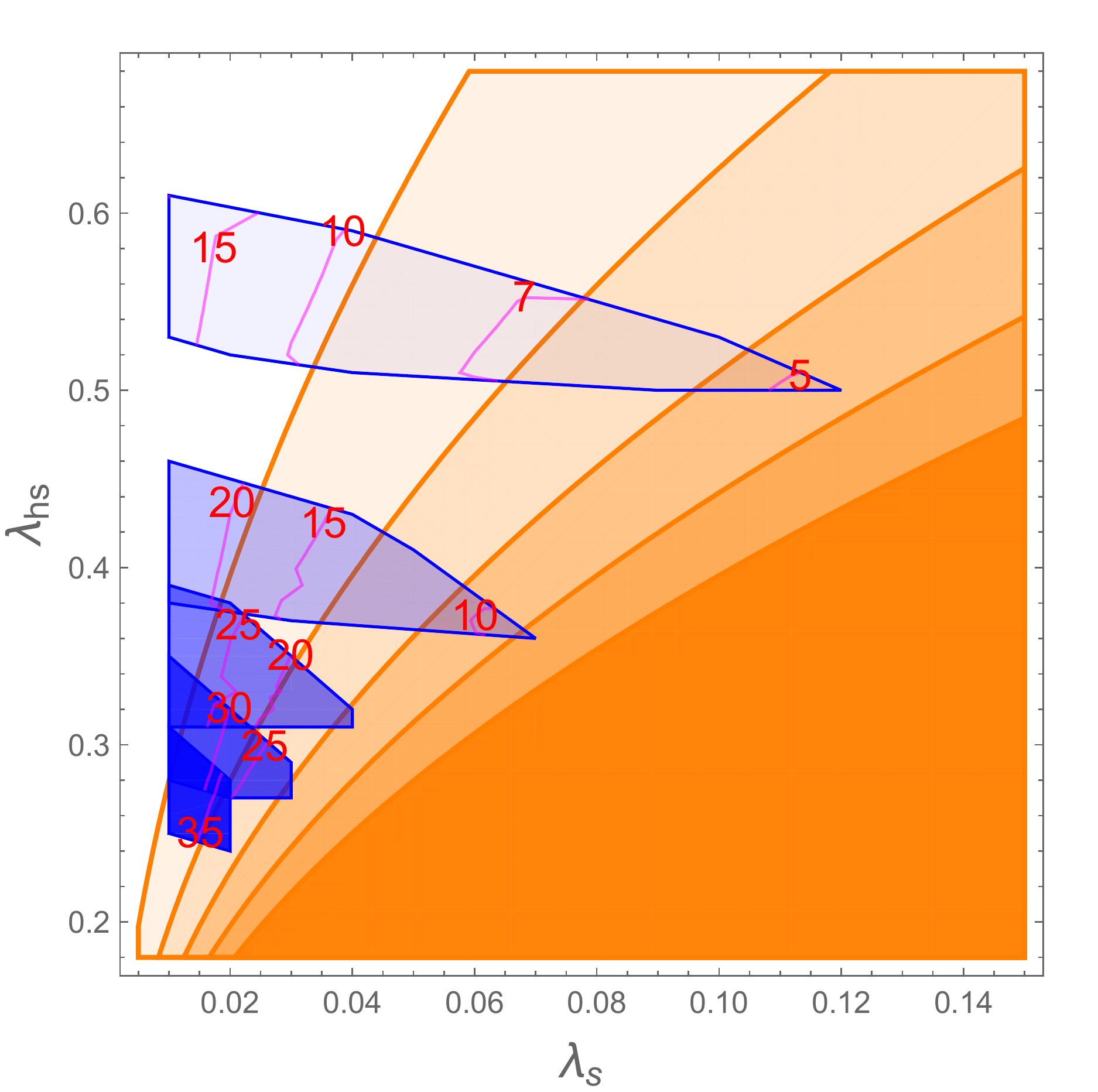}
\includegraphics[width=0.3 \textwidth]{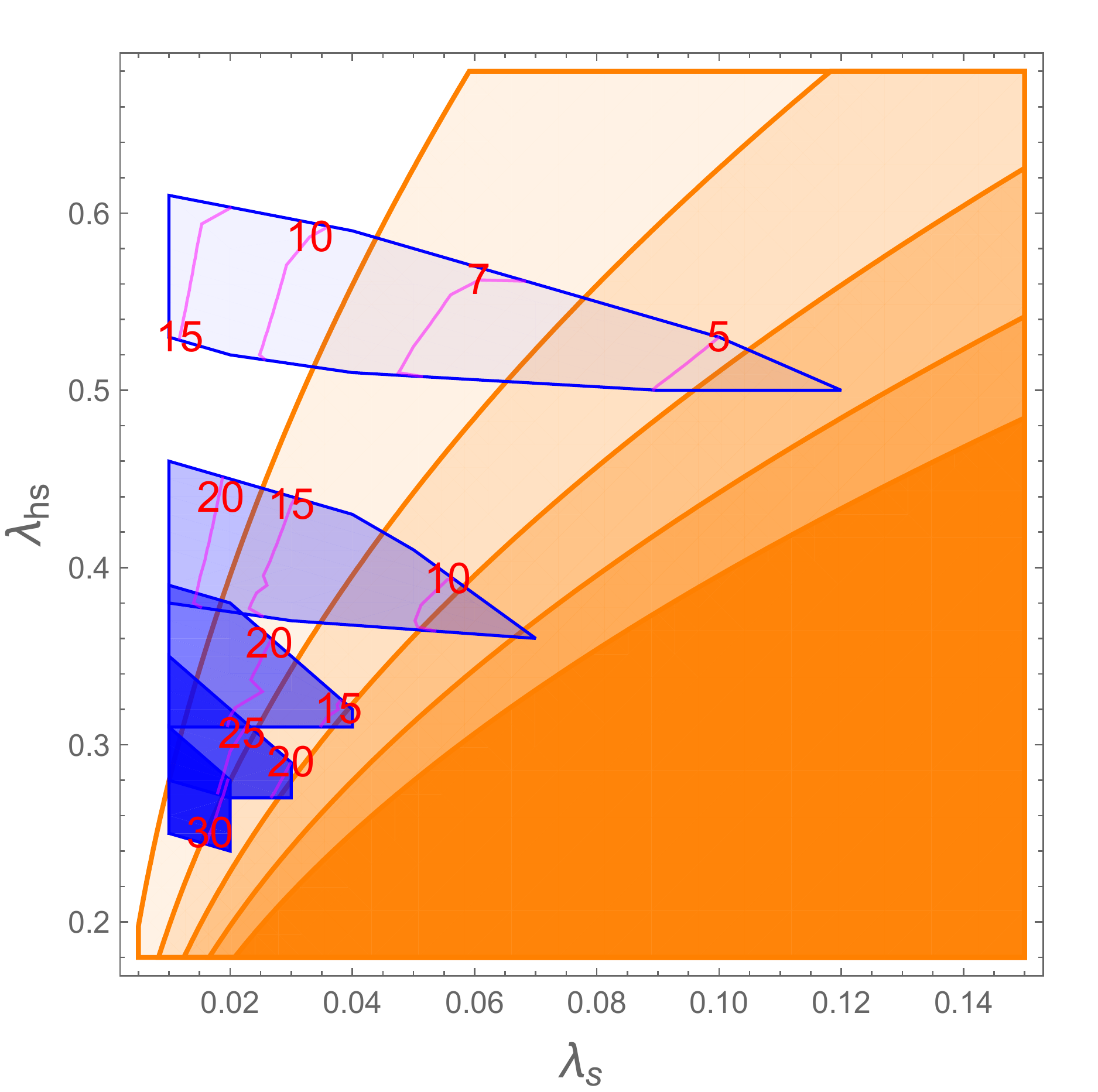}
\includegraphics[width=0.3 \textwidth]{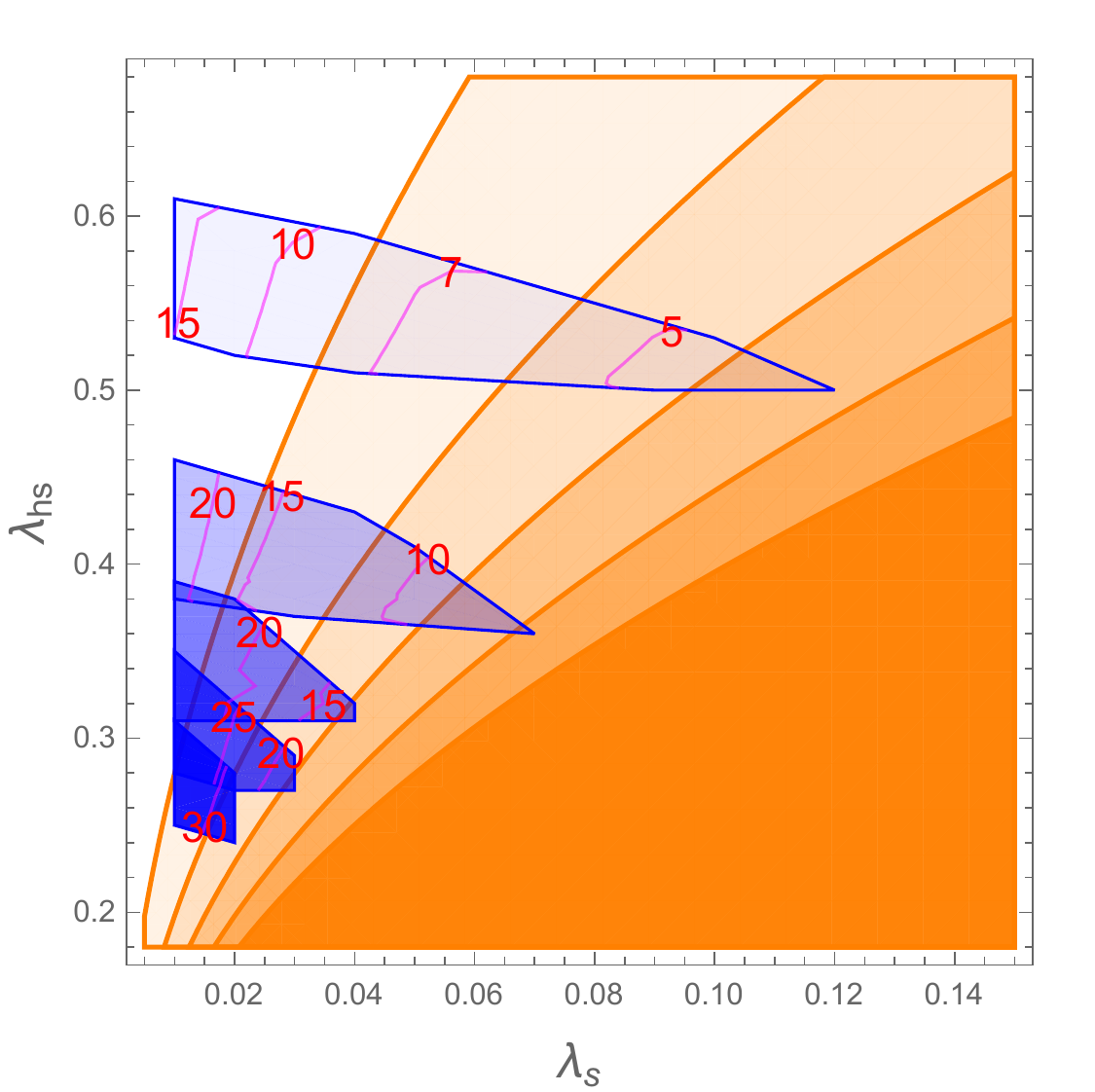}
\end{center}
\caption{Colliders, EWPOs, and inflation constraints on $m_{S_i}$ in the $(\lambda_s,\lambda_{hs})$ plane for different $N$. The magenta line is the allowed magnitude of $m_{S_i} $[TeV] by by the CEPC, ILC, and FCC-ee from left to right panel. Blue regions is for the feasible inflation within $O(N)$ model, orange regions represent the allowed regions by EWPOs confine. For both two color-codes, a deeper color corresponds to a larger $N$ with $N=1,2,3,4,5$.}\label{fig:fig20}
\end{figure}

It should be noted that the future Linear collider constraints would limit the dark matter mass to be TeV scale ($\sim O(1-10) $ TeV) in the inflation feasible region as shown in Fig.~\ref{fig:fig20}. In this case,
 the seagull diagram dominates the contributions to the dark matter pair
annihilations.  We show the annihilation cross section in Fig.~\ref{fig:msans} to illustrate that.
It's easy to see that the contributions of the Eq.~\ref{eq:annhpair} or Eq.~\ref{eq:annhpair2} would oversaturate the
 relic abundance for the inflation feasible $\lambda_{hs}$ though the highly suppress of $\sigma^{SI}$ by large $m_S$ make the mass region safe from Xenon 1T.

For $m_{DM}>10$ TeV, one obtains $T_{fs}\sim m_{DM}/x_f> 500$ GeV, therefore the dark matter freeze out happens earlier than the EWPT, and thus only the seagull diagram process can happen with the Higgs finite states have effectively $zero$ masses. Then, the annihilation cross sections of Eq.~\ref{eq:annhpair} reduces to
\bea \langle\sigma v_{rel}\rangle_{hh}& =& \frac{\lhs^{2}}{64 \pi m_s^{2}}~.\label{eq:annhpair2}
\eea
Here we point out that $m_s\sim\mu_s$ due to the Electroweak symmetry is still kept at temperature higher than $T_c$ within the framework of EWBG.
Then, if the relic abundance is partially saturated by $S_i$, one needs a larger H-S quartic coupling $\lambda_{hs}$. The large $m_S$, in fact, would decouple from the phase transition has been studied in Ref.~\cite{Espinosa:2008kw}. Furthermore, the largeness of the $\lambda_{hs}$ may result in the perturbativity and unitarity problem of quartic couplings $\lambda_{h,hs}$,
as shown in Fig.~\ref{fig:fig5} (after taking into account the RG running effects), and therefore shut down the possibility to explain inflation.

We briefly summarize this section as follows.
The numerical survey of the two step EWPT shows that a SFOEWPT requires N$ \geq$7 in the inflation favored parameter regions. Unfortunately, as can be seen from Fig.\ref{fig:fig20}, the inflation valid $N$ is bounded to be $N<4$ after imposing the constraints from EWPOs. Which therefore shout down the window to realize the SFOEWPT. The $c^N_H v^2/m_{S_i}^2$(the $c^N_H$ is given by Eq.~\ref{eq:cnh}) is bounded to be smaller than 0.0038, 0.0034, and 0.0028 by CEPC with the luminosity of 5 $ab^{-1}$, ILC with all center of mass energies, and FCC-ee with the luminosity of
 10 $ab^{-1}$. With which, we can obtain the bounds on the hidden scalar masses of $m_{S_i} \sim \mathcal{O}(1-10)$ TeV. In this mass region, the N scalars cannot explain the correct DM relic density.

\subsection{$ O(N\to N-1)$ scenario}
\label{sec:NtoN1}

\begin{figure}[!htp]
\begin{center}
\includegraphics[width=0.4 \textwidth]{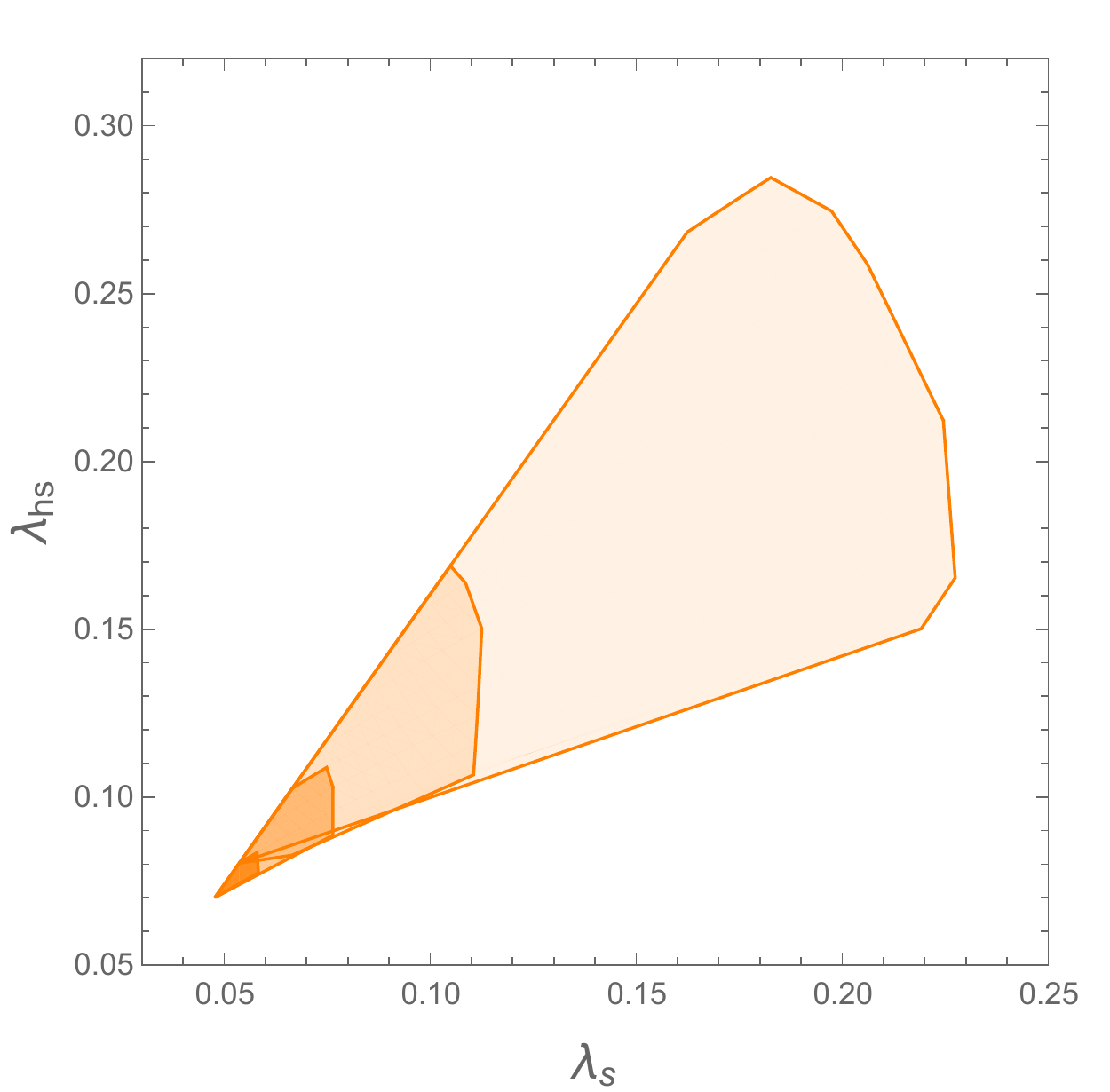}
\includegraphics[width=0.4 \textwidth]{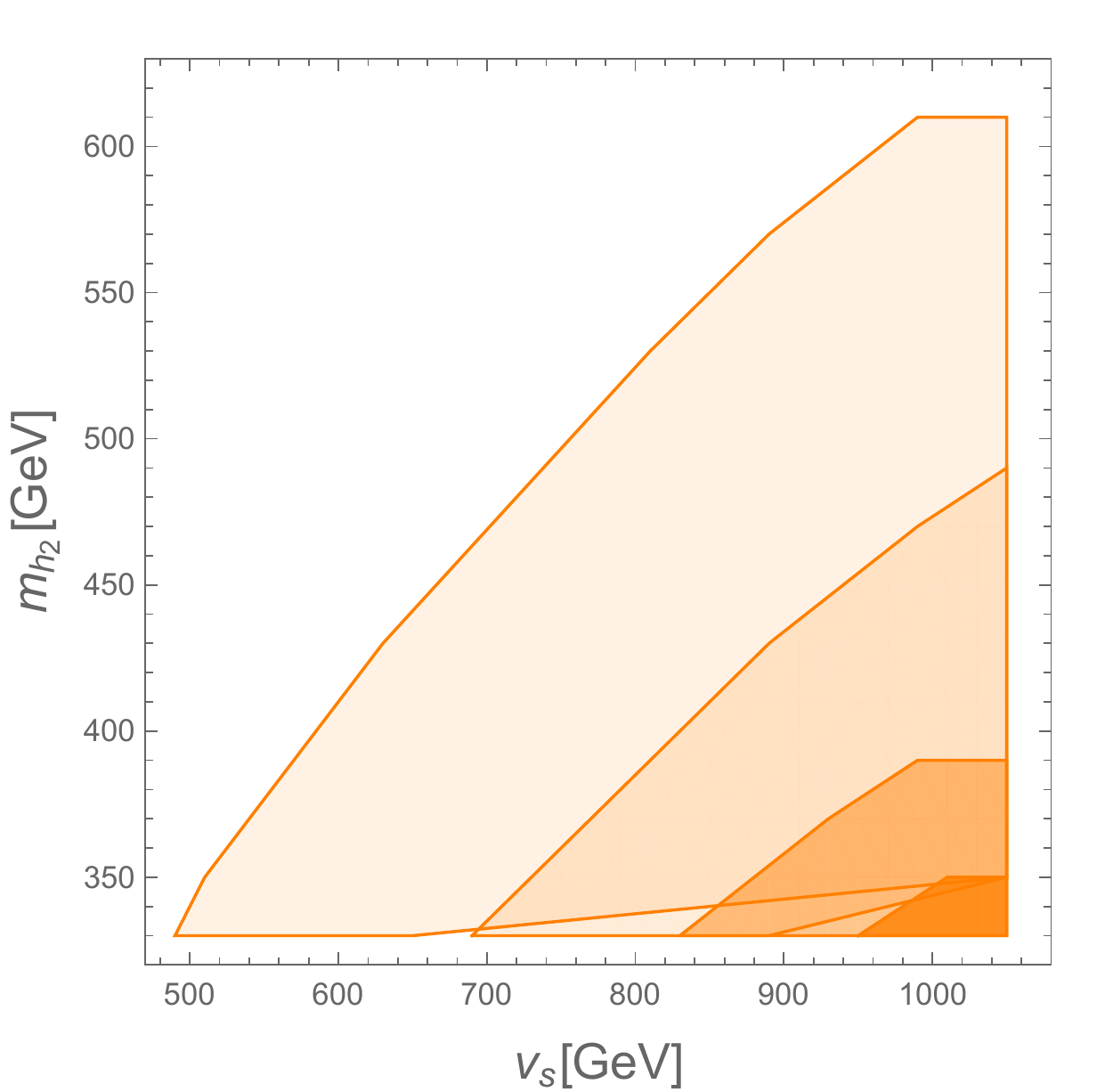}
\caption{Inflation feasible parameter planes of $(\lambda_s,\lambda_{hs})$ and $(v_s,m_{h_2})$ for different $N$ within $O(N \to N-1)$ scalar model, a deeper color corresponds to a larger $N$, the corresponding N are 1, 2, 3 and 4, respectively.}\label{fig:fig6}
\end{center}
\end{figure}

\begin{figure}[!htp]
\begin{center}
\includegraphics[width=0.8 \textwidth]{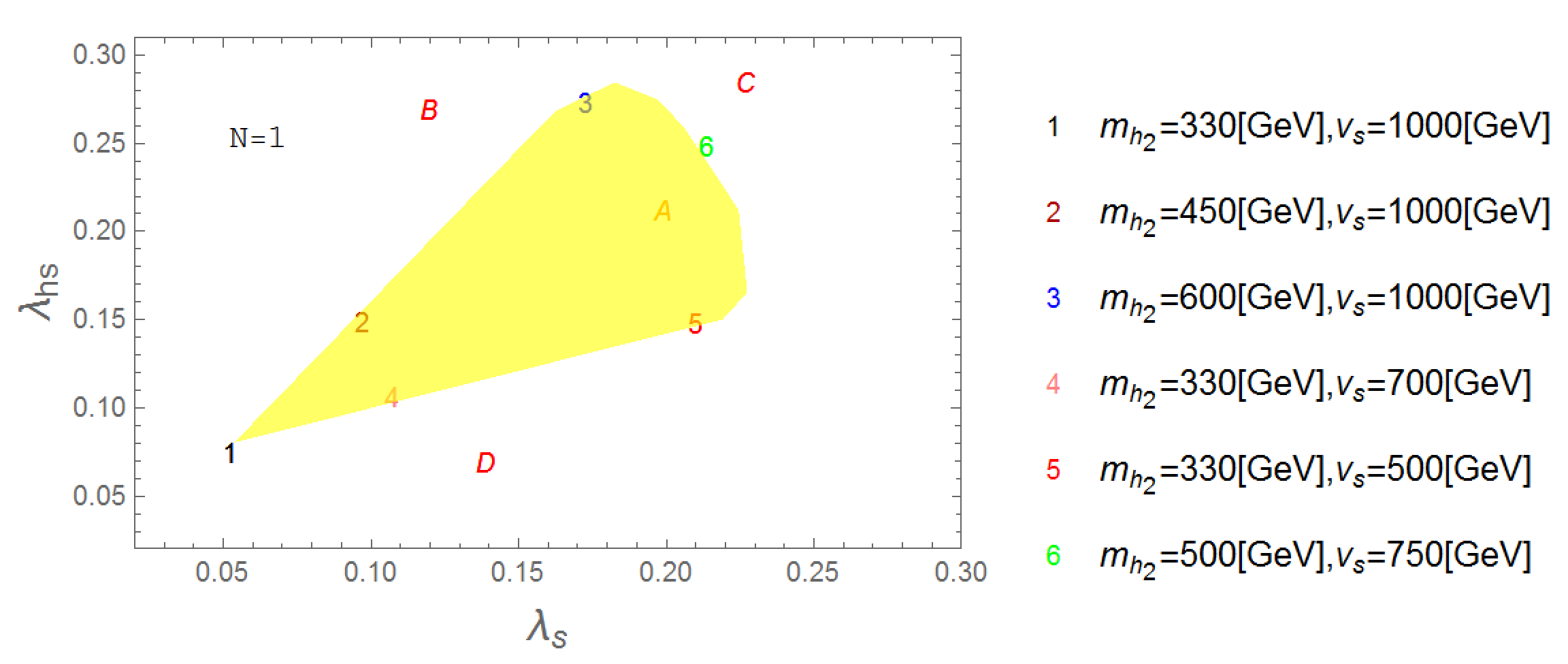}
\includegraphics[width=0.4 \textwidth]{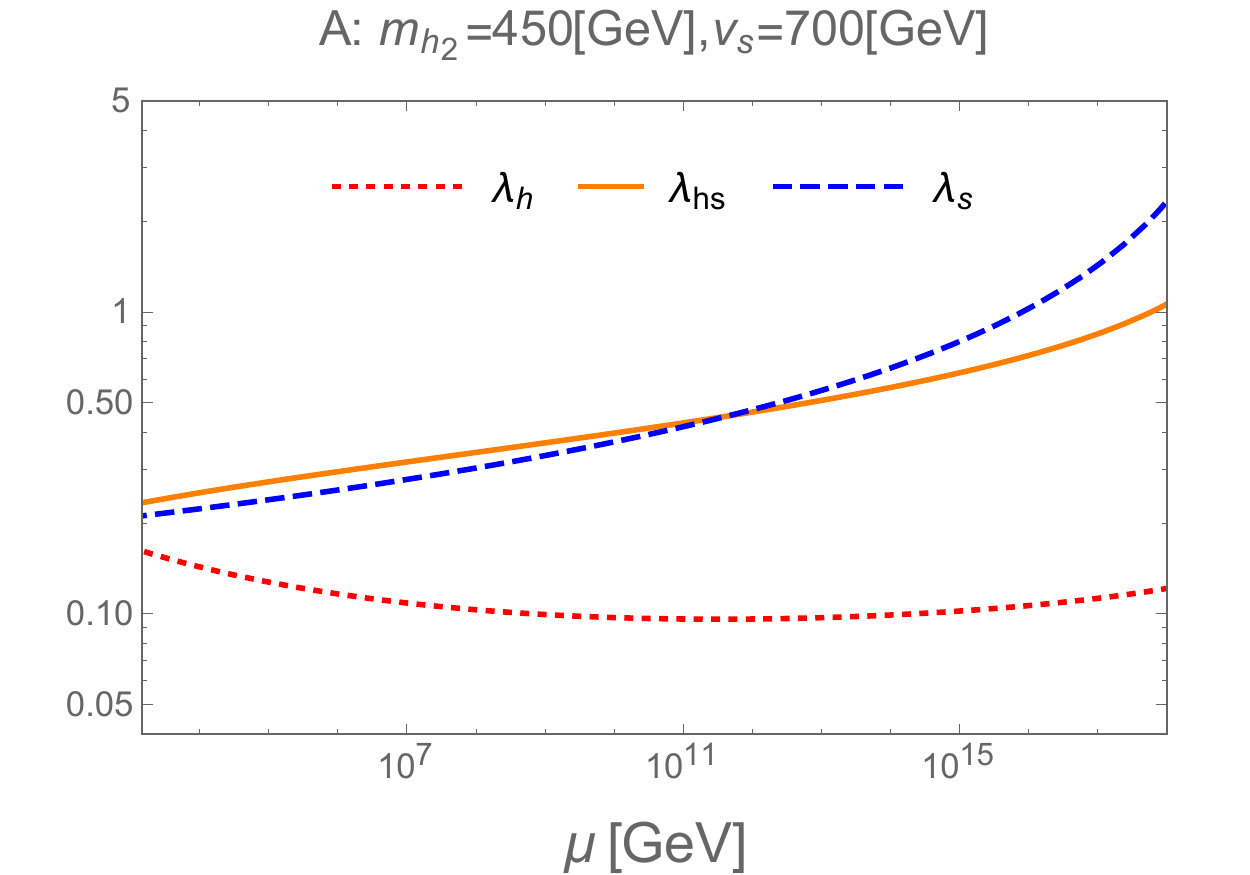}
\includegraphics[width=0.4 \textwidth]{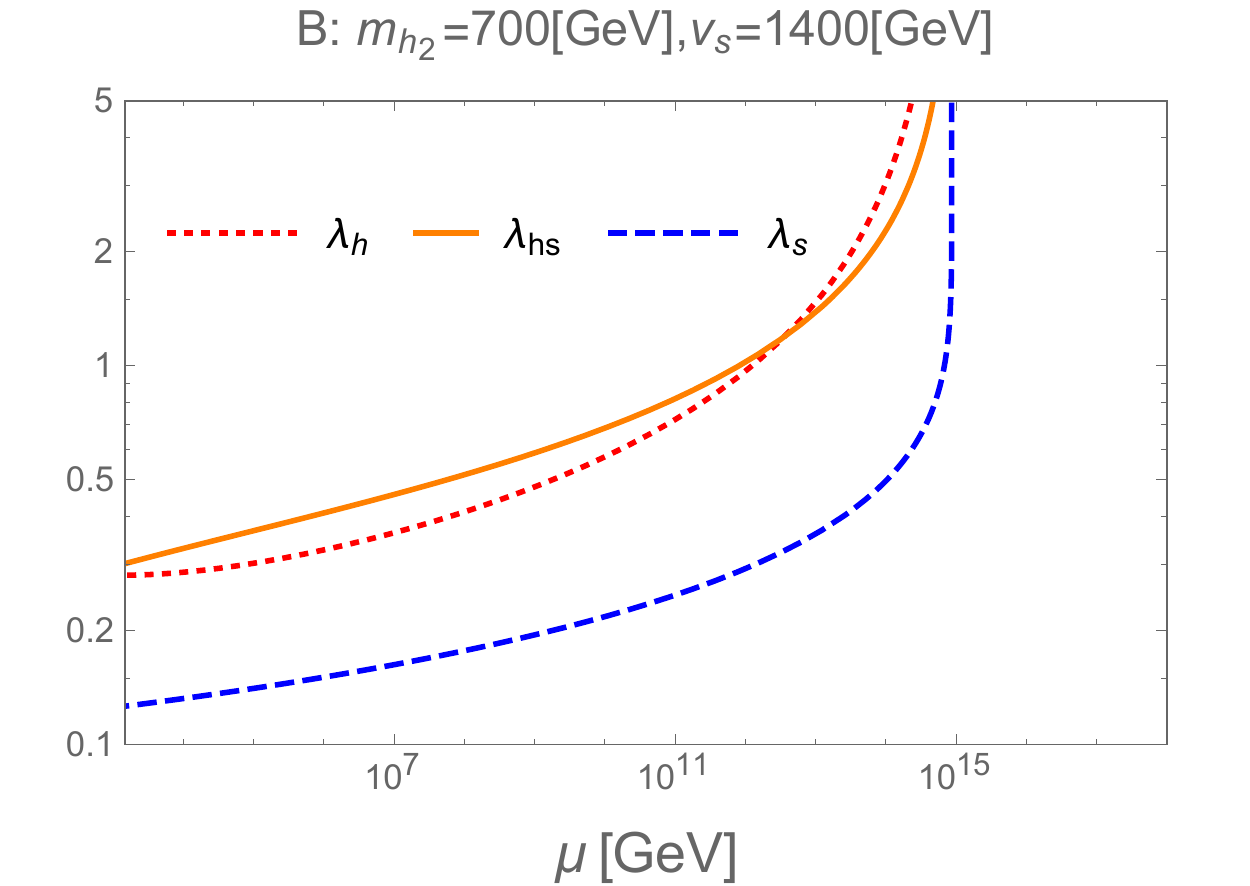}
\includegraphics[width=0.4 \textwidth]{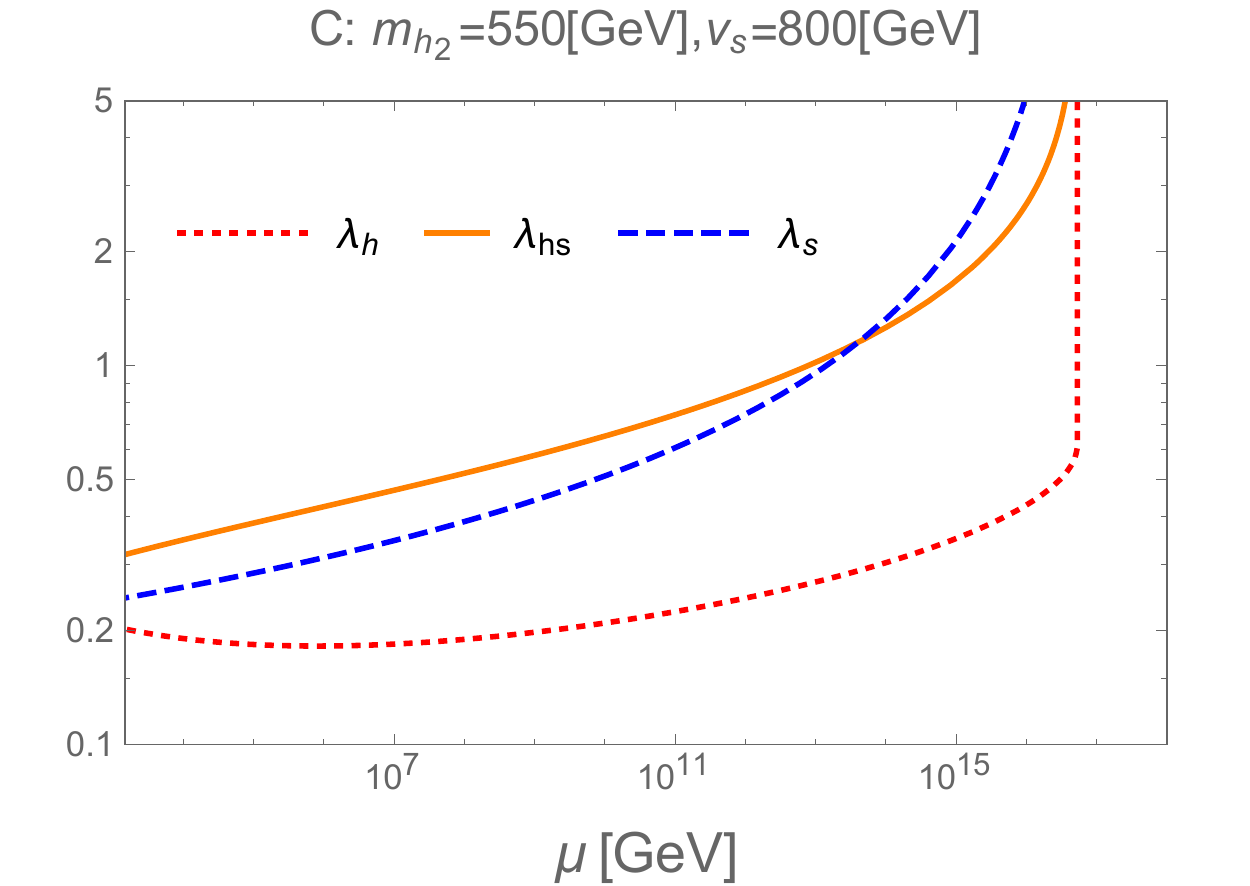}
\includegraphics[width=0.4 \textwidth]{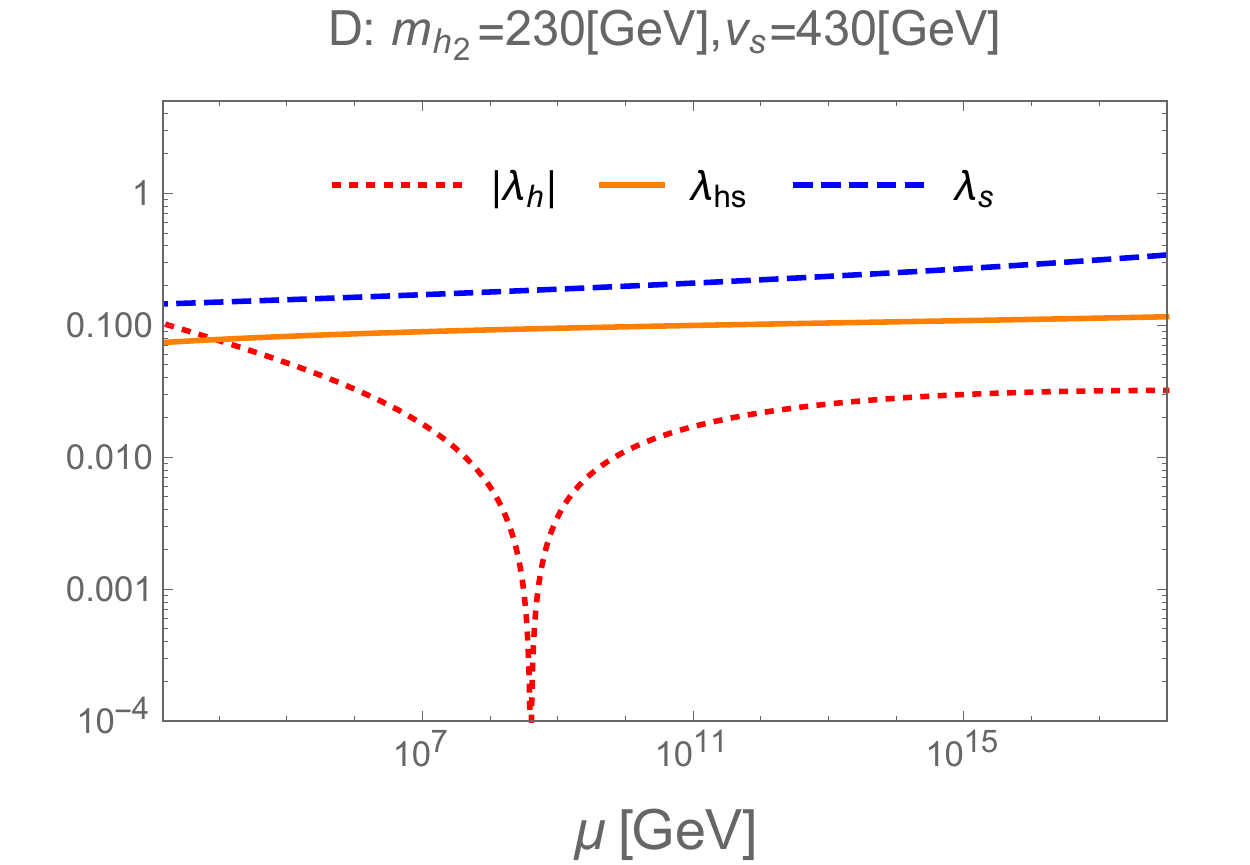}
\caption{Top panel: the Higgs inflation feasible parameter spaces of ($\lambda_s$,$\lambda_{hs}$). Middle and bottom panels: four samples of the RG running coupling ($\lambda_{h,hs,s}$).}\label{fig:inflationregion2}
\end{center}
\end{figure}

As in the previous section, here we first perform the inflation analysis before considering the Higgs precision.
We show the Higgs inflation valid parameter spaces in Fig.~\ref{fig:fig6}, the feasible ranges diminish with the increase of $N$ and are overlapped for the two contiguous $N$. The left panel indicates that the magnitude of $\lambda_{hs}$ increases with the increase of $\lambda_{s}$, which is different from the $O(N)$ scalar model scenario being explored in the previous section.
An interesting triangular shape shows up due to the bounds on $m_{h_2}$ and $v_s$ from perturbativity, unitarity, and stability, together with the relations among quartic couplings, the Higgses masses and VEVs.
Here, it should be noted that the lower bound of $m_{h_2}$ is set by the stability bounds. The scalar quartic Higgs coupling will be negative for $m_{h_2}<$330 GeV, this set the lower boundary of the inflation valid parameter region.  The upper bound is set by the magnitude of the $v_s$ assisted by the $m_{h_2}$, and the upper bound on $m_{h_2}<600$ GeV is to fulfill the perturbativity and unitarity conditions at high scale. With the increase of N, the perturbativity and unitarity conditions, together with stability requirement results in a smaller parameter region of $\lambda_{hs}$ and $\lambda_s$ as shown in Fig.~\ref{fig:fig6}. The slow-roll inflation is almost excluded for $N \ge 5$.

To explain the property more transparent, we plot the Fig.\ref{fig:inflationregion2} by taking the $N=1$ case as an example.
In the upper plot, the yellow region stands for the inflation feasible region for $N=1$ in the ($\lambda_s,\lambda_{hs})$ plane within the $O(N \to N-1)$ scalar model with the number of Goldstones being $zero$, the values of $m_{h_2}$ and $v_s$ for the numbers and alphabets can be found at the right hand of the upper, middle and bottom panels, respectively. The middle and bottom panels show the RG running of couplings (for the points A, B, C and D) from Electroweak scale to Planck scale. Note that, we use an absolute value for the coupling $\lambda_h$ in the last figure, the downward tip there means the stability is violated at the point of D. That indicates that the stability is the lower bound for ($\lambda_s,\lambda_{hs}$) plane in the $O(N \to N-1)$ scenario. For B and C points, the perturbativity and unitarity are violated due to the RG running of couplings. This indicates that the perturbativity and the unitarity set the upper bound.

\begin{figure}[!ht]
\begin{center}
\includegraphics[width=0.4 \textwidth]{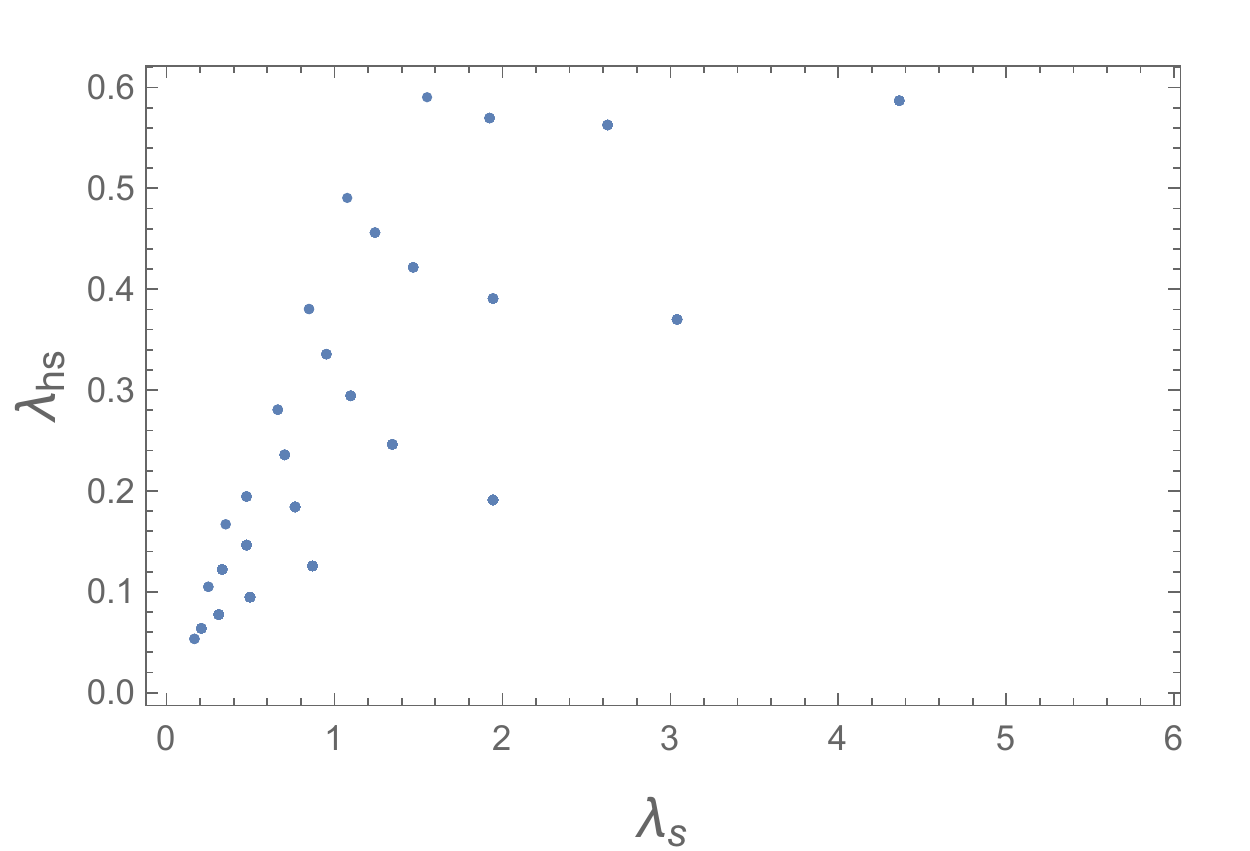}
\includegraphics[width=0.4 \textwidth]{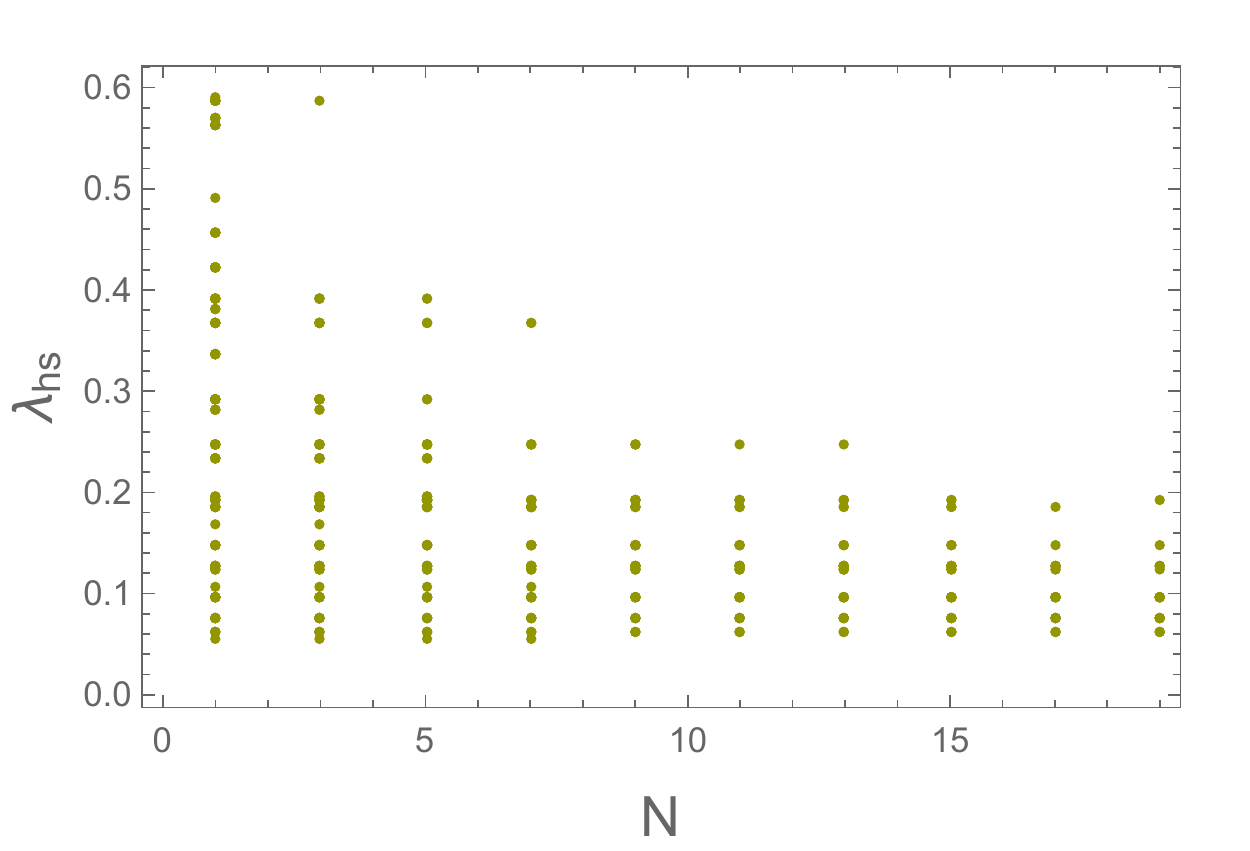}
\includegraphics[width=0.4 \textwidth]{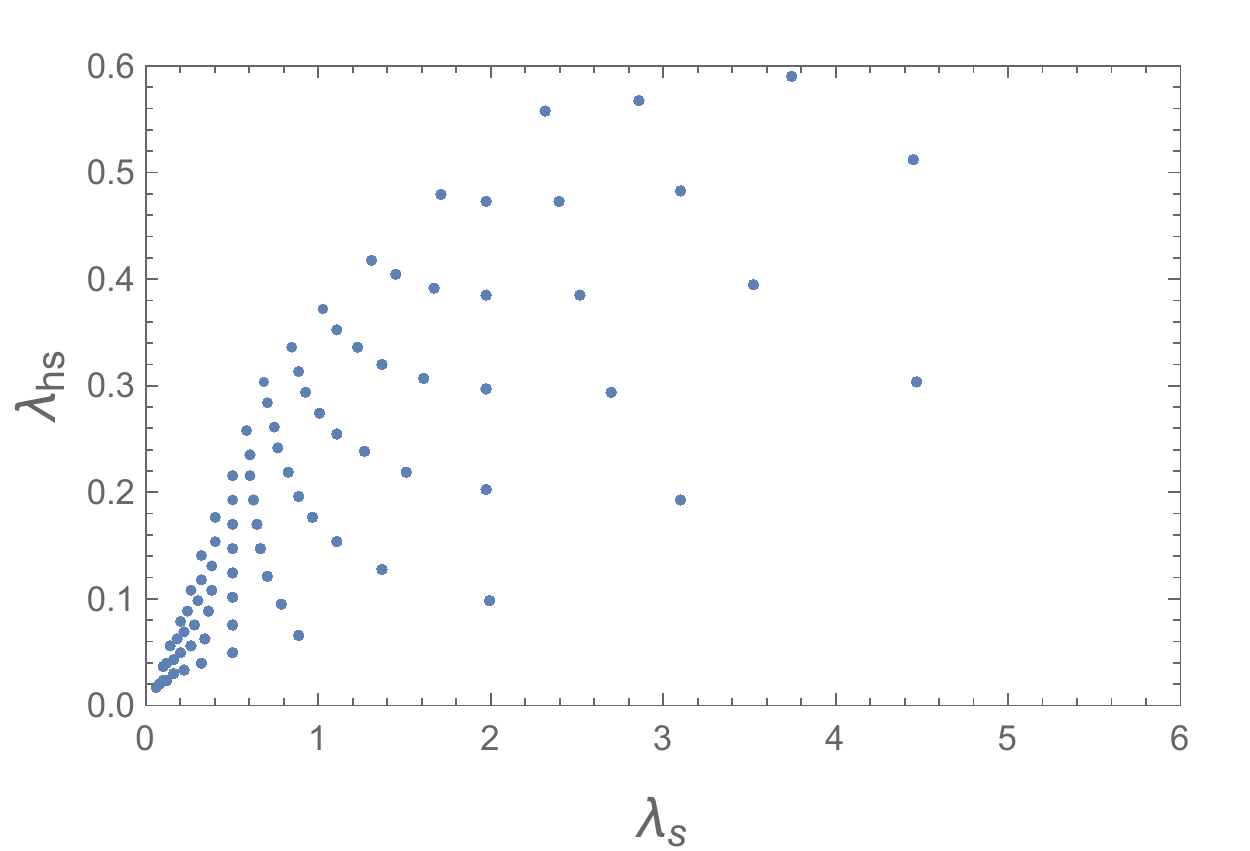}
\includegraphics[width=0.4 \textwidth]{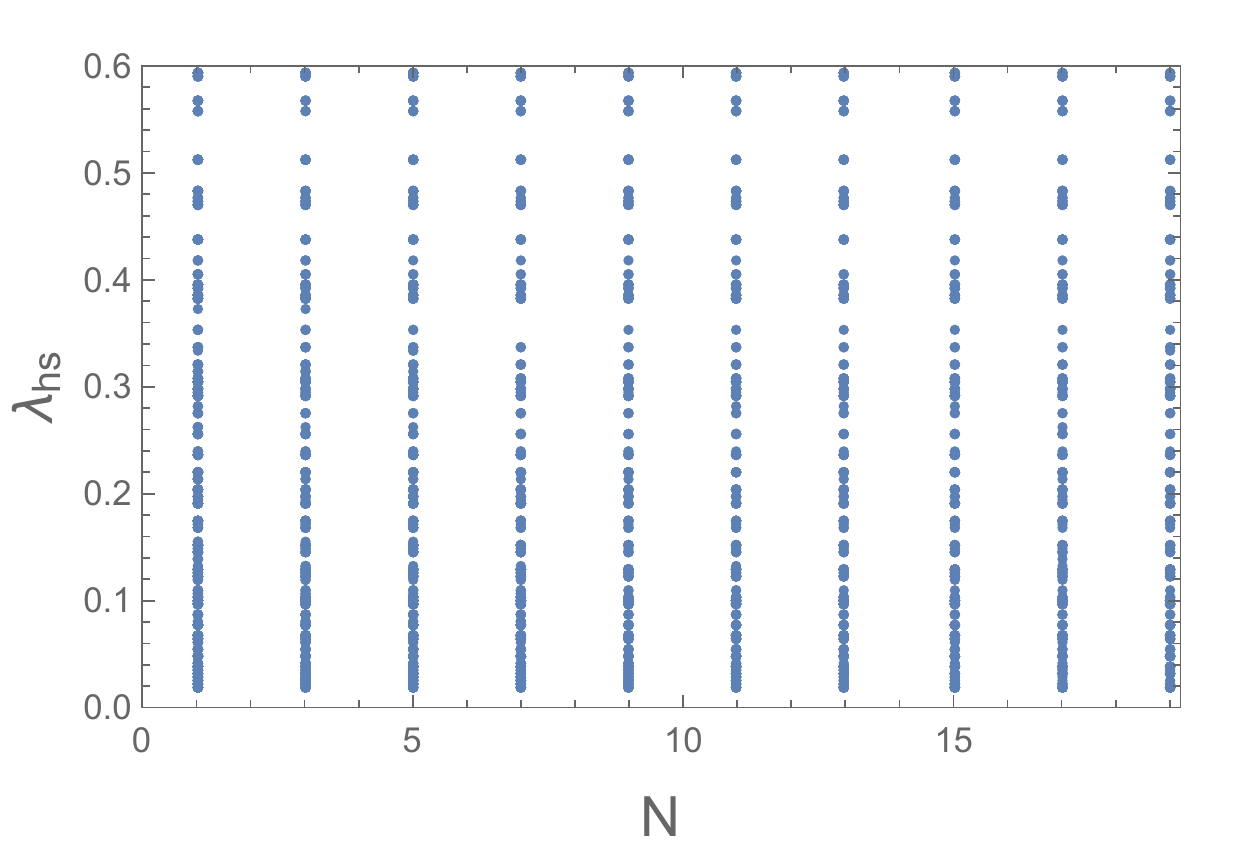}
\end{center}
\caption{ One-step (top panel) and two-step (bottom panel) SFOEWPT valid points within the $O(N \to N-1)$ scenario.}\label{fig:ONTON1EWPTone202}
\end{figure}

\begin{figure}[!htp]
\begin{center}
\includegraphics[width=0.4 \textwidth]{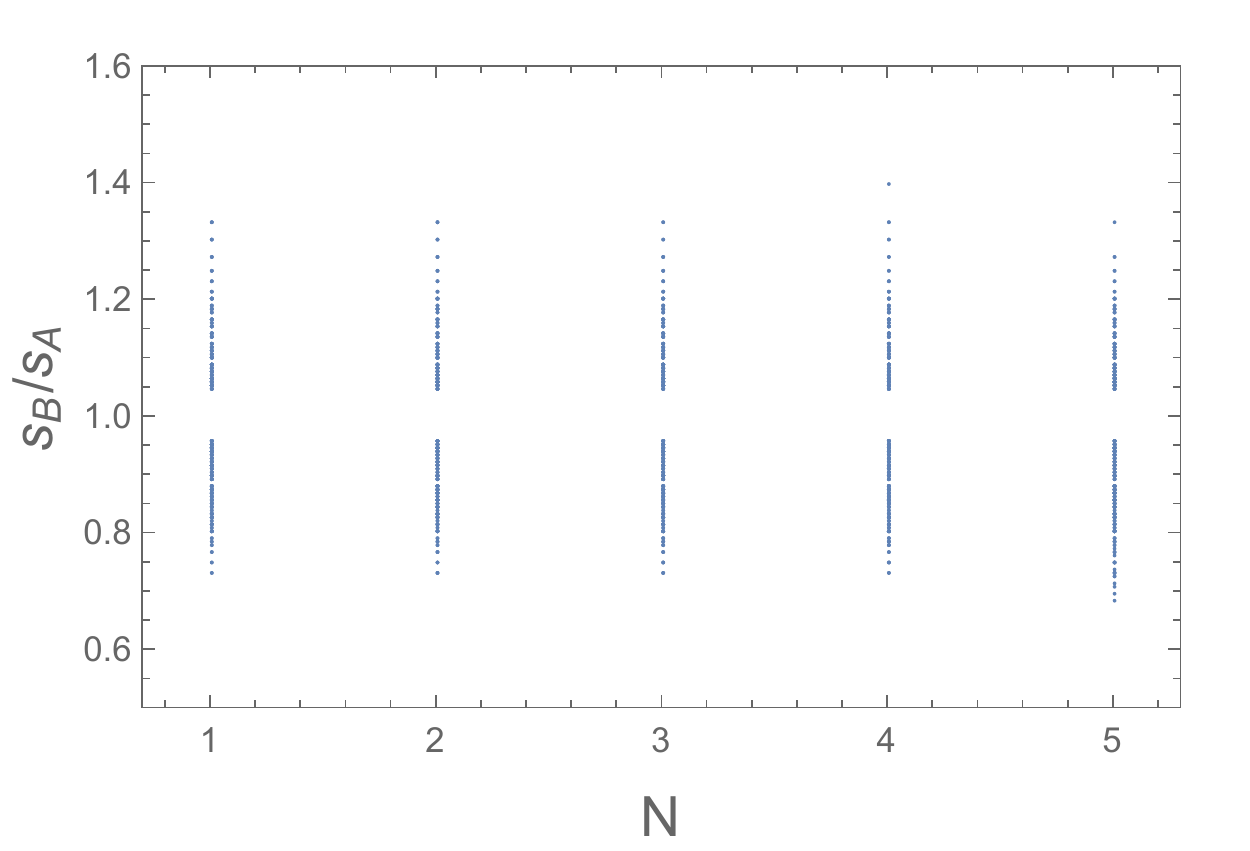}
\caption{The two-step SFOEWPT points in $O(N \to N-1)$ model.}\label{fig:SBSA}
\end{center}
\end{figure}

Now, we explore the EWPT property for the $O(N\to N-1)$ scenario.
For the one-step case, one can realize a SFOEWPT with a small $\lambda_{hs}$ with increasing of N, as shown in Fig.\ref{fig:ONTON1EWPTone202}. Which means that the Goldstones contribution to the EWPT is notable. While, this property disappears in the two-step scenario as can be seen in the bottom-right panel. The results show that different from one-step situation, the SFOEWPT occurs more easy with a relatively small value of the $\lambda_{hs}$ for the two-step case. Moreover, our study demonstrates that the rate of $s_B/s_A$ can be larger or smaller than 1 for different $N$, as shown in Fig.\ref{fig:SBSA}, which reconfirms the study of Ref.~\cite{Cheng:2018ajh}.

\begin{figure}[!htp]
\begin{center}
\includegraphics[width=0.45 \textwidth]{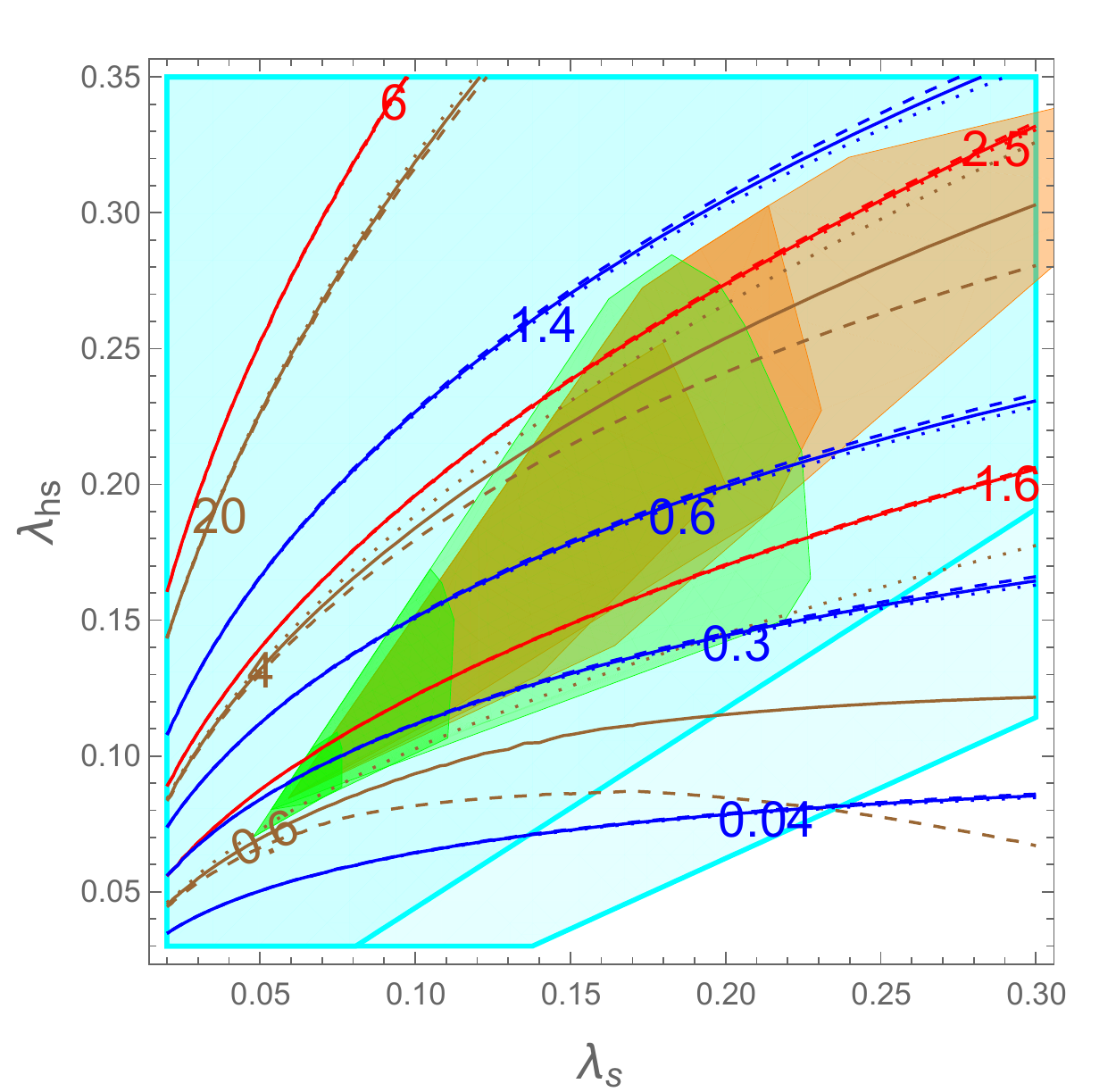}
\includegraphics[width=0.45 \textwidth]{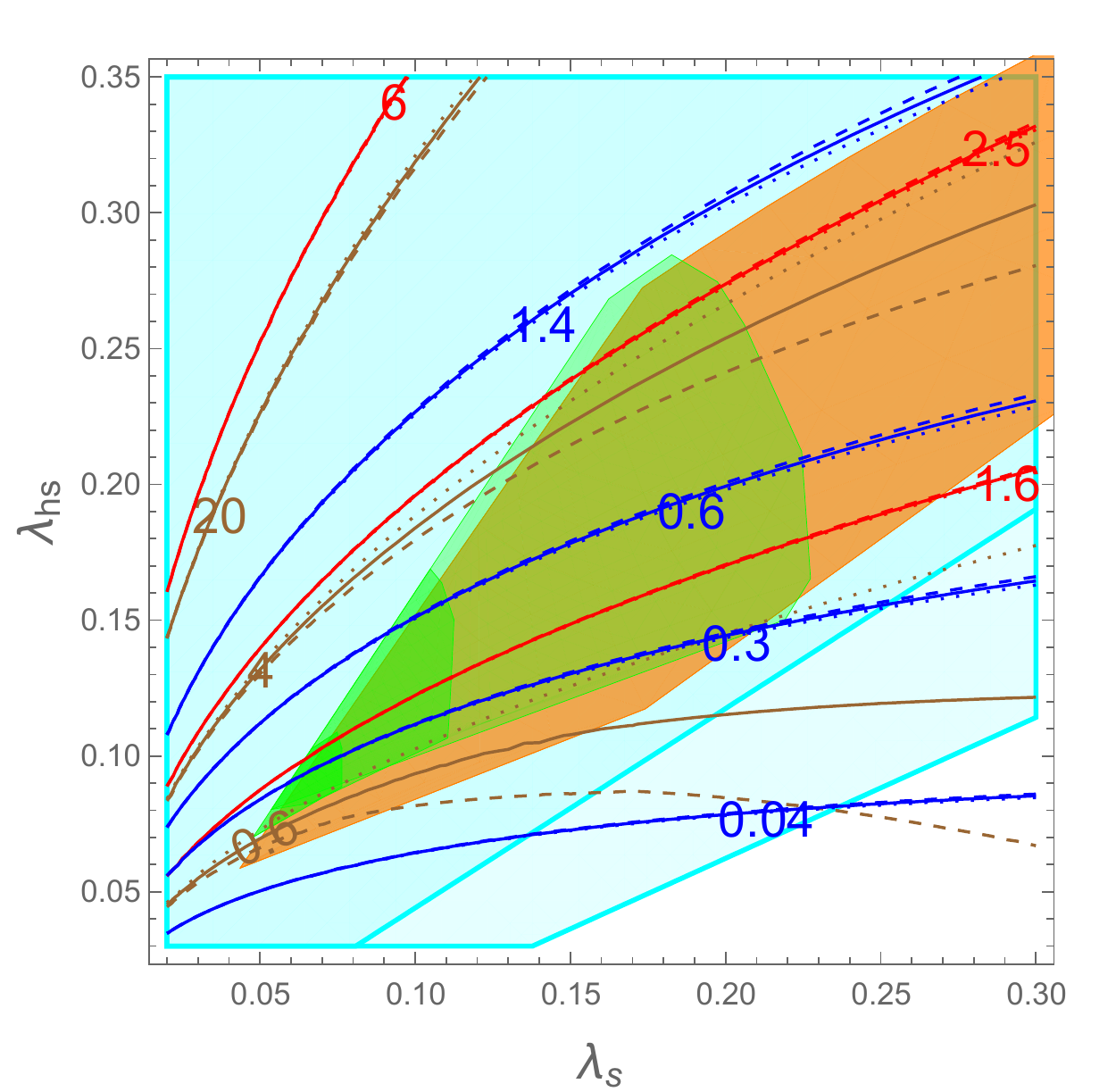}
\end{center}
\caption{The $O(N \to N-1)$ scenario. The $\lambda_{h_1h_1h_1}$, $\lambda_{h_2h_1h_1}$(both normalized by $\lambda_{hhh}^{SM}$), and $\Gamma_{h_2}^{tot}$ are shown by red, blue, and brown contours, respectively. Dashed, solid and dotted lines stand for $N=1,2,3$, respectively. Cyan regions present the allowed regions by the invisible Higgs decay bounds from $B_{inv}<0.34$~\cite{Khachatryan:2016vau}, in which the light and deep colors correspond to $N=2,3$, respectively. The feasible regions of inflation are shown by green color regions, and the orange regions represent one- and two- step SFOEWPT for the left and right panels, respectively. For those colors, a deeper color corresponds to a larger $N$ for $N=1,2,3$.}\label{fig:allEWPTINGA12TIRH12}
\end{figure}

In Fig.\ref{fig:allEWPTINGA12TIRH12}, we show the parameter regions that can accommodate successful Higgs inflation and a SFOEWPT together. The Higgs cubic couplings and Higgs decay widths in $O(N\to N-1)$ model are also shown. The inflation and one(two)-step SFOEWPT are allowed by the Higgs invisible decay bounds from LHC~\cite{Khachatryan:2016vau} for different $N$, which is marked by cyan. As shown in Fig.\ref{fig:SBSA}, the two-step SFOEWT shows no obvious relation with $N$ due to $s_A$ can be higher or lower than $s_B$. For $N>1$, the slow-roll Higgs inflation does not occur in some SFOEWPT allowed parameter spaces with relatively large quartic couplings,
this property is caused by the bound on $m_{h_2}$ and $v_s$
from perturbative unitarity and stability (from Electroweak scale to Planck scale). For 1$< N\leq$3, more parameter spaces of $(\lambda_{s},\lambda_{hs})$ are allowed by two-step SFOEWPT condition in comparison with the one-step
SFOEWPT condition.
The ratio of the triple Higgs couplings ($r_{3h_1}=\lambda_{h_1 h_1h_1}/\lambda^{SM}_{h hh}$ and $r_{h_2h_1h_1}=\lambda_{h_2h_1h_1}/\lambda^{SM}_{h hh}$) increase with the increase (decrease) of $\lambda_{hs}$ ($\lambda_s$).
With the SM Higgs resonance search using the SM Higgs pairs production process, one can estimate the cross section with respect to the SM case being $\sigma_{h_1h_1}/\sigma^{SM}_{hh}\sim \cos^2\theta\times r_{3h_1}^2\times \Gamma^{tot}_{h_1}/\Gamma_h^{SM}\sim 0.98^2\times 2^2\sim 3.8$, and therefore a large enhancement of the cross section can be expected. With  increasing of $N$, one can expect the ratio $\sigma_{h_1h_1}/\sigma^{SM}_{h_1h_1}$ decreasing due to the decrease of $r_{3h_1}$. With increasing of $\lambda_{hs}$, the triple Higgs coupling of $\lambda_{h_2h_1h_1}$ varies in the range of $[0.3 , 1.4]$ in units of $\lambda^{SM}_{3h}$. The cross section of the heavy Higgs is $\sigma^{h_2}_{h_1h_1}\sim(\sqrt{2}m_t /v)^2\sin^2\theta\times\lambda_{h_2h_1h_1}^2/(m_{h_2}\Gamma_{h_2}^{tot})$. Due to the $\Gamma_{h_2}^{tot}\ll m_{h_2}$, the resonance interference explored in Ref.~\cite{Carena:2018vpt} can be safely neglected here. We postpone the detailed collider search of the parameter spaces to a separate publication.

This part study constitutes
 one of our main results: building a connection between the inflation/EWPT and the Goldstone numbers ($N-1$) of the spontaneously broken global symmetry group.  The feature being study here can exist in other hidden sector extended SM, provided the hidden sectors respect a global symmetry that will be broken to a subset of which, wherein the remnant Goldstones
will contribute to the thermal effective potential and the RGEs of scalar quartic couplings (for the inflationary potential).

\begin{figure}[!ht]
\begin{center}
\includegraphics[width=0.4 \textwidth]{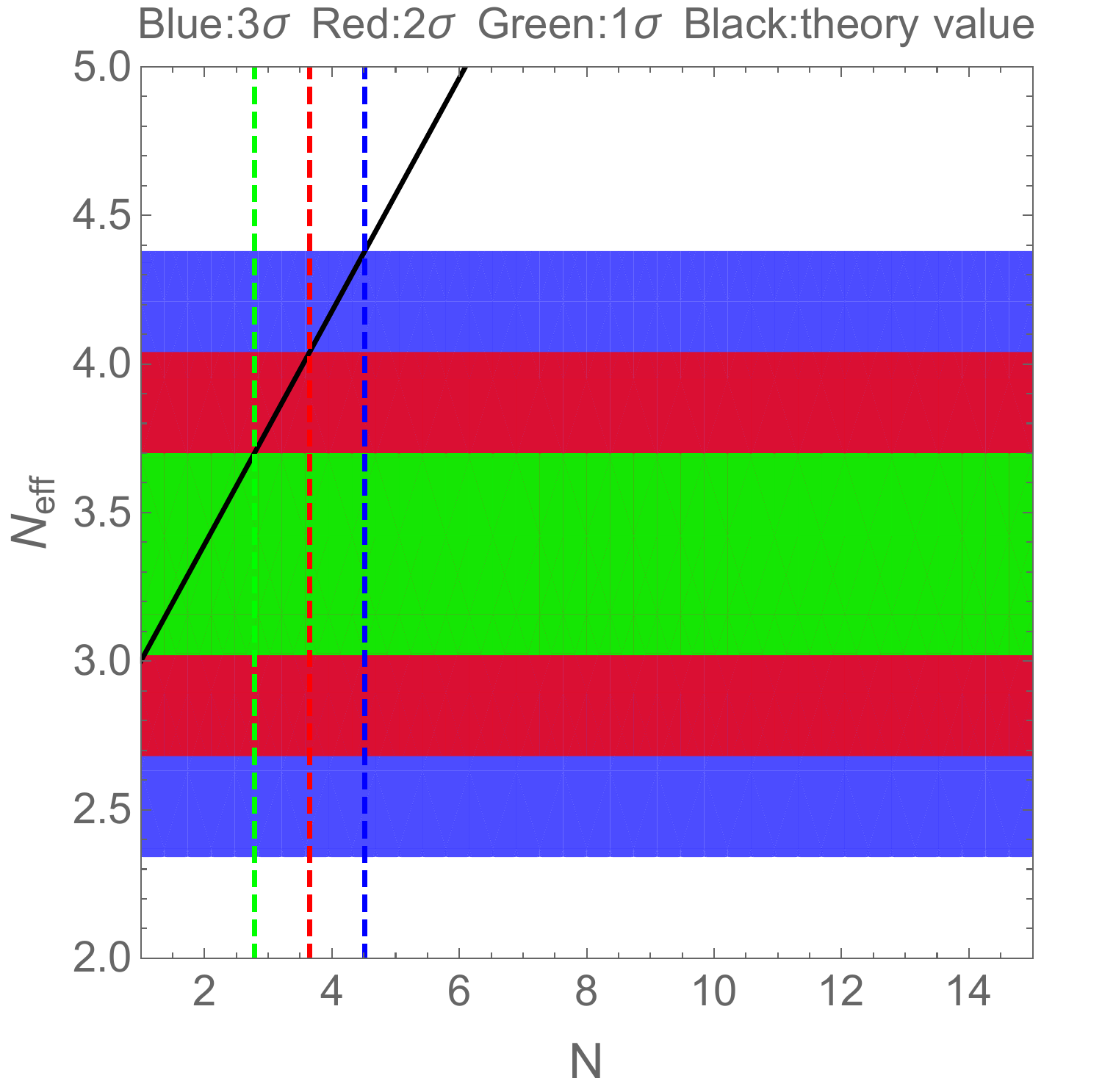}
\caption{The Goldstones faked effective neutrino number.}\label{fig:Neff}
\end{center}
\end{figure}

Since the Goldstone might fake the effective neutrino and contribute to the dark radiations, the number of Goldstones will be limited by related experiments.
 The
effective neutrino number can be expressed in terms of the Goldstone decoupling temperature as\cite{1310.6259,Steigman:2013yua},
\begin{equation}
N_{eff}=3\left(1+\frac{\Delta N_{s_{N-1}}}{3}\left(\frac{g_*(T_\nu^d)}{g_*(T_{s_{N-1}}^d)}\right)^{4/3}\right)\;,
\end{equation}
with $\Delta N_{s_{N-1}}=4(N-1)/7$ due to there is $(N-1)$ Goldstone bosons decoupled at $T>T_{s_{N-1}}^d$ and present before the recombination eras, the effective number of relativistic degrees of freedoms are $g_*(T_\nu^d)=43/4$ and $g_*(T_{s_{N-1}}^d)=57/4$
supposing that the Goldstone bosons decouple just before muon annihilation. One can constrain the
number of Goldstone as in Fig.~\ref{fig:Neff} using the recent 1$\sigma$ experimental data $N_{eff}=3.36\pm0.34$~\cite{Ade:2013zuv} . Which depicts that the number of $N\leq4$ at 3$\sigma$.

\begin{figure}[!ht]
\begin{center}
\includegraphics[width=0.4 \textwidth]{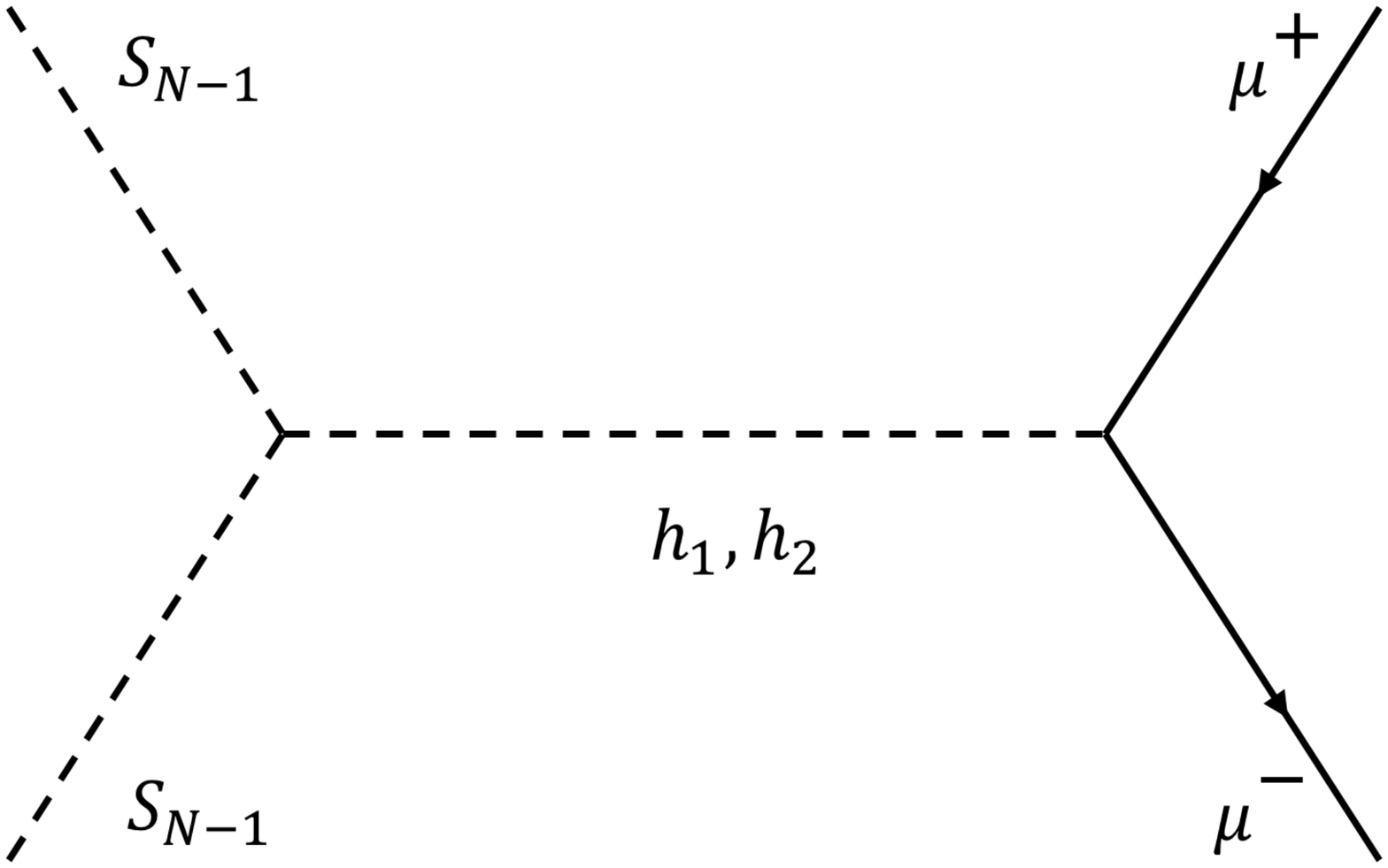}
\end{center}
\caption{Goldstone annihilation process.}\label{fig:goldstonann}
\end{figure}

\begin{figure}[!ht]
\begin{center}
\includegraphics[width=0.4 \textwidth]{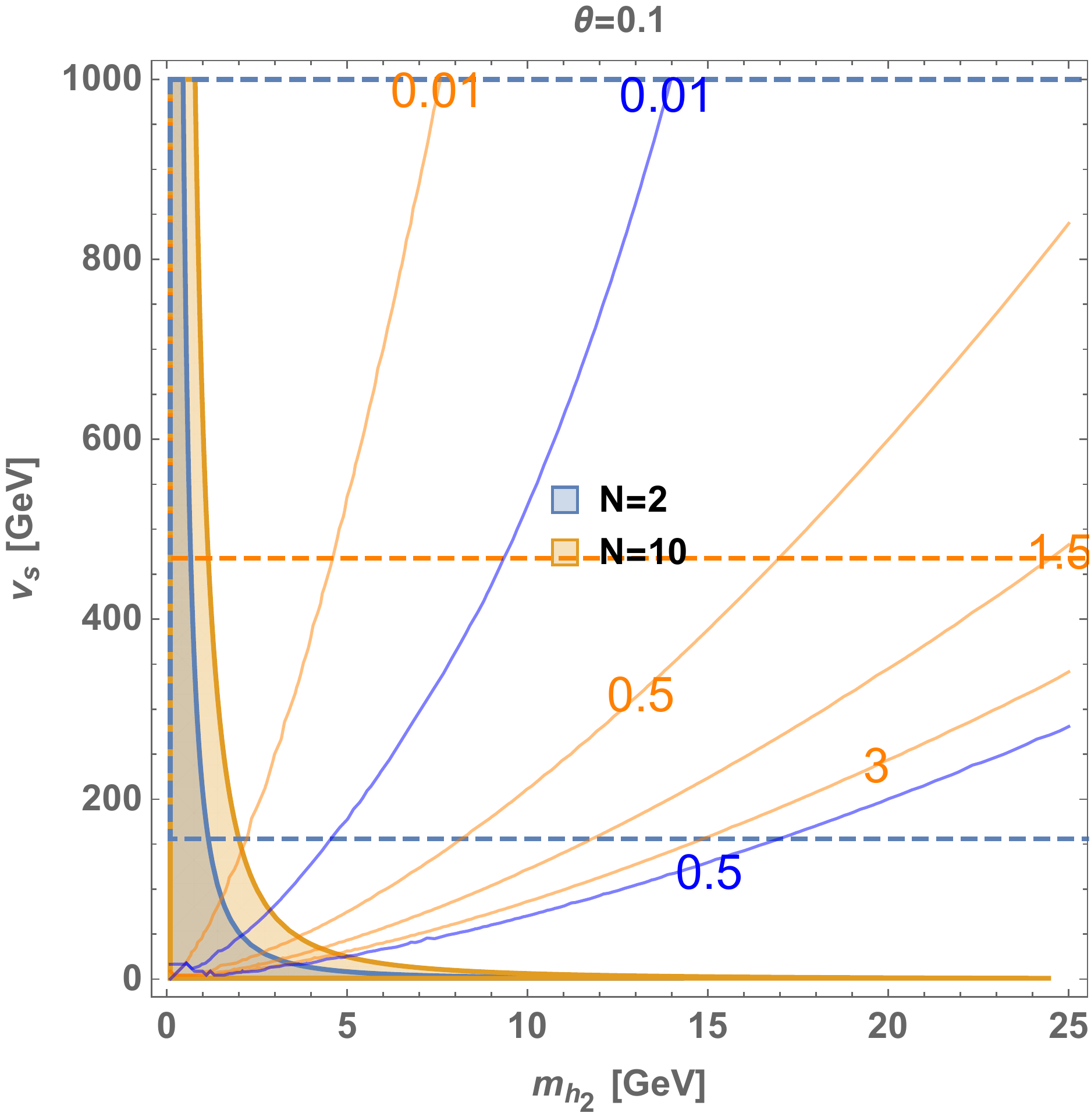}
\includegraphics[width=0.4 \textwidth]{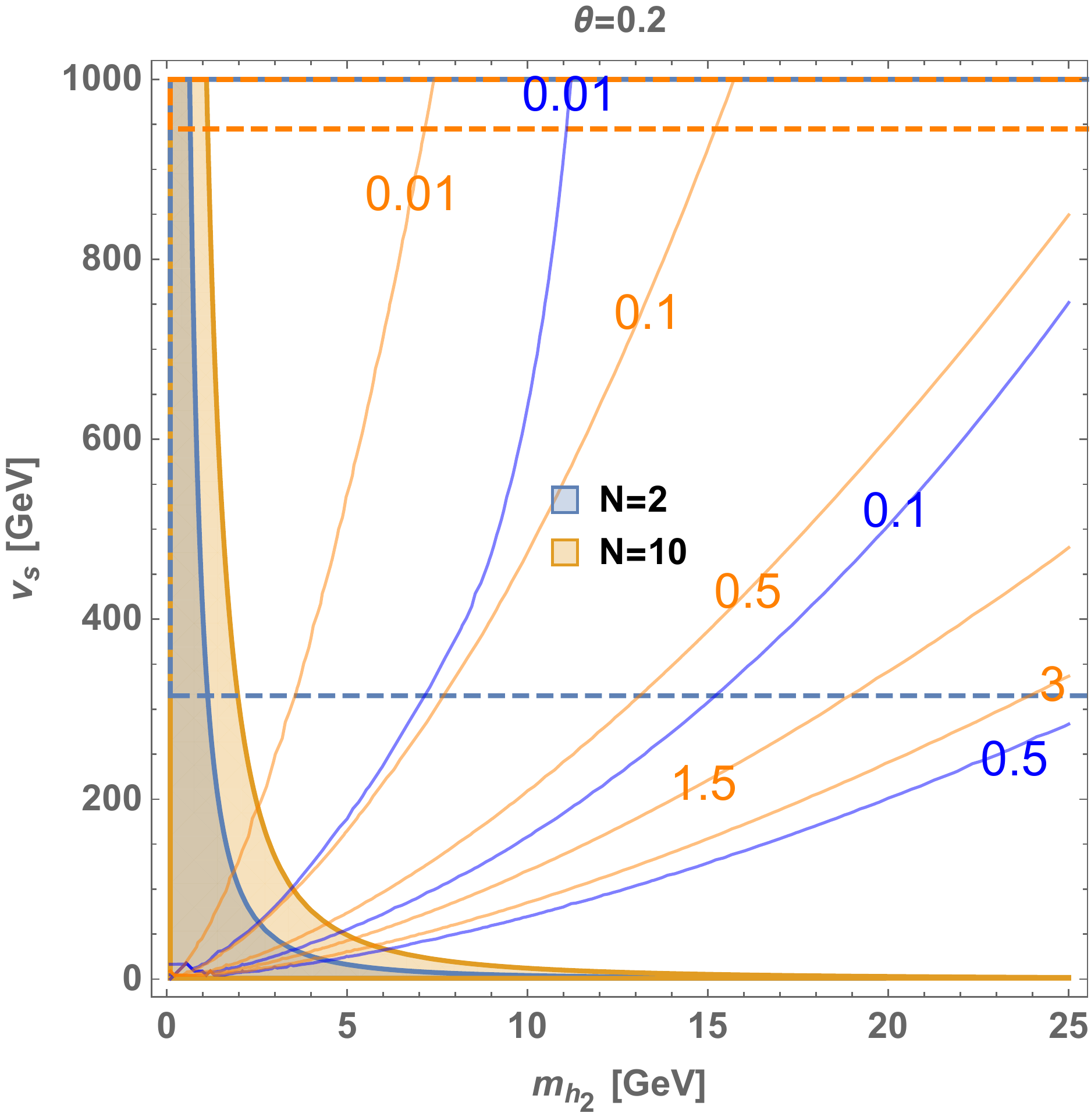}
\end{center}
\caption{Decouple conditions satisfied parameter regions in parameter spaces of $m_
{h_2}$ and $v_s$, regions in the dashed square frame are allowed by the $B_{inv}<0.34$~\cite{Khachatryan:2016vau}. The decay width of $h_2$ labeled on the blue and orange contours are shown in units of MeV.}\label{fig:decONN1sls}
\end{figure}

\begin{figure}[!htp]
\begin{center}
\includegraphics[width=0.4 \textwidth]{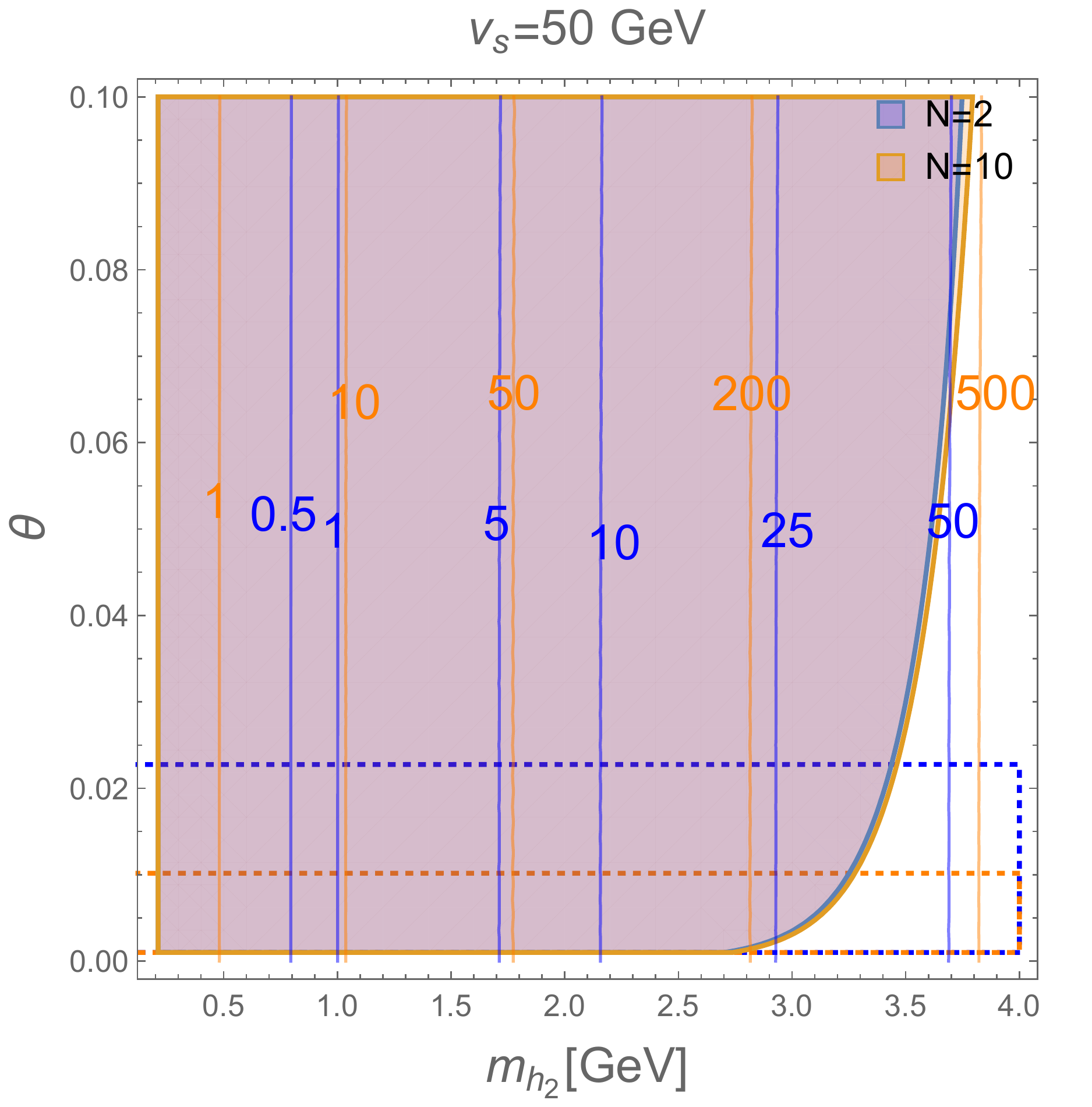}
\includegraphics[width=0.4 \textwidth]{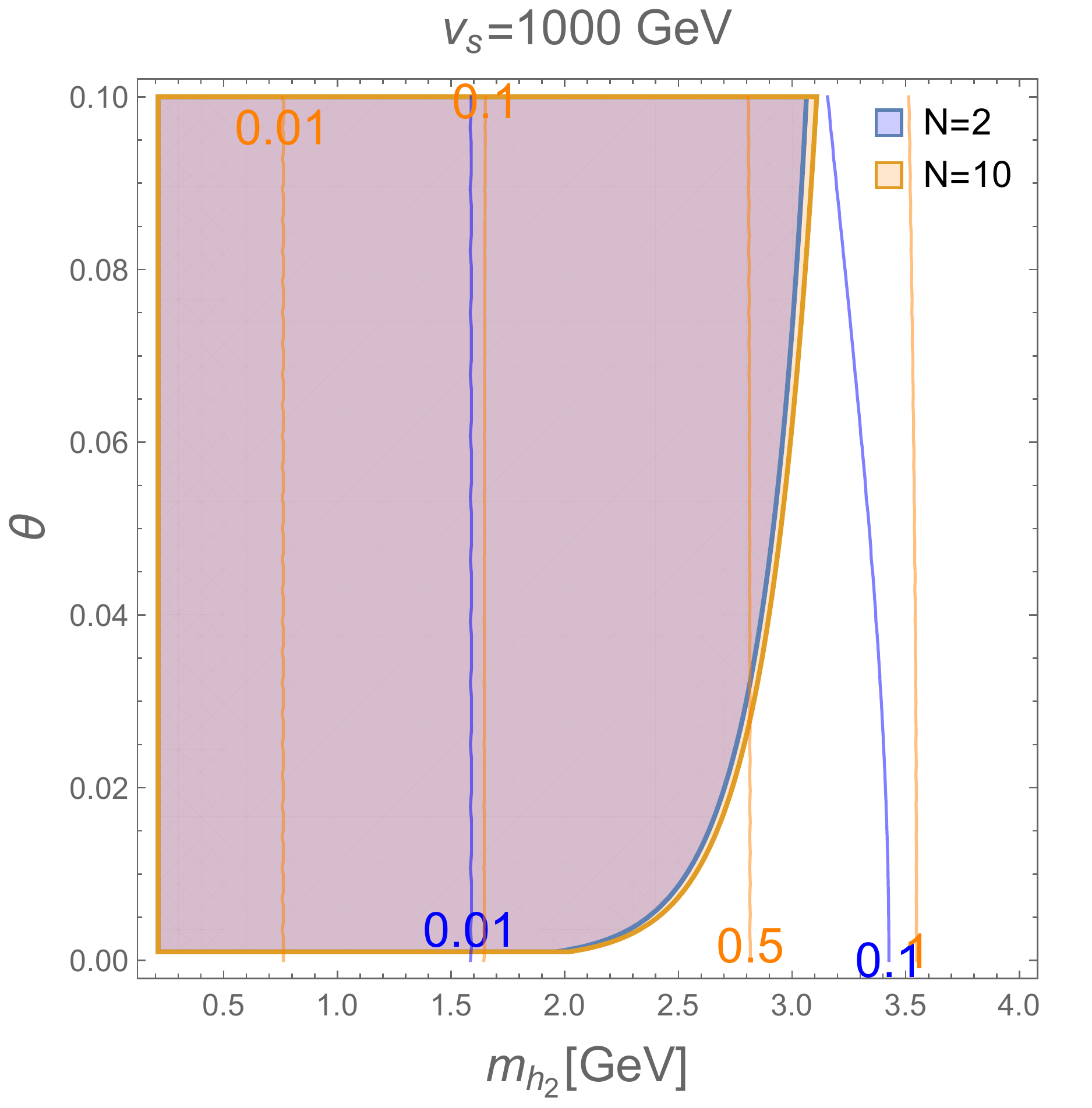}
\end{center}
\caption{Decouple conditions valid regions for $2 m_\mu<m_{h_2}<4$ GeV, regions in the dashed square frame are allowed by the $B_{inv}<0.34$~\cite{Khachatryan:2016vau}. The decay width of $h_2$( $\Gamma_{h_2}^{tot}$) labeled on the contours is shown in units of KeV.}\label{fig:decONN1th}
\end{figure}

\begin{figure}[!htp]
\begin{center}
\includegraphics[width=0.4 \textwidth]{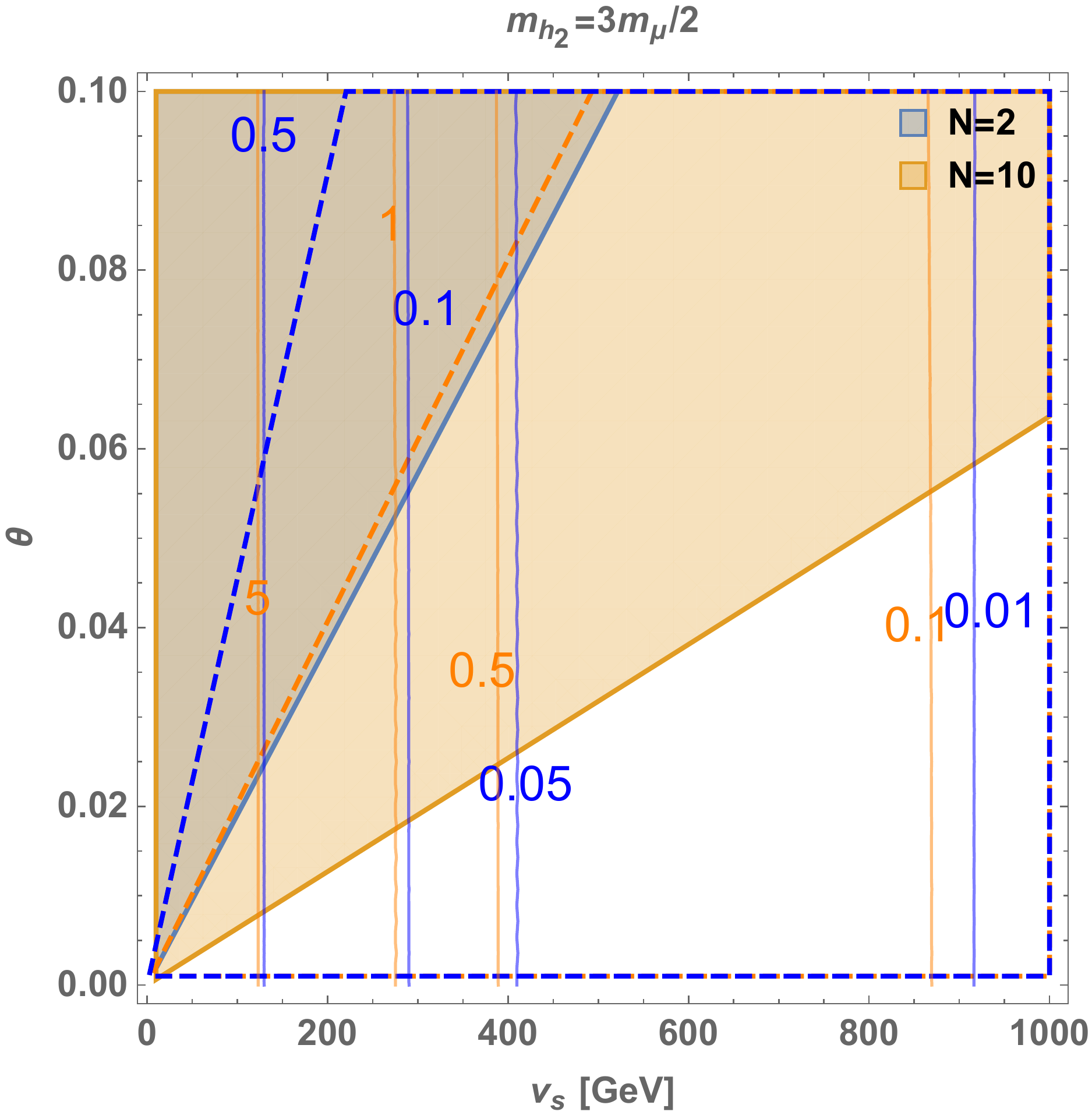}
\end{center}
\caption{Decouple conditions valid parameter regions for $m_{h_2}=3 m_\mu/2$, regions in the dashed square frame are allowed by the $B_{inv}<0.34$~\cite{Khachatryan:2016vau}. The decay width of $h_2$( $\Gamma_{h_2}^{tot}$) labeled on the contours is shown in units of KeV.}\label{fig:decONN1thvs}
\end{figure}

We study how the Goldstone decouples from the thermal bath and consider the possibility of the Goldstones contributing to the dark radiation following Ref.~\cite{Garcia-Cely:2013nin}. For the heavy Higgs contributions are typically small, one needs to focus on the light Higgs case alternatively, c.f., $m_{h_2}<2 m_{h_1}$.
 When the decay width $\Gamma_{h_2}\ll m_{h_2}$ 
in the small mass region of $m_{h_2}$, the cross-section of the Goldstone annihilating to $\mu^+\mu^-$ (as shown in~Fig.\ref{fig:goldstonann}), is given
by,
\begin{equation}
	\langle \sigma v \rangle_{s_{N-1}s_{N-1}\to \mu^{+} \mu^{-}}\;=\; \frac{\lambda_{hs}^{2}}{128\,\pi}\frac{m_{\mu}^{2}\,T^{4}}{m_{h}^{4}\,m_{h_2}^{4}}\,
	\int_{2\,m_{\mu}/T}^{\infty}\,w^{8}\,K_{1}(w)\,\text{d}w\,.
\end{equation}
 Which leads to the constraints on $v_s$ and $m_{h_2}$, as seen in Fig.~\ref{fig:decONN1sls}. The invisible decay of $h_1$ requires a small $\theta$ or a low magnitude of $N$.
 In the resonance enhanced region ($2 m_\mu<m_{h_2}<4$ GeV), using the narrow resonance conditions of $\Gamma_{h_2}\ll m_{h_2}$, one obtains,
\begin{eqnarray}
	\langle \sigma v \rangle_{s_{N-1}s_{N-1}\to \mu^{+} \mu^{-}}&=&
	 \frac{\lambda_{hs}^{2}}{256}\,\frac{m_{\mu}^{2}\,m_{h_2}^{6}}{T^{5}\,m_{h}^{4}\,\Gamma_{h_2}}\,
	 \left(1-\frac{4\,m_{\mu}^{2}}{m_{h_2}^{2}}\right)^{3/2}\,K_{1}(m_{h_2}/T) \label{Wcon2}\,.
\end{eqnarray}
Which set lower bounds on the mixing angle of $\theta$, see Fig.~\ref{fig:decONN1th}. For a small $v_s$, the invisible decay of the $h_1$
set the upper limits on $\theta$ depending on the number of $N$ as shown in the left panel of Fig.~\ref{fig:decONN1th}. For the case of $m_\mu<m_{h_2}<2m_\mu$, we have,
\begin{equation}
	\langle \sigma v \rangle_{s_{N-1}s_{N-1}\to \mu^{+} \mu^{-}}=\frac{\lambda_{hs}^{2}}{128\pi}\,\frac{m_{\mu}^{2}\,}{m_{h}^{4}}\,
	\int_{2 m_\mu/T}^\infty w^4 K_{1}(w) dw\,.\label{sinthetanalytic}
\end{equation}
Requiring the Goldstone bosons annihilation process contribute to the equivalent neutrino numbers, we obtain the bounds on mixing angle and the
$v_s$, see Fig.~\ref{fig:decONN1thvs}. A large N requires a small $\theta$ to meet the decoupling conditions.

\section{Conclusions and discussions}
\label{sec:conc}

In this work, we studied the slow-roll Higgs inflation and the possibility to realize a SFOEWPT with the N scalars extended SM.
The condition of the successful inflation and a SFOEWPT set a stringent bound on the number of the singlet scalars (or Goldstones) when the symmetry respected by the scalars is exact $O(N)$ symmetry (or the symmetry is broken to $O(N-1)$). The stability problem can easily be remedied up to the inflation scale with the assistance of the $N$ scalars that couple to the SM Higgs through the Higgs-portal interactions. Meanwhile, the perturbativity and the unitarity set severely upper bounds on the scalar quartic  couplings with the increasing of the energy scale, especially with the increasing of the number of $N$. The Higgs inflation valid parameter regions of scalar quartic  couplings diminish with the increase of the number $N$ mostly due to the perturbativity and the unitarity bounds.

The EWPOs severely constrain the parameters spaces of both $O(N)$ and $O(N\to N-1)$ scenarios. Further improvement on EWPOs constraints would restrict the $N$ more rigorous. 
For the $O(N)$ case, the number of $N$ validating the inflation is bounded to $N\leq3$, which make both one-step and two-step SFOEWPT unreachable. 
Though all of the $N$ scalars can serve as WIMP DM candidates, no way to expect the $N$-scalar WIMP DM can saturate the correct relic density here. This is because the masses of the N-scalars is of $\sim\mathcal{O}(1-10)$ TeV scale considering the future $e^+ e^-$ colliders (such as ILC, FCC-ee, and CEPC) bounds. Here, the freeze out happens earlier than the EWPT process, thus the only relevant DM annihilation process is $S_iS_i\to h h$ with $m_h(T\geq T_{fs})\sim 0$.

When the $O(N)$ symmetry is spontaneously broken to $O(N\to N-1)$, one obtains $N-1$ Goldstones, and the one extra Higgs.
Therefore, the invisible Higgs decay is very powerful to set the bound on the Goldstone number $N-1$ and the mixing angle $\theta$. With one moderate $\theta=0.2$ allowed by EWPO and Higgs precisions as well as invisible Higgs decay bounds set by LHC, we explore the possibility to realize Higgs inflation and a SFOEWPT through one-step and two-step types. In the parameter regions where a SFOEWPT can occur, the perturbativity problem appears at high scales might preclude the possibility to reach a successful inflation. In the parameter regions where one can obtain successful slow-roll Higgs inflation and a SFOEWPT, the triple Higgs couplings $\lambda_{h_2h_1h_1}$ and $\lambda_{h_1h_1h_1}$ increase with the increase of $\lambda_{hs}$. The decay widths of the two Higgses are not large enough to introduce significant interference effect in the resonant mass regions of Higgs pair productions.
The future ILC, FCC-ee, and CEPC are more powerful to test the mixing angle and the number of Goldstones $N-1$ in comparison with the LHC, and can probe the inflation and SFOEWPT valid parameter regions.
With the increase of the Goldstone number, we obtain a decrease parameter space of the scalar quartic couplings that can address successful inflation and a SFOEWPT. The gravitational wave signals search can tell the number of Goldstones for the one-step SFOEWPT, and the phase transition can be the two-step one if there is no relation with the number of Goldstones. We left the study to the future works. The dark radiations calculations indicate that Goldstones decouple from the thermal bath at mass ranges of a small $m_{h_2}$.

 One can expect the feature being explored in this work is general, the representation of a global or local symmetry respected by a hidden sector might be highly restricted if its contribution to the thermal potential and/or the inflationary scalar potential (directly or indirectly) is noticeable.

\section*{ACKNOWLEDGMENTS}

LGB thank Shufang Su, Zhen Liu, Wei Su, Jiayin Gu,
Honglei Li for communications and discussions on future collider searches of Higgs physics.
LGB is grateful for helpful discussions on dim-six operator and EWPT from Marcela Carena, Zhen Liu, and Marc Riembau.
LGB is also grateful for helpful discussions on the Electroweak phase transition in the $O(N)$ scalars model from Toshinori Matsui. LGB appreciates Hyun Min Lee for helpful discussion and comments on this paper.
This work is supported by the National Natural Science Foundation of China (under Grant No.11605016 and No.11847301), Basic Science Research Program through the National Research Foundation of Korea (NRF) funded by the Ministry of Education, Science and Technology (NRF-2016R1A2B4008759), and Korea Research Fellowship Program through the National Research Foundation of Korea (NRF) funded by the Ministry of Science and ICT (2017H1D3A1A01014046).

\newpage

\appendix
\section{ One-loop renormalization group equations}
\label{sec:RGEs}
Following the Ref.~\cite{Lerner:2009xg}, beta functions for the N-singlet scalars model are given bellow,
\begin{eqnarray}
\beta_{g_s}& = & \frac{g_s^3}{(4\pi)^2}(-7) + \frac{g_s^3}{(4\pi)^4}\left(\frac{11}{6}g^{\prime 2} + \frac{9}{2}g^2 - 26g_s^2 - 2x_h y_t^2\right)\;, \\
\beta_g& = & \frac{g^3}{(4\pi)^2}\left(-\frac{39-x_h}{12}\right) + \frac{g^3}{(4\pi)^4}\left(\frac{3}{2}g^{\prime 2} + \frac{35}{6}g^2 + 12g_s^2 - \frac{3}{2}x_h y_t^2\right)\;,\\
\beta_{g'}& = & \frac{g^{\prime 3}}{(4\pi)^2}\left(\frac{81+x_h}{12}\right) + \frac{g^{\prime 3}}{(4\pi)^4}\left(\frac{199}{18}g^{\prime  2} + \frac{9}{2}g^2 + \frac{44}{3}g_s^2 - \frac{17}{6}x_h y_t^2\right)\;, \\
\beta_{\lambda_h} &= & \frac{1}{(4\pi)^2}\left(6(1+3x_h^2)\lambda_h^2 - 6y_t^4 + \frac{3}{8}(2g^4 + (g^2+g^{\prime2})^2) + \lambda_h\gamma_h + \frac{Nx_s^2}{2}\lambda_{hs}^2\right)\;,\label{eq:beta}\\
\beta_{\lambda_{hs}} &= & \frac{\lambda_{sh}}{(4\pi)^2}\left(6(x_h^2+1)\lambda_h + 4x_h x_s\lambda_{hs} + 6 N x_s^2\lambda_s + 6y_t^2 - \frac{9}{2}g^2 - \frac{3}{2}g^{\prime2}\right)\;,\label{eq:beta2}\\
\beta_{\lambda_s} &= & \frac{1}{(4\pi)^2}(18 N x_s^2\lambda_s^2 + \frac{1}{2}(x_h^2+3)\lambda_{hs}^2)\;,\label{eq:beta3}\\
\beta_{y_t} &= & \frac{y_t}{(4\pi)^2}\left[ -\frac94g^2 - \frac{17}{12}g^{\prime 2} - 8g_s^2 + \frac{23+4{x_{S}}}{6}y_t^2 \right]\;.
\end{eqnarray}
with $\gamma_h=(-9g^2 - 3g^{\prime2} + 12y_t^2)$, $g$, $g'$ and $y_t$ are the standard model $SU(2)$, $U(1)$ and top-quark Yukawa couplings, and
\begin{eqnarray}
x_h & = & \frac{1 + \xi_h h^2/M_{\rm p}^2}{1 + \xi_h h^2/M_{\rm p}^2 + 6 \xi_h^2 h^2/M_{\rm p}^2},\\
x_{S} & = & \frac{1 + \xi_{S} {S}^2/M_{\rm p}^2}{1 + \xi_{S} {S}^2/M_{\rm p}^2 + 6 \xi_{S}^2 {S}^2/M_{\rm p}^2}.
\end{eqnarray}
Following Ref.~\cite{Lerner:2009xg,Aravind:2015xst}, The $\xi_S$ is set to zero at EW scale to ensure the kinematical mixing term canonical, and the $\xi_h$ is determined by CMB observations, see Eq.~\ref{eq:cmb}. The beta functions of $\xi_{h,S}$ can be found in Ref.~\cite{Lerner:2009xg,Aravind:2015xst}. The effects of $\xi_S$ here is negligible. The $\xi_h$ can lead to a tinny enhancement of the quartic couplings before the RG scale (here it is $h$) is comparable with the $M_{\rm p}$ since it's contribution to the beta functions is additive, see Eq.~(\ref{eq:beta},\ref{eq:beta2},\ref{eq:beta3}).

\section{Ingredients for electroweak phase transitions  }
\label{sec:EWPT}

The tree level scalar potential for $O(N)$ scenario is obtained directly from Eq.~\ref{eq:lag},
\begin{eqnarray}
V_0(h, S) &=& -\frac{\mu^2h^2}{2}+\frac{\lambda h^4}{4}+\frac{\mu_s^2 S^2}{2}+\frac{\lambda_s S^4}{4}+
\frac{\lambda_{hs}h^2S^2}{4}\;,
\end{eqnarray}
Here, we drop the subscript since all N directions are the same and we assume only one direction get VEV during the EWPT process.
The ``S" should be the ``s" for the $O(N\to N-1)$ scenario to indicate the possible symmetry breaking direction with other directions $s_i$ ($i=1,...,N-1$) do not get VEV during the EWPT process.
For the $O(N)$ scalar model, the one-loop Coleman-Weinberg potential for the scalar parts is given by
\begin{eqnarray}
V_{CW}(h,S,N) &= &\frac{1}{64\pi ^2}\Big[m_h^4(h,S)\bigg(log\frac{m_h^2(h,S)}{Q^2} - c_i\bigg)+ 2m_{G^+}^4(h,S)\bigg(log\frac{m_{G^+}^2(h,S)}{Q^2} - 3/2\bigg)\nonumber\\
&+&m_{G_0}^4(h,S)\bigg(log\frac{m_{G_0}^2(h,S)}{Q^2} - 3/2\bigg)+ N\,m_S^4(h,S)\bigg(log\frac{m_s^2(h,S)}{Q^2} - 3/2\bigg)\Big]\;.
\end{eqnarray}
If the $O(N)$ is broken to $O(N-1)$ we have,
\begin{eqnarray}
V_{CW}(h, s,N)&=&\frac{1}{64 \pi^2}\Big[ m_{h_1}^4(h,s)\bigg(log\frac{m_{h_1}^2(h,s)}{Q^2} -\frac{3}{2}\bigg) + m_{h_2}^4(h,s)\bigg(log\frac{m_{h_2}^2(h,s)}{Q^2} - \frac{3}{2} \bigg) \nonumber\\
&&+2m_{G^+}^4(h,s)\bigg(log\frac{m_{G^+}^2(h,s)}{Q^2} - \frac{3}{2} \bigg) + m_{G^0}^4(h,s)\bigg(log\frac{m_{G^0}^2(h,s)}{Q^2} - \frac{3}{2} \bigg)\nonumber\\
&&+(N-1)m_s^4(h,s)\bigg(log\frac{m_s^2(h,s)}{Q^2} - \frac{3}{2}\bigg)\Big] \;.
\end{eqnarray}
For other gauge bosons contributions and fermions contributions we refer to Ref.~\cite{Bernon:2017jgv}.
%
The running scale $Q$ is chosen to be $Q=246.22$ GeV in the numerical analysis process.
The field dependent masses are given as follows for both $O(N-1)$ and $O(N)$ cases (in this case one need do ``s"$\rightarrow$``S" ) ,
\begin{eqnarray}
m_{hs}(h,s)&=& \lambda_{hs}h s\;,\\
m_{h}^2(h,s)&=& 3\lambda{h^2}  - \mu^2 + \frac{\lambda_{hs}s^2}{2}\;,\\
m_{s}^2(h,s)&=& \frac{\lambda_{hs}h^2}{2} + \mu_s^2 + 3\lambda_s{s^2}\;,\\
m_{G^+}^2(h,s)&=& \lambda {h^2} - \mu^2 + \frac{\lambda_{hs}s^2}{2}\;,\\
m_{G^0}^2(h,s)&=& \lambda {h^2} - \mu^2 + \frac{\lambda_{hs}s^2}{2}\;,
\end{eqnarray}
and when $O(N)$ is broken to $O(N-1)$, we need to diagonalization the field dependent mass matrix of $M=\left((m_{h}^2,m_{hs}),(m_{hs},m_{s}^2)\right)$
to obtain the mass eigenvalue, i.e., $(m_{h_1}^2,m_{h_2}^2)$.
The finite temperature corrections to the effective potential at one-loop are given by~\cite{Dolan:1973qd},
\begin{eqnarray}
V_{T}(h, S,N, T)&=&\frac{T^4}{2\pi^2}\bigg[J_{B}\bigg(\frac{m_{h}^2(h,S, N, T)}{T^2}\bigg) +J_{B}\bigg(\frac{m_{G^0}^2(h, S,N, T)}{T^2}\bigg)+2J_{B}\bigg(\frac{m_{G^+}^2(h, S,N, T)}{T^2}\bigg)\nonumber\\
&&+ N J_{B}\bigg(\frac{m_{S}^2(h, S,N, T)}{T^2}\bigg) \bigg]\; ,
\end{eqnarray}
and
\begin{eqnarray}
V_{T}(h, s,N, T)&=&\frac{T^4}{2\pi^2}\bigg[J_{B}\bigg(\frac{m_{h_1}^2(h, s,N, T)}{T^2}\bigg) +J_{B}\bigg(\frac{m_{G^0}^2(h, s,N, T)}{T^2}\bigg)+2J_{B}\bigg(\frac{m_{G^+}^2(h, s,N, T)}{T^2}\bigg) \nonumber \\
&&+J_{B}\bigg(\frac{m_{h_2}^2(h,s, N, T)}{T^2}\bigg)+ (N - 1)J_{B}\bigg(\frac{m_{s}^2(h,s, N, T)}{T^2}\bigg) \bigg]\; ,
\end{eqnarray}
for $O(N)$ and $O(N-1)$ scenarios respectively. Where the functions $J_{B}(y)$ are
\begin{eqnarray}
J_{B}(y) = \int_0^\infty\, dx\, x^2\, \ln\left[1-{\rm exp}\left(-\sqrt{x^2+y}\right)\right],
\end{eqnarray}
In addition, the above integral $J_{B}$ can be expressed as a sum of the second kind modified Bessel functions $K_{2} (x)$,
\begin{eqnarray}
J_{B}(y) = \lim_{N \to +\infty} - \sum_{l=1}^{N} {(1)^{l}  y \over l^{2}} K_{2} (\sqrt{y} l)\;.
\end{eqnarray}

The thermal masses/corrections are given by,
\begin{eqnarray}
m_{h}^2(h, S,N, T)&=& m_{h_2}^2 +\frac{1}{16}T^2(g_1^2 +3g_2^2 + 4g_t^2)+ T^2(\frac{\lambda}{2} + \frac{N\lambda_{hs}}{12})\;,\\
m_{G^+}^2(h,S, N, T)&=&m_{G^+}^2 + \frac{1}{16}T^2(g_1^2 + 3g_2^2 +4g_t^2) +T^2(\frac{\lambda}{2} + \frac{N\lambda_{hs}}{12})\;,\\
m_{G^0}^2(h,S,N, T)&=&m_{G^0}^2 + \frac{1}{16}T^2(g_1^2 + 3g_2^2 + 4g_t^2) +T^2(\frac{\lambda}{2} +\frac{N\lambda_{hs}}{12})\;,\\
m_S^2(h, S,N, T)&=& m_S^2 +T^2\bigg(\frac{(N + 2)\lambda_s}{4} +\frac{\lambda_{hs}}{3}\bigg)\;,
\end{eqnarray}
for the $O(N)$ case, and for the $O(N \to N-1)$ case one needs to replace the ``S" by ``s" and replace the thermal mass of the Higgs fields by
\begin{eqnarray}
m_{h_1}^2(h, N, T)&=& m_{h_1}^2 +\frac{1}{16}T^2(g_1^2 +3g_2^2 + 4g_t^2)+ T^2(\frac{\lambda}{2} + \frac{N\;\lambda_{hs}}{12}),\\
m_{h_2}^2(h, N, T)&=&m_{h_2}^2+\frac{1}{16}T^2(g_1^2 + 3g_2^2 + 4g_t^2) +T^2(\frac{(N + 2)\lambda_{s}}{4} +\frac{\lambda_{hs}}{3})\; ,
\end{eqnarray}
%
Last but not least, the resummation of \textit{\ daisy} diagrams are also crucial for the evaluation of $v_C$ and $T_C$ with the finite temperature effective potential~\cite{Anderson:1991zb,Carrington:1991hz}, which is given by
\begin{eqnarray}
V_{\textit{\ daisy}}(h, S,N, T)&=&\frac{T}{12} \bigg[(m_{h}^3 - m_{h}^3(h,S, N, T)) +(m_{G^0}^3 -m_{G^0}^3(h, S,N, T)+2(m_{G^+}^3 -m_{G^+}^3(h, S,N, T)) \bigg]\nonumber \\
&& + N (m_S^3 -m_S^3(h, S,N, T)) \bigg]\;,
\end{eqnarray}
and
\begin{eqnarray}
V_{\textit{\ daisy}}(h, s,N, T)&=&\frac{T}{12} \bigg[\big(m_{h_1}^3 - m_{h_1}^3(h, s,N, T)\big) +3\big(m_{G^0}^3 -m_{G^0}^3(h,s, N, T)\big) +\big(m_{h_2}^3-m_{h_2}^3(h, s,N, T)\big)\nonumber \\
&&+ (N - 1)\big(m_s^3 -m_s^3(h,s, N, T)\big)\bigg]\;,
\end{eqnarray}
for $O(N)$ and $O(N\to N-1)$ cases.
Here, again, we list only the contributions of scalar contributions for $V_{T}$ and $V_{\textit{\ daisy}}$, the other particle fields contributions are the same as the SM, see Ref.~\cite{Carrington:1991hz,Bernon:2017jgv}.
It should be noted that, the counter terms can keep the VEVs of the potential from shift caused by the $V_{CW}$, we add that parts follow Ref.\cite{ Bernon:2017jgv}.

\section{Dark matter calculation approach of $O(N)$ case}
\label{ap:DM}

For SM Higgs pair final states,  the annihilation cross section is given by,
\bea \langle\sigma v_{rel}\rangle_{hh}& =& \frac{\lhs^{2}}{64 \pi m_s^{2}}
\left[ 1 + \frac{3 m_{h}^2}{\left(4 m_s^2 - m_h^2\right) } + \frac{2 \lhs v^2}{\left(m_h^2 - 2 m_s^2\right) }\right]^2  \times \left(1-\frac{m_{h}^{2}}{m_s^{2}}\right)^{1/2}\; ,\label{eq:annhpair}
\eea
the cross section for gauge boson final states are,
\bea
\langle\sigma v_{rel}\rangle_{WW}& =& 2\Bigg[1+\frac{1}{2}\left(1-\frac{2 m_s^{2}}{m_{W}^{2}}\right)^{2} \Bigg] \left(1-\frac{m_{W}^{2}}{m_s^{2}}\right)^{1/2} \times \frac{\lhs^{2} m_{W}^{4}}{8 \pi m_s^{2}\left(\left(4 m_s^{2} -m_{h}^{2}\right)^{2}+m_{h}^{2}\Gamma_{h}^{2}\right)  }
~,\eea
\bea
\langle\sigma v_{rel}\rangle_{ZZ}& =&  2\Bigg[1+\frac{1}{2} \left(1-\frac{2 m_S^{2}}{m_{Z}^{2}} \right)^{2}\Bigg] \left(1-\frac{m_{Z}^{2}}{m_S^{2}}\right)^{1/2}  \times \frac{\lhs^{2} m_{Z}^{4}}{16 \pi m_S^{2} \left(\left(4 m_S^{2} -m_{h}^{2}\right)^{2}+m_{h}^{2}\Gamma_{h}^{2}\right)  }
\eea
and the fermion pair final states cross section is given by,
\bea  \langle \sigma v_{rel}\rangle_{\overline{f}f} =\frac{m_{W}^{2}}{\pi g^{2}} \frac{\lambda_{f}^{2}
\lhs^{2} }{\left(4m_S^{2}-m_{h}^{2}\right)^{2}+m_{h}^{2}\Gamma_{h}^{2} } \Bigg(1-\frac{m_{f}^{2}}{m_S^{2}}\Bigg)^{3/2}\;.
~.\eea
The formula of spin independent cross section is given by~\cite{Cline:2013gha}
\begin{equation}
\sigma_{SI}^{S}=\lambda_{hs}^2\frac{f_N^{2}}{4\pi}\left(\frac{m_N m_{S}}{(m_N+m_{S})}\right)^{2}\frac{m_N^2}{m_H^4 m_{S}^2}\;,
\label{eq:sip}
\end{equation}
\noindent where $m_{N}=0.946$ GeV is the neutron mass and $m_H=126$ GeV is the SM-Higgs mass.
\noindent The strengths of the hadronic matrix elements, $f_N=0.35$.
The dark matter direct detection constrains the dark matter masses and the quartic Higgs-DM couplings after taking into account the rescale effects supposing the evaluated dark matter relic density will not oversaturate the DM relic abundance,
\bea
\sigma_{SI}=\sigma_{SI}^{S_i}\times\Sum_{i=1,...,N}\frac{\Omega^{S_i} h^2}{\Omega_{DM} h^2} \;.\label{eq:dd}
\eea

\end{document}